\documentclass[12pt]{article}
\pdfoutput=1

\usepackage{amsmath,amssymb,amsfonts,epsfig,cite,setspace,bigstrut,framed,eufrak,float}
\usepackage[all]{xy}
\usepackage{color}
\usepackage{pifont}
 \usepackage{ulem}

\makeatletter \@addtoreset{equation}{section} \makeatother
\renewcommand{\theequation}{\thesection.\arabic{equation}}

\begin{document}

\vskip 0.25in

\newcommand{\todo}[1]{{\bf\color{blue} !! #1 !!}\marginpar{\color{blue}$\Longleftarrow$}}
\newcommand{\nn}{\nonumber}
\newcommand{\comment}[1]{}
\newcommand\T{\rule{0pt}{2.6ex}}
\newcommand\B{\rule[-1.2ex]{0pt}{0pt}}

\newcommand{\cG}{{\cal G}}
\newcommand{\cV}{{\cal V}}
\newcommand{\cP}{{\cal P}}
\newcommand{\cL}{{\cal L}}
\newcommand{\cO}{{\cal O}}
\newcommand{\cI}{{\cal I}}
\newcommand{\cM}{{\cal M}}
\newcommand{\cW}{{\cal W}}
\newcommand{\cN}{{\cal N}}
\newcommand{\cR}{{\cal R}}
\newcommand{\cH}{{\cal H}}
\newcommand{\cK}{{\cal K}}
\newcommand{\cT}{{\cal T}}
\newcommand{\cZ}{{\cal Z}}
\newcommand{\cQ}{{\cal Q}}
\newcommand{\cB}{{\cal B}}
\newcommand{\cC}{{\cal C}}
\newcommand{\cD}{{\cal D}}
\newcommand{\cE}{{\cal E}}
\newcommand{\cF}{{\cal F}}
\newcommand{\cA}{{\cal A}}
\newcommand{\cX}{{\cal X}}
\newcommand{\IA}{\mathbb{A}}
\newcommand{\IP}{\mathbb{P}}
\newcommand{\IQ}{\mathbb{Q}}
\newcommand{\IH}{\mathbb{H}}
\newcommand{\IR}{\mathbb{R}}
\newcommand{\IC}{\mathbb{C}}
\newcommand{\IF}{\mathbb{F}}
\newcommand{\IS}{\mathbb{S}}
\newcommand{\IV}{\mathbb{V}}
\newcommand{\II}{\mathbb{I}}
\newcommand{\IT}{\mathbb{T}}
\newcommand{\IW}{\mathbb{W}}
\newcommand{\IZ}{\mathbb{Z}}
\newcommand{\re}{{\rm Re}}
\newcommand{\im}{{\rm Im}}
\newcommand{\tr}{\mathop{\rm Tr}}
\newcommand{\ch}{{\rm ch}}
\newcommand{\rk}{{\rm rk}}
\newcommand{\ext}{{\rm Ext}}
\newcommand{\bi}{\begin{itemize}}
\newcommand{\ei}{\end{itemize}}
\newcommand{\beq}{\begin{equation}}
\newcommand{\eeq}{\end{equation}}
\newcommand{\bea}{\begin{eqnarray}}
\newcommand{\eea}{\end{eqnarray}}
\newcommand{\ba}{\begin{array}}
\newcommand{\ea}{\end{array}}

\newcommand{\CN}{{\cal N}}
\newcommand{\y}{{\mathbf y}}
\newcommand{\z}{{\mathbf z}}
\newcommand{\C}{\mathbb C}\newcommand{\R}{\mathbb R}
\newcommand{\CA}{\mathbb A}
\newcommand{\CP}{\mathbb P}
\newcommand{\tmat}[1]{{\tiny \left(\begin{matrix} #1 \end{matrix}\right)}}
\newcommand{\mat}[1]{\left(\begin{matrix} #1 \end{matrix}\right)}
\newcommand{\diff}[2]{\frac{\partial #1}{\partial #2}}
\newcommand{\gen}[1]{\langle #1 \rangle}

\newcommand{\half}{\frac{1}{2}}

\newtheorem{theorem}{\bf THEOREM}
\newtheorem{proposition}{\bf PROPOSITION}
\newtheorem{observation}{\bf OBSERVATION}

\def\theequation{\thesection.\arabic{equation}}
\newcommand{\setall}{
	\setcounter{equation}{0}
}
\renewcommand{\thefootnote}{\fnsymbol{footnote}}

\begin{titlepage}
\vfill
\begin{flushright}
{\tt\normalsize KIAS-P19053}\\

\end{flushright}
\vfill

\begin{center}
{\Large\bf Generalized Euler Index, \\ \vskip 3mm
Holonomy Saddles,  and Wall-Crossing }

\vskip 1.5cm
Dongwook Ghim$^{1}$\footnote{\tt dghim@kias.re.kr},
Chiung Hwang$^{2}$\footnote{\tt chiung.hwang@unimib.it}
and Piljin Yi$^{1}$\footnote{\tt piljin@kias.re.kr}
\vskip 8mm

$^{1}${\it School of Physics,
Korea Institute for Advanced Study, Seoul 02455, Korea \\}
\vskip 2mm
$^{2}${\it Dipartimento di Fisica, Universit\`a di Milano-Bicocca \& INFN, Sezione di Milano-Bicocca, I-20126 Milano, Italy}

\end{center}
\vfill

\begin{abstract}
We formulate Witten index problems for theories with two
supercharges in a Majorana doublet, as in $d=3$ $\cN=1$
theories and dimensional reduction thereof.
Regardless of spacetime dimensions, the wall-crossing occurs
generically, in the parameter space of the real superpotential
$W$. With scalar multiplets only, the path integral reduces to
a Gaussian one in terms of $dW$, with a winding number
interpretation, and allows an in-depth study of the wall-crossing.
After discussing the connection to well-known mathematical approaches
such as the Morse theory, we move on to Abelian gauge theories.
Even though the index theorem for the latter is a little more
involved, we again reduce it to winding number countings of the neutral
part of $dW$. The holonomy saddle plays key roles for both
dimensions and  also in relating indices across dimensions.

\end{abstract}

\vfill
\end{titlepage}

\tableofcontents
\renewcommand{\thefootnote}{\#\arabic{footnote}}
\setcounter{footnote}{0}
\vskip 2cm

\section{Motivation}

Since the appearance of the index theorem \cite{Atiyah:1963zz}
and its adaptation in physics \cite{Witten:1982df}, such
topological characterizations became important tools of the trade
for studying supersymmetric theories. The path integral
reformulation of the index theorem started with the non-linear
sigma model by Alvarez-Gaume \cite{AlvarezGaume:1983at}, hence suitable for
geometric problems. Since then, numerous generalizations to
gauge theories have appeared in literature, and  perhaps the
most recent and the most systematic descendent are
Refs.~\cite{Benini:2013nda, Benini:2013xpa} for $d=2$
and Ref.~\cite{Hori:2014tda} for $d=1$ gauge theories.
The former computes the elliptic genera while the latter
computes $\chi_y$ genus, when the theory flows to a compact
nonlinear sigma models, and may be considered a sweeping
generalization of the Atiyah-Singer index theorems and
Alvarez-Gaume's path integral reformulation thereof,
to a certain class of gauged dynamics with complex supersymmetries.

These recent computations of topological quantities
often rely on the so-called localization \cite{Moore:1998et,Pestun:2007rz}.
Although one often attributes the resulting simplification
to the presence of a BRST operator constructed out of selected
supercharges, the real mechanism behind the simplification
can be traced to the massive deformation of the theories, which
of course is also loosely connected to the choice of the
BRST operator. Various
flavor chemical potentials as well as flavor masses are
typical such, and also immensely helpful is the $R$-symmetry
chemical potential, which combined with angular momentum
is capable for preserving some supersymmetry. This reduces
the problem to a functional Gaussian integral,
leaving behind a finite number of integrals, often
parameterizing the Cartan directions.
Another batch of the simplification arises from the
holomorphicity\cite{Seiberg:1993vc}, namely how the
supersymmetric partition functions depend on
external parameters in holomorphic combinations and also
how the final phase of the finite integrals reduces to
a contour integral of a meromorphic function.

On the other hand, theories with supersymmetries in a
Majorana doublet in $d=3$ sense, are qualitatively different.
There is no $R$-symmetry in $d=3$ $\cN=1$, and although
the dimensional reduction to $d=1$ creates one $U(1)_R$,
one cannot introduce its chemical potential because both
supercharges transform under it. We could try to turn on
flavor chemical potentials, as one would have done for
$d=2$ $\cN=(0,2)$ theories. The flavor chemical potentials
do invoke a fixed point theorem \`a la Lefschetz \cite{Atiyah:1967fx},
but, as we will see later in this note, tend to disappear
from the final form of the twisted partition functions.
No notion of the refinement of the index seems possible
with the supersymmetry in Majorana doublet.

For geometric theories, the absence of the refinement can be
understood also from how $d=2$ $\cN =(1,1)$ does not impose
any constraint on the target holonomy. With $\cN=(0,2)$,
for example, the target would be complex if the theory
flows down to a nonlinear sigma model \cite{Witten:1993yc}, whereby
the cohomology decomposes into a Hodge diamond.
No such further decomposition of the cohomology is implied
with $\cN=(1,1)$, so, for example, the Euler number is the
only topological quantity that can be constructed out of
the cohomology.

This brings us to the question: While we are all familiar with
how the Euler number is defined and computed via cohomology when
the target is geometric and compact, what would be its analog
for general supersymmetric theories? In other words, what are
the general features of the Witten index when the supersymmetry
is in a Majorana doublet. The subject is hardly new, as this
was one of the first class of problems studied by Alvarez-Gaume  \cite{AlvarezGaume:1983at}
in his path integral reformulation of Atiyah-Singer index theorem \cite{Atiyah:1963zz}.
For the Euler index, Witten also expanded the discussion into
the Morse theory\cite{Witten:1982im}, again mostly in the
context of compact non-linear sigma model. In this note,
we wish to explore a generalized notion of the Euler index
for $d=3$  $\cN=1$ theories, and for dimensional reductions
thereof, by expanding these pioneering works.

One reason why we come back to this system is the wall-crossing,
which turns out to be quite prevalent, and surprisingly so
regardless of the spacetime dimensions. Recall how the wall-crossing
happens in the more familiar $d=1$ $\cN=4$ \cite{Hori:2014tda}
but not in its uplift to $d=2$ $\cN=(2,2)$\cite{Witten:1993yc}.
The $d=1$ refined Witten index may experience a sudden jump
as a Fayet-Iliopoulos constant $\zeta$ changes the sign
\cite{Hori:2014tda,Cordova:2014oxa,Hwang:2014uwa}, yet the
elliptic genus, its $d=2$ analog, does not. This difference
can be understood in many different perspectives, but one
such is to realize that the phenomenon arises from gapless
Coulombic flat direction. The ground state sector of the
Coulombic side is notoriously dimension-dependent, yielding
such differences between $d=1$ and $d=2$.

On the other hand, with the current, reduced supersymmetry
content, the FI-term becomes a term in the superpotential.
The would-be Coulombic direction, which is responsible for
the wall-crossing, is also spanned by a scalar multiplet
\cite{Gates:1983nr}. The familiar wall-crossing phenomena
must be now attributed to asymptotics along the matter side;
as such the strong dependence on spacetime dimensions
is no longer necessary.  The wall-crossing occurs in the
parameter space of the real superpotential. The simplest
example is a theory with a single real scalar field $X$
with a quadratic superpotential $W=mX^2/2$, whose index
proves to be
\bea
\cI={\rm sgn}(m) \,,
\eea
and the wall of marginal stability occurs at $m=0$. This
already hints at how rampant the wall-crossing phenomena would be
with this reduced supersymmetry.

Although the compact (geometric) theories admit topologically
robust Euler numbers, computable by several well-known methods,
study of the Witten index for theories that are more familiar to
physics, such as the above free massive thoery, is surprisingly
sparse. Furthermore, a systematic study of  wall-crossing is much needed.
In this note, we will resort to the heat kernel expansion \cite{heat}
and formulate the twisted partition functions and Witten indices,
with emphasis on $d=1$ and on $d=3$. For the latter
we restrict our attention to massive theories with discrete
classical vacua along the matter side, since for theories with
nontrivial classical moduli space of vacua the $d=3$ Witten index
itself would be ill-defined. While the latter might be sometimes
circumvented by introducing flavor chemical potentials, we will
not pursue this here as this in principle represents a heavy-handed
deformation of the dynamics \cite{Lee:2016dbm}.

When the theory has
scalar multiplets only and a superpotential that admits
discrete vacua only, the $d=3$ Witten index is easily
computed by $d=1$ Witten index of the dimensionally reduced
theory, as the former can be
reduced to $d=1$ by putting theory on spatial $\IT^2$.
These cases will be discussed in Section 2, where we
reduce the relevant path integral down to a set of
ordinary Gaussian integrals. The result connects the
Witten index, or a generalized Euler index, to various
topological interpretations such as the winding number associated
with the superpotential, the Morse theory,  and the
Lefschetz fixed point theorem. Our main concern in
Section 2, however, will be the wall-crossings and
also finding a formulation that will generalize naturally
to the gauged dynamics in the subsequent Sections.

With the gauge sector present, vacuum physics is more
sensitive to the dimensions. Section 3 deals with $d=1$
$SO(2)$ theories, where we again reduce the path integral
to a purely Gaussian form, very similar to the matter-only
cases in Section 2, although the indices, or the twisted
partition functions, have even simpler form than the
matter only cases. The gauge sector often offers additional
asymptotic flat directions, which can in principle cause
a new type of wall-crossing, but we show that in fact
this does not happen. All wall-crossings arise entirely due to
the asymptotics of the $d=1$ $\cN=2$ matter sectors, instead.
This unexpected behavior can be explained by the mere logical
possibility of uplifting the theory to a $d=3$ Chern-Simons
theory, as we will discuss at the end of Section 3.

Whenever one considers a supersymmetric partition function
of a gauge theory on a circle, the holonomy saddles
\cite{Hwang:2017nop,Hwang:2018riu} enter
the story. For $SO(2)$ theories, this concept is related
to how a gauge-symmetry breaking by charge $q>1$ leaves
behind the $\IZ_q$ discrete subgroup, and how the vacua would
be labeled by the discrete holonomies valued in such $\IZ_q$.
In the Hamiltonian view, the relevant holonomies would be
those associated with spatial circles in $\IT^{d-1}$.
In path integrals, however, the $\IZ_q$ holonomies along
the Euclidean time circle also enter, which multiply the
above Gaussian integrals by a factor $|q|^d$ for theories
put on $\mathbb T^d$. Throughout Sections 3 and 4,
this factor will play a crucial role.

In Section 4, we move on to computation of Witten indices
of $d=3$ $\cN=1$ $SO(2)$ Chern-Simons-Matter theories.
Note that we should expect immediate and qualitative differences
between $d=3$ $\cN=1$ and the better studied $d=3$ $\cN=2$ \cite{Aharony:1997bx,Intriligator:2013lca}.
For example, we will see proliferation of wall-crossing in
$d=3$ $\cN=1$, while it is well known that the wall-crossing
is absent for the latter \cite{Intriligator:2013lca}. As
we emphasized earlier, this can be easily understood from
its $d=1$ cousins and how the wall-crossing enters the story
via the (real) superpotential.

In this last section, we will again manage to reduce the $d=3$ path
integral such that the entire Witten index is expressed
again in terms of certain winding numbers associated with the
superpotential $W$. In particular, much of these winding numbers
can be attributed to those of neutral matter multiplets on
par with the results in Section 2. This in particular
means that if the gauge theory has no neutral scalar
multiplet, the index takes a rather universal form.
We formulate an index theorem complete with a computational
routine and close with several examples, as well as a simple
check of proposed $d=3$ $\cN=1$ dualities.

We close this summary with a comment on the notation. The
main purpose of this note is to reduce path integrals to
ordinary Gaussian integrals in the end. This involves a scaling
limit of the Euclidean time span $\beta\rightarrow 0$, and
an accompanying rescaling of bosonic fields and the superpotentials. We will
need to talk about the original field variables, the zero modes,
and also $\beta$-rescaled version of the latter. The first two
would be denoted by capital alphabet, such as $X$, while the last
by the lowercase $x$. However, in the later part of the note, we
will take such rescalings for granted and be less strict about
the distinction between the original zero mode $X$ and the
rescaled one $x$. As such,  $X$'s would be used in place of
$x$'s in many of the middle steps. We made sure to retain
$x$ for the final integral formulae, however, as a reminder of
$\beta$-rescaling performed to get there.

\subsection{Theories with Two-Component Majorana Supercharges}

We consider a non-chiral supersymmetry, responsible  for $d=3$ $\cN=1$ theories
or  for $d=2$ $\cN=(1,1)$ theories \cite{Gates:1983nr}. As for $d=1$, there is a unique superalgebra with two real supercharges, yet one can construct different types of supersymmetric theories \cite{Coles:1990hr}. Two notable
examples have well-recognizable geometric interpretations, sometimes 
referred to as $\cN=2a$ and $\cN=2b$ \cite{Gibbons:1997iy}: These are related
to dimensional reductions of $d=2$ $\cN=(1,1)$ and $d=2$ $\cN=(0,2)$, 
respectively, and, as such, constructed from very different forms of
supermultiplets. For instance, the scalar multiplet in the former has, 
on shell, two real fermions per a real boson while its counterpart 
in the latter has a complex fermion matching a complex boson. We will 
develop techniques for computing supersymmetric partition function
for the former class, to be called $d=1$ $\cN=2$ theories in this note 
for simplicity,\footnote{We refer readers to Ref.~\cite{Hori:2014tda} 
for a thorough treatment of $d=1$ supersymmetric partition functions 
with the latter ``complex" realization.} and build
back up to $d=2$ $\cN=(1,1)$  and to $d=3$ $\cN=1$.

The supercharges for the theories we wish to study in this note
belong to a Majorana doublet, with
\bea
\gamma^0=i\sigma_2\ , \quad \gamma^1= \sigma_3, \quad \gamma^2= \sigma_1 \,.
\eea
For the Euclidean signature with which all computations here will proceed,
the fermions are pseudo-real, and a convenient basis is
\bea
\gamma^1= \sigma_1\ , \quad \gamma^2= \sigma_2\ ,\quad \gamma^3=\sigma_3\ .
\eea
We can deduce the supermultiplet content of $\cN=1$ from those of
the more familiar $\cN=2$ supersymmetries by halving the supermultiplet.

The basic scalar supermultiplet is literally a  half of the more familiar
$d=3$ $\cN=2$ chiral
multiplet,
\bea
\Phi= X + \theta^a\psi_a - \theta^2 F \,,
\eea
where the spinor indices are raised and lowered, respectively,
by the left and the right multiplication by $C\equiv i\sigma_2$,
and we also define
\bea
\theta^2=-\frac12\,\theta^a\theta_a \,.
\eea
With the supercovariant derivative
\bea
D_a\equiv \partial_a+i\theta^b\,\partial_{ab}
\eea
where
\bea
\partial_{ab}\equiv \gamma^i_{\;ab}\,\frac{\partial}{\partial x^i} \,,
\eea
the scalar multiplet has the following superspace
Lagrangian,
\bea
\cL&=&\frac14\int d^2\theta \;\left(D^a\Phi D_a\Phi\right) + \int d^2\theta\; W(\Phi) \,,
\eea
which produces upon the Grassmanian integral,
\bea
\frac12\partial_i X \partial^i X + \frac12 \psi^a i\partial_a^{\;\;b}\psi_b
+\frac12 (\partial_X W)^2+ \psi^2\partial_X^2W(X) \,,
\eea
where $W$ is a real superpotential. With more than one
scalar multiplet $\Phi^I$, we have
\bea
\sum_\mu \left(\frac12\partial_i X^\mu \partial^i X^\mu + \frac12 \psi^a_\mu i\partial_a^{\;\;b}\psi_b^\mu
+\frac12 (\partial_\mu W)^2\right)+ \sum_{\mu,\nu} \psi^\mu\psi^\nu\;\partial_\mu\partial_\nu W(X) \,,
\eea
where $\partial_\mu$ stands for the derivative with respect to a real scalar $X^\mu$.
For matter multiplets, note that we literally divided an $\cN=2$ chiral into two equal parts
and end up with real scalar multiplets with Majorana fermions.

For vectors, on the other hand, the $\cN=2$ vector is split into
an $\cN=1$ vector $(A_i,\lambda)$ and an $\cN=1$ real scalar multiplet
$(\Sigma, \Lambda, D)$. The real auxiliary $D$ field belongs to the
latter, so one qualitatively new feature is that the would-be the
Fayet-Iliopoulos term is actually part of the real superpotential. This
fact is rather significant since the FI constant often plays
a central role in determining infrared behavior of supersymmetric
gauge theories, and in particular controls wall-crossings for
$d=1$ theories. As such, the role of superpotential will prove
to be central when we consider the Witten  index and wall-crossing
for theories with a supercharge in a two-component Majorana spinor.

We refer the readers to  \cite{Gates:1983nr} for a complete detail
of the vector supermultiplet and here merely write down a bare-bone
feature of the Lagrangian. In addition to the standard kinetic term
for $A$ and $\lambda$, a unit-charged complex scalar multiplet
$(\Phi,\Psi,F)$ would couple to an $SO(2)$ vector as follows, for example,
\bea
&&\frac12\left(\partial_i X \partial^i X +  \psi^a i\partial_a^{\;\;b}\psi_b\right)\cr\cr
&\Rightarrow& -\bar \Phi (\partial-iA)^2 \Phi
+ i\bar\Psi^a (\partial_a^{\;\;b}-iA_a^{\;\;b})\Psi_b +i( \Psi^a\lambda_a \bar \Phi -\bar \Psi^a\lambda_a \Phi) \,,
\eea
where we now took the matter multiplets to be complex,
$\Phi=(X^1+iX^2)/\sqrt{2}$, etc.  In Section 2,
Abelian vector multiplets will make a brief appearance as
external objects that make the theory massive. In Sections 3 and 4,
on the other hand, we will consider gauge theories where
the vector multiplet becomes dynamical.

\section{Index and Wall-Crossing: Matter Only}
\label{sec:matter}

In this section, we will explore the usual heat kernel expansion
\cite{heat} further, with an aim at handling theories
with wall-crossing phenomena or continuum sector or both.
We start with $d=1$ $\cN=2$ theories with two real supercharges \cite{AlvarezGaume:1983at}.
The canonical Hamiltonian for scalar multiplets is
\bea
\cH\equiv \sum_{\mu =1}^N \left(\frac12\pi_\mu \pi_\mu +\frac12 (\partial_\mu W)^2\right)+ \sum_{\mu,\nu}^N \psi^\mu\psi^\nu\;\partial_\mu\partial_\nu W(X) \,,
\eea
with the canonical conjugate momenta $\pi_\mu $ to $X^\mu$ and
the two sets of Grassman variables each spanning a Clifford algebra,
\bea\label{Clifford}
\{\psi^\mu_a,\psi^\nu_b\}=\delta^{\mu\nu}\delta_{ab} \,.
\eea
The quantity we wish to compute is
the twisted partition function,
\bea
{\rm Tr}\; (-1)^\cF e^{-\beta\,\cH}
\eea
with two interesting limits,
\bea
\begin{split}
\cI&=&\lim_{\beta\rightarrow +\infty}{\rm Tr}\; (-1)^\cF e^{-\beta\,\cH} \,, \\
\Omega&=&\lim_{\beta\rightarrow 0^+}{\rm Tr}\; (-1)^\cF e^{-\beta\,\cH} \,.
\end{split}
\eea
The former is the desired Witten index \cite{Witten:1982df}, while the latter, so-called
bulk part, is the one that  can be compute systematically.\footnote{As
has been emphasized elsewhere \cite{Lee:2016dbm}, the so-called localization
method for supersymmetric partition functions in the presence of a circle
often computes the analog of $\Omega$, even though
there it may appear that no particular limit of $\beta$ is
taken. Since $\beta$ is dimensionful, it is really a
dimensionless combination
of $\beta$ with some other scaling parameter that is
implicitly taken to zero in such a localized path integral.
Insertion of the chemical potential can retain some
memory of finite circle size \cite{Hwang:2018riu}, yet the parameter sitting in front
of $H$ or more precisely $\cQ^2$ is effectively taken to zero
at the end of the localization process.}

The trace over the Hilbert space reduces to the following
integral,
\bea
{\rm Tr}\; (-1)^\cF e^{-\beta\cH}=\int d^N X\;
{\rm tr}_\psi\left((-1)^\cF\langle X\vert
\;e^{-\beta\, \cH}\;\vert X\rangle\right) \,,
\eea
where the ${\rm tr}_\psi$ means the trace over the spinor
representation of the algebra (\ref{Clifford}), where $(-1)^\cF$ is
nothing but the usual chirality operator.
The small $\beta$ limit of this expression computes the
so-called bulk index $\Omega$, or the twisted partition function.
The small $\beta$ expansion proceeds as
\bea\label{start}
\Omega\equiv \lim_{\beta\rightarrow0^+} {\rm Tr}\;
(-1)^\cF e^{-\beta\,\cH}= \lim_{\beta\rightarrow0^+}
\frac{\beta^{N}}{(2\pi\beta)^{N/2}}\int d^N X \;
{\rm det}(\partial_\mu\partial_\nu W)\; e^{-\beta\, (\partial W)^2/2} \,,
\eea
where we used the standard Heat kernel expansion with the familiar zero-th order part,
\bea\label{heat0}
\langle \tilde X\vert \;e^{\;\beta\,\partial_X^2/2}\;
\vert X\rangle =\frac{1}{(2\pi \beta)^{N/2}}\;e^{-(\tilde X-X)^2/2\beta} \,,
\eea
in the $N$ dimensional Euclidean space ${\mathbb R}^N$.
With sufficiently gapped $W$, on the other hand, \eqref{start} should
in turn equal the integral index $\cI$.

Along the way to (\ref{start}), it is instructive to note that
the determinant prefactor is actually a Pfaffian,
\bea
{\rm det}(\partial_I\partial_J W)\quad\Leftarrow\quad {\rm Pf}(\epsilon_{ab}\partial_\mu\partial_\nu W)
\eea
since the fermion bilinear  responsible
for this contribution has the form
\bea
\epsilon_{ab}\psi^{\mu a}\psi^{\nu b}\partial_\mu \partial_\nu W
\eea
with Majorana $\psi$'s. When we introduce the gauge symmetry,
it is the latter that will be generalized straightforwardly.

\subsection{A Single Scalar: $\cI$ vs. $\Omega$}\label{single}

For the simplest illustration,  consider a single scalar
multiplet with a polynomial $W$ with the leading power
$k$, $W=c_k X^k/k!+\cdots $. Rescaling $X$ by
$x\equiv \beta^{1/2(k-1)}X$, we remove the power of
$\beta$ in the prefactor; this also allows us to ignore lower
powers of $X$ in $W$. Then we end up with
\bea
\Omega&=&
\frac{1}{(2\pi)^{1/2}}\int_{-\infty}^{\infty}
dx \; \frac{c_k x^{k-2}}{(k-2)!}\; e^{-(c_k x^{k-1}/(k-1)!)^2/2} \,.
\eea
For odd $k$, this is an integral of odd function, hence we find
\bea \cI=0 \,.
\eea
With $k$ even, the integral reduces to a Gaussian integral
we will encounter repeatedly below,
\begin{align}
\frac{1}{(2\pi)^{1/2}}\int_{-\infty\cdot{\rm sgn}(c_k)}^{\infty\cdot{\rm sgn}(c_k)}
dy \;  e^{-y^2/2} \,,
\end{align}
with $y\equiv c_k x^{k-1}/(k-1)!$,
so that
\bea \cI=\Omega={\rm sgn}(c_k) \,.
\eea

Another way to understand these results is to recall
that the supersymmetric wavefunction in this simple theory
is proportional to
\bea\label{wvfn}
\Psi_\pm(X) &\sim &e^{\pm W(X)} \,,
\eea
where the sign in the exponent is correlated with the chirality
$(-1)^\cF$ of the wavefunction. Clearly $W(X)$ must diverge at
the two asymptotic infinities $X\rightarrow\pm\infty$ with a
common same sign, for one of  these two states to be physically
acceptable, so $k$ should be even for a ground state to exist.
This also tells  us the above result goes well beyond the
polynomial form of the superpotential we assumed for the sake
of convenience.

It is instructive to redo this computation with slightly different
scaling, which sheds more light. Suppose we chose to scale coefficients
of the lower powers of $X$ such that $a_{k-\delta}\equiv \beta^{\delta/2(k-1)}c_{k-\delta}$
kept finite,
\bea
\widetilde W(x)\equiv W(X)\biggr\vert_{X\rightarrow x\,;\; c_n\rightarrow a_n}
\eea
so that the integrand is
\bea
\int dx \; \widetilde W''(x)\; e^{-(\widetilde W'(x))^2/2}=  dy \; e^{-y^2/2}
\eea
with $y\equiv \widetilde W'(x)$. Note how all the information is now
transferred to the integral range via the map $x\rightarrow \widetilde W'(x)$.
In fact this can be done for any real function $W$,  so we may as
well consider  $\widetilde W(x)$ an arbitrary smooth function.

Since one can generally find places where $\widetilde W''(x)=0$, the
integral range of $y$ is in general composed of multiple
overlapping segments.
Suppose the range of $y$ integration decomposes into
$[y_L\equiv \widetilde W'(-\infty), y_1]\cup[y_1,y_2]\cup\cdots\cup[y_{k-2},y_R\equiv \widetilde W'(\infty)]$.
The adjacent segments will cancel, at least  partially, and the net integration range is
$[y_L, y_R]$,
\bea \label{Om1scalar}
\Omega=
\frac{1}{(2\pi)^{1/2}}\int_{y_L}^{y_R} dy \; e^{-y^2/2} \,.
\eea
This non-integral result can be understood from the fact
that $y_{L,R}$ control the asymptotic flat directions along
$X=\pm\infty$. If these numbers are finite, it means that
we can have a continuum sector contribution to $\Omega$,
only to be suppressed by $e^{-y_{L,R}^2/2}$.

In order to count the genuine ground states, unaffected
by the finite values of these asymptotic potentials, one may
scale up the entire potential $W$ by an arbitrary large positive
number, say $C$. As long as neither of $y_{L,R}$ is zero,
the problem is Fredholm, and the true index would be robust
under such a deformation. Then,
\bea
\Omega_C \equiv
\frac{1}{(2\pi)^{1/2}}\int_{y_L\times C}^{y_R\times C} dy \; e^{-y^2/2}
\eea
produces
\bea\label{I_W}
\cI= \lim_{C\rightarrow \infty}
\Omega_C&=&\left\{\begin{array}{rr} 1 & \qquad y_L<0<y_R \\
0 & \qquad y_L\times y_R >0 \\
-1 & \qquad y_L> 0>y_R
\end{array}\right. \,,
\eea
which of course reduces to the standard result, in accord
with the explicit wavefunction construction (\ref{wvfn}).

{}From this simple exercise we learn a few valuable things about
$\Omega$. First of all, if the asymptotic potential approaches a
finite value, $\Omega$ won't be in general integral but will be
contaminated by continuum, exponentially suppressed by
$e^{-\beta W'(\pm \infty)^2/2}$. It is important that we keep
$\beta$ small but finite, despite the usual heat kernel expansion,
since, otherwise, we would lose control over the continuum contributions.
Since the index is robust as long as the asymptotic dynamics is
qualitatively unaltered, we should take care to keep the
functional form of the combination $\beta (W')^2$ intact and finite
in the middle steps, even as $\beta\rightarrow 0$. As long as the problem is
Fredholm, we always have a choice to scale up $W$ in the end
without affecting the true ground state sector, producing the true
index $\cI$ at the very end of the process.

Note that, already for this simplest of theories, the system can undergo
wall-crossing at a codimension-one wall where one of the two asymptotic
values $W'(\pm\infty)$ crosses zero. In order to catch this, it is advantageous to postpone
the scaling (\ref{I_W}) such that $\Omega$ remains a smooth function of
coupling constants: again, we must not be hasty in deforming
away details of the superpotential, even if naive topological
robustness seems to allow it.
A simple corollary of (\ref{I_W}) is,
\bea\label{1dw}
 \cI_W  -\cI_{-W} =\pm 2 \,,
\eea
where $\cI_W$ is
the index for the theory of a single scalar multiplet
with the superpotential $W$. This
gives the prototype wall-crossing behavior of all.
The superpotential $W$, or more precisely its asymptotic
leading part crosses a co-dimoension-one wall, and this
forces the generalized Euler index will jump by $\pm 2$
somewhere in the middle.

\subsection{Multiple Scalar Prototype: $\cI$ vs. $\Omega$}\label{proto}

With more-than-one scalar multiplets, the quantity that
controls the continuum contributions and the potential
wall-crossing behavior is the smallest possible asymptotic
value of $(\partial_\mu W)^2$. We will again concentrate on a
prototype where this quantity is finite, and see what should be
done to deal with the continuum sector. Take $n$ $X_i$'s,
collectively denoted by $\mu$-component vector $\vec X$, and
a single $P$ such that
\bea
W(P;\vec X) = \frac12 \,P \, (\vec X^2  -\xi) \,,
\eea
which leads to
\bea
\partial_P W= \frac{(\vec X^2-\xi)}{2}\,, \quad
\partial_\mu W= PX_\mu \,,
\eea
and
\bea
\partial_P\partial_P W=0\,, \quad
\partial_P\partial_\mu W= X_\mu \,, \quad
\partial_\mu \partial_\nu W= P\delta_{\mu \nu} \,.
\eea
The integral becomes, after rescaling the $n+1$ fields upon which
$\xi$ is effectively scaled down to 0,
by $\beta^{1/4}$,
\bea\label{finite}
\Omega=
\frac{1}{(2\pi)^{(n+1)/2}}\int dp\,d^n x \;p^{n-1}\, \vec x^{\, 2}\; e^{- (p^2\vec x^{\, 2} + (\vec x^{\, 2}/2)^2)/2} \,,
\eea
which already shows $\cI=0$ for even $n$. For odd $n$, we have
\bea
\Omega=
\frac{V_{n-1}\Gamma(n/2)}{\pi^{(n+1)/2}\sqrt{2}}\int_0^\infty d(r^2/2)
\; e^{- ( (r^2/2)^2)/2}=1 \,,
\eea
where we integrated out $p$ and the angular part of $\vec x$ first,
which led to the angular volume $V_{n-1}$ of a unit $(n-1)$-sphere.

Note that, however, the anticipated index $\cI$ differs from this
$\Omega$. If $\xi>0$, the vacuum manifold is an $(n-1)$-dimensional
sphere, $\hat\IS^{n-1}$ defined by $\vec X^2=\xi$ and $P=0$, with
the Euler index equal to 2. If $\xi<0$, on the other hand, the vacuum manifold is absent
entirely, so the index should be zero. The above result, $\Omega=1$,
even though it is suggestively integral, does not really count
the true ground states for any value of $\xi$.

Again this misleading answer appears because the path integral can admit
contributions from continuum sectors. In this prototype,
even though the asymptotic potential grows quadratically
for generic direction, there is a valley of $\vec X=0$, where the
bosonic potential equals  $V(P; \vec X =0)= \xi^2/8$ for any
value of $P$. The twisted partition function can be
contaminated by continuum, suppressed as
\bea
&\sim & e^{-\beta \xi^2/8} \,.
\eea
The suppression factor happens to 1 as $\beta\rightarrow 0$,
resulting in an integral $\Omega$, which does not equal the desired
index $\cI$. The continuum sector conspires to contribute to $\Omega$
a nonzero integer value on top of $\cI$, potentially causing much
confusions. The above small $\beta$ computation, which would have
been acceptable in more ideal circumstances with fully discrete
Hilbert space, cannot be trusted to distinguish the continuum
contribution from the true index.

Therefore, scaling away details of $W$
is not wise; we should have kept the combination
$\beta\xi^2$ finite until the very end of the computation.
Instead of (\ref{finite}), therefore, we are
lead to consider
\bea\label{Snbulk2}
\Omega&=&\frac{V_{n-1}\Gamma(n/2)}{\pi^{(n+1)/2}\sqrt{2}}\int_0^\infty d(r^2/2)
\; e^{- ( (r^2/2-\sqrt{\beta}\xi/2)^2)/2} \,.
\eea
The deviation of $\Omega$ from these integral values are of
order $\sim e^{-\beta \xi^2/8}$ as expected.
As long as $\xi\neq 0$, the problem is Fredholm and
we are allowed to take the $\beta\xi^2\rightarrow \infty$ limit
while preserving the true ground state contributions,
from which we find
\bea\label{Index}
\cI\;\;=\;\;\lim_{\beta\xi^2\rightarrow\infty} \Omega&=&\left\{\begin{array}{lr} 2 & \qquad \xi >0  \\
0 & \qquad \xi<0
\end{array}\right. \,,
\eea
and the correct integral index emerges at the end of the process.

This behavior is reminiscent of an early confusion and the subsequent
resolution \cite{Hori:2014tda} with the localization for gauged quantum
mechanics, with complex supersymmetries. The localization naively claimed
that the index is inert under the continuous change of the FI constant
$\xi$, which is a BRST trivial deformation, yet this was  clearly at odd
with the well-known wall-crossing phenomena. As here,  the
resolution came from the asymptotic flat direction with gap $\sim \xi^2$,
which induces wall-crossing at $\xi=0$.
With half and non-chiral supersymmetry we are considering, the
would-be FI constants are now part of the superpotential for real
scalar multiplets, and yet still lead to wall-crossing. This
means that the wall-crossing now occurs quite generically
for continuous deformations of the superpotential. The
question is whether there is a similar universal scheme
for computing the integral index for non-chiral two-supercharge
problems at hand.

\subsection{$d=1$  Index Theorem}\label{1dindexthm}

What we have seen so far tells us to keep the functional
form of $W$ intact in the $\beta\rightarrow 0$ limit.
One easy way to achieve this is to redefine the
 bosonic zero modes somewhat differently than the above,
\bea\label{universal-scaling}
\beta^{-1/2}X \quad\rightarrow \quad x
\eea
after the integration over the fermions. Note that this
is a dimensionless combination in $d=1$. Another
advantage of this choice is that $W$ need not be
restricted to a polynomial form; Any real and
twice-differentiable $W$ will do.
This transforms (\ref{start}) into
\bea\label{thm0}
\Omega&=&\frac{1}{(2\pi)^{N/2}}\int d^N x \;
{\rm det}(\partial_\mu \partial_\nu w(x))\; e^{-(\partial_\mu w(x))^2/2} \,,
\eea
with replacement
\bea
w(x)&=& W(X)\biggr\vert_{X\rightarrow x}
\eea
Note that, in view of (\ref{universal-scaling}), this last
replacement is possible only with some complicated
rescalings of parameters that go into the
definition of $W$, much as in the previous subsection
but now including an overall multiplication.

One may view this a new definition of $\Omega$.
Of course, the point is that any such rescaling
is harmless as long as computation of $\cI$ goes.
This exponent is still proportional to $\beta$, hidden
in $\partial_\mu$, but the point is that we will take
$\partial_\mu w$ as the new integration variables and
transfer all information, including the $\beta$
dependence, to the integration range information.
In the end, the only relevant information,
as far as the final index $\cI$ goes, would be
whether or not the integration range includes
the origin $\partial w=0$ with what multiplicity.

To be more explicit, one introduces new variables
$y_\mu=\partial_\mu w(x)$ and finds
\bea \label{thm1}
\Omega&=&\Omega_Y\equiv \int_Y\mu \,, \\\cr
\mu&=&\frac{1}{(2\pi)^{N/2}} \left(e^{-y_1^2/2}dy_1\right)\wedge
\left( e^{-y_2^2/2}dy_2\right)\wedge\cdots\wedge\left(e^{-y_N^2/2}dy_N\right) \,, \nonumber
\eea
where all the nontrivial information is now transferred
into the integration domain
\bea\label{Y}
Y=\cup_\Delta Y_\Delta\ ,
\qquad Y_\Delta\equiv \{\;  \partial_\mu w( x)\;\vert\; x_\mu \in \Delta \subset \IR^N\} \,,
\eea
where each $\Delta$ is defined so that $\IR^N=\cup \Delta$
and each $\Delta$ is a maximal domain in which the map
$x_\mu \rightarrow  \partial_\mu w(x)$ induces a unique
orientation in the target $Y_\Delta \subset \tilde\IR^N$
from that of $\IR^N$. The integral over $Y_\Delta$ will
naturally encode this orientation via a $\pm$ sign.

The general ideas behind the index, and our treatment of the
path integral so far, say the following:
{\bf (1)} If the asymptotic flat directions are absent
altogether, we will find $\Omega$ count
the number of ground states correctly and
\bea\label{I=Omega}
\cI=\Omega_Y
\eea
by itself;
{\bf (2)} If there exists an asymptotic flat direction
with a positive energy gap, one will find deviation
from $\cI$ by an exponentially suppressed term, which can be
removed by a further scaling $w(x)\rightarrow C\cdot w(x)$ by an
arbitrarily large positive number $C$
\bea\label{CW}
\cI=\lim_{C\rightarrow +\infty}\Omega_{CY}
\eea
where $CY$ is the scale-up version of $Y$ by the factor $C$;
{\bf (3)} Finally, if there is an asymptotic flat direction
with no energy gap, $\Omega$ can be generally non-integral
and such continuum contributions cannot be removed within
this straightforward path integral setup by itself; more
refined approaches need to be invoked.

\subsubsection*{The (Non-)Integrality}

To show that these assertions indeed hold with (\ref{thm1}),
we start by noting  that  each Gaussian integral produces
$\pm\sqrt{2\pi}/\sqrt{2\pi}=\pm 1$ if the range is over
the entire real line. Therefore, the integral there would
be an integer  if each $Y_\Delta$ equals $\tilde\IR^N$,
up to the orientation. But of course the life is not that
simple.

The claim that (\ref{CW}) will give an integer
amounts to the assertion that,
\bea
CY=\cup_\Delta CY_\Delta\ ,
\qquad CY_\Delta\equiv \{\;C\cdot\partial_\mu w(x)\;\vert\; x^\mu \in \Delta \subset \IR^N\}
\eea
asymptotes to a multiple cover of $\tilde \IR^N$, after
the orientation is taken into account, as $C\rightarrow +\infty$.
This will happen, clearly, if and only if all asymptotic flat
directions, if any, come with nonzero energy gap.
It is clear that a gapped asymptotic flat direction will create
boundaries in some of $Y_\Delta$, which however do not intersect
the origin. As such, the scaled version, $CY_\Delta$, would
push out such boundaries by the factor $C$. If the origin
is contained in any open subset of $Y_\Delta$, $CY_\Delta$
would expand and fill the entire $\tilde\IR^N$, so that
\bea
\lim_{C\rightarrow \infty} \int_{CY_\Delta}\mu=\int_{\pm\tilde\IR^N}\mu=\pm 1 \,.
\eea
If the origin lies outside of $Y_\Delta$, $CY_\Delta$
would be pushed out to the asymptotic region, so that
\bea
\lim_{C\rightarrow \infty} \int_{CY_\Delta}\mu=0 \,.
\eea
This leads to
\bea
\lim_{C\rightarrow\infty}\int_{CY}\mu=\lim_{C\rightarrow\infty}\sum_\Delta \int_{CY_\Delta}\mu \quad\in\quad \IZ \,,
\eea
which should equal the index $\cI$. This confirms {\bf (2)} above,
and thus the index formula (\ref{CW}).

Boundaries of individual $Y_\Delta$ could come either from the asymptotic
behavior of $w(x)$ or due to a linear combination of $\partial_\mu w(x)$
bouncing at finite values of $x$, where one necessarily expect
$0={\rm det}(\partial\partial w)$. So far, we overlooked the
latter boundaries but as long as such a boundary does not
meet the origin, i.e., $0=(\partial w)^2$, the scaling
by $C$ will remove it as well. What happens if such a boundary
happens to meet the origin of $\tilde\IR^N$? One easy way
to see how boundaries of $Y_\Delta$'s at finite $x$
are irrelevant for the ground state counting is the following:
Since we are really after the integer quantity $\cI$,  rather
than $\Omega$, and since the former is robust  under a ``small''
deformation of $W(X)$, one can easily deform $w(x)$ such that
the troublesome boundary is shifted slightly away from the origin,
$y=0$, while maintaining the boundaries due
to $\vec x^{\; 2}\rightarrow \infty$ intact.
This will of course affect both of $Y_\Delta$'s
in question but in such a manner that the integral part
$\cI$ is unaffected. Afterward we are free to scale the deformed
$w$ by $C$ and obtain $\cI$ via (\ref{CW}).

By the same token,
we see how the integrality can fail for {\bf (3)}. If there
is a gapless asymptotic direction, this means
\bea
(\vec\partial w)^2\rightarrow 0 \,,
\eea
along some asymptotic direction, $\vec x^{\;2}\rightarrow\infty$.
This allows a boundary of some $Y_\Delta$ to meet the origin,
which can, generically, neither fill $\tilde\IR^N$ nor be pushed
out to infinity by a simple overall scaling by $C$. Invoking
a perturbation of $W$ at such a place can lift the problematic
direction in principle but produces multiple different outcomes
for $\cI$. In fact, the wall-crossing happens precisely because
different deformations that lift such gapless flat directions
generally lead to different integral indices.

Finally we need to show the integrality of $\Omega$ itself
for the case {\bf (1)}, where there is no asymptotic flat direction
at all. This condition implies
\bea
(\vec\partial w)^2\;\rightarrow\;\infty
\qquad \hbox{as} \; \;\; \vec x^{\;2}\;\rightarrow\; \infty \,,
\eea
for all possible directions $\vec x$. The integrality
would follow
immediately if $Y$, after possible cancelation between
$Y_\Delta$'s due to mutually opposite orientations, is
itself a multiple-cover of $\tilde\IR^N$.  Since a
boundary of each component can come only from finite $x$,
there should be necessarily a pair, say $Y_{\Delta_1}$
and $Y_{\Delta_2}$ meeting at such a boundary. These pairs,
by definition, overlap in $\tilde \IR^N$ with mutually
opposite orientations. The two can then be combined to
a third domain $Y_{\Delta_3}$ in $\tilde \IR^N$.
If the latter has a boundary from finite $x$, we then
find another adjacent and canceling domain $Y_{\Delta_4}$,
and so on. In the end, all boundaries originating from
finite $x$ will effectively disappear. Since those
from the asymptotic $x$, $\vec x^{\, 2} \rightarrow \infty$ are absent to begin with,
$Y$ has to be a multiple cover $\tilde \IR^N$ and
$\Omega$ has to be integral and thus equal to $\cI$,
resulting in (\ref{I=Omega}).

\subsubsection*{Back to the Examples}

Let's redo the two examples we started with, using this language.
If one starts with $\CN=1$ with $k$-th order polynomial
$w(x)$, we will generically have $k-1$ $\Delta$'s and
$Y_\Delta$'s, divided by $0=w''(x)$. Ordered by the
natural ordering of the real line $x$, each pair of adjacent
$Y_\Delta$'s comes with mutually opposite orientations,
so cancels out partially. For instance, $w(x)=x^3/3$
gives $y=x^2$, with $\IR=\Delta_- \cup\Delta_+$ and
$\Delta_\pm =\pm(0,\pm\infty)$ where the sign $\pm$
in front denotes the orientation. Mapping to $y$
variables, this gives
\bea
Y_{\Delta_\pm} =\pm(0,+\infty) \,,
\eea
which cancel each other. Proceeding similarly for
arbitrary power $k$, one can show easily that
\bea
Y={\rm sgn}(c_k)\,\tilde\IR\;\; \hbox{for even} \;k \quad ;\quad
Y=\emptyset\;\; \hbox{for odd} \;k
\eea
with the orientation for the former, fixed by the sign of the
coefficient $c_k$ of $x^k$. Hence $\cI={\rm sgn}(c_k)$ for
even $k\ge 2$ and $0$ otherwise.

For the multi-scalar prototype above, the map
\bea
y_p=(\vec x^{\;2}-\xi)/2 \,, \quad y_i= p x_i \,,
\eea
admits two $\Delta$'s such that
\bea
\Delta_1 = -[0,-\infty)_{p}\times \IR^n_x
\ ,&\qquad &Y_{\Delta_1}=-[-\xi/2,+\infty)\times (-1)^n\tilde\IR^n \,, \cr\cr
\Delta_2 =[0,+\infty)_{p}\times \IR^n_x
\ ,&\qquad &Y_{\Delta_2}=[-\xi/2,+\infty)\times \tilde\IR^n \,.
\eea
For even $n$, the two $Y_\Delta$'s cancel out each other precisely.
For odd $n$, the two scaled-up domains,
\bea
CY_{\Delta_1}= CY_{\Delta_2} =[-C\xi/2,+\infty)\times
\tilde\IR^n \;\;\rightarrow\;\;\tilde\IR^{n+1}\quad \hbox{as}\;\;C\rightarrow +\infty \,,
\eea
contribute 1 each for $\xi > 0$, again resulting in $\cI=2$.
For $\xi < 0$, $\cI = 0$ because the two domains are pushed out
to infinity under the $C$ scaling.

\subsection{Alternatives for Integral $\Omega=\cI$}\label{sec:alternatives}

When the twisted partition function $\Omega$ actually
equals $\cI$ and thus integral, there are several
alternative forms of this index (\ref{thm1}) under
additional assumptions, well known in the relevant
literatures. In particular, the
Morse theory \cite{Witten:1982im} is a direct consequence of (\ref{thm1}),
while the winding number interpretation will later
prove to be very useful for polynomial  $W$.
Both of these alternatives rely on generic form of
the superpotential, which could fail if certain
global symmetries are imposed. For the latter,
the Lefschetz fixed point theorem comes to the
rescue as usual, however.

\subsubsection*{The Winding Number Interpretation}

The integration formula (\ref{thm1}) counts in the end
how many times the map $ X\rightarrow dW$ covers $\tilde \IR^N$,
with the orientation taken into account. Recall that $\Omega=\cI$
is guaranteed if $|dW|^2\rightarrow \infty$ along
all asymptotic directions in the $X$ space. Suppose further
that the superpotential and its derivatives are regular
everywhere at finite $X$. The continuity of the map then
implies that $\Omega=\cI$ can be also counted by
the winding number of the map,
\bea
\vec \varphi_R(\hat X) \equiv \frac{\vec \partial W(X)}{|\vec\partial W(X)|}\biggr\vert_{X=R\hat X}
\eea
with $|\hat X|=1$,  from $\IS^{N-1}$ to $\hat \IS^{N-1}$,
which is well-defined for sufficiently large $R$. More
explicitly, we have
\bea\label{winding}
\cI\;\;\cdot\;\; \int_{\hat \IS^{N-1}} \hat \cV\;\;= \;\;\int_{\IS^{N-1}}\varphi_R^*\hat \cV
\eea
at some large $R$, where $\hat\cV$ is the volume form
of $\hat \IS^{N-1}$ and $\varphi_R^*$ is the pull-back.

While this is not a very practical form, e.g., for numerical
computation, it offers an ultraviolet perspective,
opposite of the Morse theory interpretation below,
and instructs us that the asymptotic $W$ suffices
for the enumerative $\cI=\Omega$. This also implies,
as we study  in the next subsection, that the
codimension-one wall-crossing is also controlled
entirely by the asymptotic form of $W$.

For instance, suppose $N$ is odd and $W(-X)=-W(X)$.
For such cases, (\ref{thm0}) immediately implies $\cI=0$
identically; the Pfaffian contribution flips the sign
under $X\rightarrow -X$, so the integral is odd
under the parity. Note that the same vanishing theorem manifests in (\ref{winding})
in how $\varphi^*\cV$ contributions cancel between
the antipodal points pairwise for odd $N$.
On the other hand, (\ref{winding}) does not depend
on details of $W$ at finite $X$. Only the asymptotic part
matters. Therefore, for odd $N$, whenever one can
identify an approximate $W^{\rm asymp}$, such that
$W^{\rm asymp}(-X)=-W^{\rm asymp}(X)$ and that
\bea\label{oddW}
\frac{W^{\rm asymp}-W}{W}\;\rightarrow \;0\quad \hbox{as } |X|\rightarrow \infty
\eea
along all directions, we may conclude
\bea\label{odd}
\cI_W=0\ ,
\eea
regardless of details of $W$ at finite $X$ region.

One obvious example of this phenomenon occurs when
the superpotential has a polynomial form.
Suppose that the superpotential starts with the
leading power $p+1$,
\bea
W(X)=W^{(p+1)}(X)+\cdots \, .
\eea
We can say immediately that, with odd $N$, $\cI$ computed
from (\ref{winding}) is identically zero if $p+1$ is also odd,
whenever the homogeneous $W^{(p+1)}(X)$ is nondegenerate. The latter
is in turn guaranteed
by the genericity of the superpotential in the absence of
global symmetries. When a global symmetry exists, on the other
hand, we may resort to a Lefschetz fixed point theorem outlined
below.

\subsubsection*{The Morse Theory Interpretation}

Looking at (\ref{thm1}), or more precisely at
its $C$-scaled version (\ref{CW}), one realizes that it
also counts the critical points of the superpotential, $dW=0$,
provided that the latter are all isolated and nondegenerate.
As we scale up $W(x)$ by a large multiplicative factor $C$,
the integral is localized to its critical points,
$dW(x_*)=0$, in the $\vec x$ space. The Gaussian integral
over a single $\Delta$ domain would converge to
$(-1)^{\gamma(x_*)}$ where $\gamma(x_*)$ is the so-called
Morse index \cite{Witten:1982im} of $x_*$, i.e., the number
of negative eigenvalues of the Hessian $\partial_\mu \partial_\nu W(x_*)$.
Each Gaussian integral gives $\pm 1$, depending on whether the relevant range
is $(-\infty,\infty)$ or $(\infty,-\infty)$, and the
combined sign at each $x_*$ is dictated by the
sign of the determinant of the Hessian.

Naturally, if critical points of $W$  are
discrete and nondegenerate and all distributed over a
closed  subregion of $\tilde \IR^N$, we find
\bea
\cI=\lim_{C\rightarrow \infty}\Omega_{CY}=\sum_{x_*} (-1)^{\gamma(x_*)} \,.
\eea
Here again we can see that for odd $N$, $W(X)=W^{\rm asymp}(X)+\cdots$
with odd $W^{\rm asymp}(X)$ in the sense of (\ref{oddW}) leads to $\cI_W=0$.
Consider $W_1(X)=-W(X)$ and $W_2(X)=W(-X)$. If $x_*$ is a
critical point of $W$, it is also a critical point of $W_1$
and vice versa, and $-x_*$ is obviously a critical point
of $W_2$. The sign of the Hessian for $W_1$ at $x_*$ and that
of $W_2$ at $-x_*$ have the sign $(-1)^{N}$ and $(-1)^{2N}$
relative to that of $W$ at $x_*$, point by point, which give
$$
 \cI_{W_2}=\cI_W =-\cI_{W_1}
$$
for odd $N$. On the other hand, $W_1$ and $W_2$ share a common
leading asymptotic form $W^{\rm asymp}_1(X)=W^{\rm asymp}_2(X)=-W^{\rm asymp}(X)$, for odd $W^{\rm asymp}$,
so should have the same index $\cI_{W_1}=\cI_{W_2}$ whenever
$W^{\rm asymp}(X)$ is nondegenerate. Therefore we find again that
$$\cI_W=0$$ outside codimension-one walls, where $W^{\rm asymp}(x)$ is degenerate, in the parameter space of $W$,
bringing us back to (\ref{odd}).

\subsubsection*{The Lefschetz Fixed Point Theorem}

When the theory admits a global symmetry, the index
theorem can be substantially simplified into a form
of the Lefschetz fixed point theorem\cite{Atiyah:1967fx}.
Introducing external vector multiplets $A_i$ that
couple to such global symmetries, the zero modes
thereof, $a_i$, enter the Hamiltonian, schematically,
as follows:
\bea
H \;\rightarrow \;H(a)=H + \cdots + a_0F \,,
\eea
where $F$ is the global symmetry charge. The ellipsis
denotes complex mass terms due to $a_{1,2}$ while the
combination $\beta a_0$ will play the role
of the chemical potential.

In turn, $F$ will have the form
\bea
-F=\cL_F + \hbox{fermion bilinear} \,,
\eea
where $\cL_F$ rotates the bosonic variables. When we proceed
with the Heat kernel expansion the latter has the effect
of
\bea
\langle \,X\,\vert\; e^{-\beta H(a)}\;\vert\, X\,\rangle
= \langle \,X\,\vert\; e^{-\beta (H(a) +a_0\cL_{F})}\;\vert\, X(\beta a_0)\,\rangle \,,
\eea
where $X(\beta a_0)=e^{\beta a_0 \cL_F}X$.

As we expand in small $\beta$, as in previous section,
this shift creates a term in the exponent
of type $(X(\beta a_0)-X)^2/\beta$,  from (\ref{heat0}),
which in turn translates to $\sim(\beta a_0 \cL_F[X])^2/\beta$
unless $a_0$ becomes as large as $1/\beta$. Thus, the rotation by small but
finite $\beta a_0$ makes fields that rotate under $F$ arbitrarily
massive. Since the bosonic and the fermionic degrees of
freedom match precisely, their contributions cancel out,
leaving behind the invariant loci only. In the end,
one merely collect the field content that spans the
invariant locus and use (\ref{thm1}) for this subsystem.

If the dynamics has no asymptotic flat directions, the usual
topological robustness of index ensures that the addition of
this mass term will not affect the outcome $\cI=\Omega$.
As such, the index theorem further reduces to a sum of
smaller index problems, each of which is defined over the
common invariant loci of all global symmetries.
Labeling these loci by $l$, we find
\bea
\Omega = \sum_l  \, \Omega(l) \,,
\eea
where $\Omega(l)$ is the Euler index of the locus $l$.

In particular, it is worthwhile to note that the ``flavor chemical potential''
$a_0$ does not appear anywhere in the final expression. This
is in a stark contrast against the localization computation
with complex supersymmetries in recent literatures.
This appears to be related to the fact that with our minimal
supersymmetry there is no $R$-symmetry chemical potential available.
Indeed for many of existing twisted partition functions with larger
supersymmetries, such as the $(2,2)$ elliptic genus and the refined
index for $\cN=4$ quantum mechanics, the dependence on flavor chemical
potentials tends to disappear altogether as one takes the limit
of the vanishing $R$-charge chemical potential.

When asymptotic and gapped flat directions are present in the
original problem but entirely lifted by this mass deformation,
(\ref{thm1}) will naturally compute the desired index $\cI$
instead of $\Omega$. If such a lift is not quite complete,
$(\ref{thm1}) $ will again compute some $\Omega$, from which
reading off $\cI$ is again a matter of scaling $W$ by
a large positive number. If there are gapless asymptotic
flat directions in the original problem, the resulting
$\Omega$ computed by our fixed point theorem needs to be
treated more carefully; for one thing, the original
theory might be sitting at a wall of marginal stability,
in which case the mass deformation of this kind may not
be very innocuous.

Again with the multi-scalar prototype of the previous subsection,
even $n$ admits $n/2$ $SO(2)$'s that rotate $X$'s pairwise,
say $X_1$ and $X_2$, $X_3$ and $X_4$, etc. The invariant
locus is $\vec X=0$, where ${\rm det}(\partial\partial W)=0$.
Once the path integral localizes to such a locus, $\Omega$ vanishes,
and so we find $\cI=0$ as well.  For odd $n$, $(n-1)/2$ $SO(2)$
rotation kills all $X$'s but the last: Hence the computation
reduces, effectively, to $n=1$, and we again find $\cI=2$ in the end.

A word of caution is needed here. Although  the Lefschetz
fixed point theorem is applicable to theories with compact
dynamics, one also encounters examples where the dynamics
prior to turning on the chemical potential and the real mass
is not compact while, thereafter, it becomes compact. In such
situation, the index computed via the fixed point theorems
will be integral, yet it is not obvious what it computes.
In many localization computations with larger supersymmetries,
the chemical potentials do survive in the final form
of the twisted partition functions, yet one usually encounters
a pole at the $a=0$ limit, giving us an ample warning in interpreting
such results. For the generalized Euler index of ours,
such poles are absent since $a$ itself disappears in the
end; it remains to be seen if this independence implies
no adverse contamination of $\cI$ by such localizing masses
or merely misleading results.

\subsection{Wall-Crossing}

The wall-crossing phenomenon, familiar from $\cN=4$ gauged
quantum mechanics, is discontinuity of the (refined) index
which occurs as one continuously changes FI constants. With
the current $d=1$ $\cN=2$
real supersymmetries, the auxiliary $D$ field of the $d=1$ $\cN=4$ vector
multiplet now becomes an auxiliary and real $F$ field of the
real scalar multiplet. This means that the wall-crossing would
be ubiquitous in $d=1$ $\cN=2$ theories, as one deforms the superpotential,
with or without the gauge sector. Here we wish to characterize the
wall-crossing phenomena in the absence of the gauge sector, in
terms of the superpotential $W$. We will find that the basic
path-integral form (\ref{thm1}) for $\Omega$ and the winding
number interpretation (\ref{winding}) are most useful for
understanding the wall-crossing in this reduced setting.

On the other hand, the topological robustness of the index must
still hold; as such the index should remain piecewise constant
in the space of the superpotential and can change only when we
change the asymptotics of $W$ qualitatively. We will mostly work
with a polynomial form of $W$, and characterise how the primary
wall-crossing happens even while we deform the leading
homogeneous part, call it $W^{(p+1)}$, not necessarily changing the power.
This is different from $\cN=4$ Landau-Ginzburg theories
where the leading power of the holomorphic $W$ fixes the
index once and for all.

The simplest example of wall-crossing occurs with
the quadratic $W$. Suppose we have
\bea
W(X)=\frac12 \sum_{\mu,\nu} m_{\mu\nu}X^\mu X^\nu \,.
\eea
If $m$ is non-singular, the superpotential  admits
a single critical point $\vec X=0$, and as such one
immediately learns that $|\cI|=1$. While the sign of $\cI$ is
in principle ambiguous for a given theory due to the
same ambiguity with $(-1)^\cF$, the relative sign of $\cI$ between a pair of theories
that can be continuously connected is not ambiguous.

It is clear that the sign of $\cI$ is dictated by
the fermionic Pfaffian, or in this case the determinant
of $m$, so that\footnote{We have in effect fixed the
would-be ambiguous sign convention for $(-1)^\cF$,
in favor of this sign choice, and will follow the
same convention throughout this note.}
\bea
\cI= {\rm sgn}({\det}(m))\ .
\eea
As is clear from the index formula (\ref{thm1}),
the wall-crossing here is from how the map
\bea
X^\mu \quad\rightarrow \quad  m_{\mu\nu}X^\mu
\eea
reverses the orientation as one crosses codimension-one wall
${\det}(m)=0$ in the space of the mass matrix.
at a point where exactly one eigenvalue of $m$ vanishes.
At such $m$, the superpotential becomes degenerate and
a flat and gapless asymptotic direction opens up, pushing the
vacuum to infinity. In this simple prototype, one can
see that a vacuum is pushed out to infinity as one approaches
such a wall, while on the other side of the wall a different
vacuum with the opposite chirality moves in from the infinity.
If one jumps across two such walls, the two wall-crossing
would cancel each other. Therefore, the wall-crossing
in this simple toy model is such that ${\det}(m)=0$
divides the parameter space into two regions, each consisting
of many disconnected wedges, for which $\cI=1$ or $\cI=-1$.

What happens if we allow subleading linear terms? With
\bea
W(X)=\frac12 \sum_{\mu,\nu} m_{\mu\nu}X^\mu X^\nu +\sum_\mu  k_\mu X^\mu \,,
\eea
what changes qualitatively for wall-crossing? When det$(m)\neq0$
the asymptotics are dominated entirely by $m_{\mu\nu}$, so linear terms
do little. To see what happens at the wall det$(m)=0$, we diagonalize
the quadratic term and re-label the fields
\bea
W(X,Y)=\frac12 (m_1 (X_1)^2  +m_2 (X_2)^2)+ k_1 X_1+k_2 X_2 \,,
\eea
so that the vacuum conditions are
\bea
m_1 X_1+k_1=0\,, \qquad m_2 X_2+ k_2=0 \,.
\eea
When one of the eigenvalues vanishes, say $m_2=0$, the
vacuum  again runs off to infinity, $X_2=-k_2/m_2 \rightarrow \infty$.
Note that this could not be seen clearly if we sat at $m_2=0$;
one sees only that no finite vacuum exists.
If $k$'s were absent, this runaway vacuum phenomenon would have
been manifested as a flat gapless direction. However, with or without
$k$'s the end result for $\cI$ is the same: A wall appears
when the leading quadratic terms become degenerate, and a
codimension-one wall-crossing occurs across such a wall.

This simple pattern is really a universal feature of codimension-one
wall-crossings for polynomial $W$. Consider
\bea
W(X)&=&W^{(p+1)}(X) +W^{(p)}(X)  +\cdots \,, \cr\cr
W^{(q)}(X) &=& \sum_{\sum n_I = q} C^{(q)}_{n_1\cdots n_N }
X_{1}^{n_1}\cdots X_N^{n_N} \,,
\eea
with coefficients $C$'s assumed to be generic.
Since a flat gapless asymptotic direction or a runaway vacuum is
necessary, the wall-crossing is possible if and only if the leading
homogeneous $W^{(p+1)}(X)$ admits a nonzero critical point. Due to
the scaling  $W^{(p+1)}(\lambda X) =\lambda^{p+1}  W^{(p+1)}(X)$,
a nonzero solution to $0=\vec\partial\, W^{(p+1)}(X) $ generates
an $\IR$'s worth of vacua of $W^{(p+1)}(X)$. This means that a
codimension-one wall-crossing occurs precisely at
parameters where $W^{(p+1)}(X)$ becomes degenerate, or in other
words, when
\bea
dW^{(p+1)}=0
\eea
admits a straight line of solutions passing through the origin
and extending out to the asymptotic region. Note that the
homogeneity implies $W^{(p+1)}=0$ as well. When we take into
account subleading terms, the same codimension-one condition
can instead translate to a vacuum that runs away to infinity,
and the location of these codimension-one walls are not
affected by such subleading terms in $W$.

It is important to note here that the codimension-one walls
defined by the leading power
$W^{(p+1)}$ of $W$ are invariant under the overall
scaling, such that the homogeneous rescaling of the
coefficients, power by power, does not change the
index. This observation is important since we performed scalings
of coefficients in $W$, different for different powers
of $X$'s, to reach at $w$ with the same functional
form as $W$. Under such power-dependent scalings of
coefficients, no wall-crossing happens as long as the
leading power remains non-degenerate in the above sense.
This justifies the main result of this Section, (\ref{thm1}),
despite potential wall-crossing threats.

Recall that, away from the codimension-one walls, the index
$\cI$ is also entirely determined by the asymptotic superpotential,
$W^{(p+1)}$, as is obvious from the  winding number interpretation
(\ref{winding}). Sometimes, this by itself  produces indices and
wall-crossing rather quickly. For example, consider theories
with two scalar multiplets, starting with a cubic model,
\bea
W^{(3)}={\rm Re}\left( A_0 \Phi^3+ A_1 \Phi^2\bar \Phi\right) \,,
\eea
with $\Phi=(X_1+iX_2)/\sqrt{2}$ and complex numbers $A_0$ and $A_1$. The effective
number of times $dW^{(3)}$ covers $\tilde\IR^2$ equals
the winding number of
\bea
(\partial_1+i\partial_2) W^{(3)}= 3 \bar A_0 \bar \Phi^2+ 2 \bar A_1 \Phi \bar \Phi +  A_1  \Phi^2 \,,
\eea
which says immediately that the absolute value of the winding
number cannot exceed $2$, although the number of critical points
can be as many as 4. With $\Phi\rightarrow Re^{i\phi}$ and
$\phi_1$ the phase angle of $A_1$,
\bea
(\partial_1+i\partial_2) W^{(3)}\;\;\rightarrow \;\; R^2\left[3 \bar A_0 e^{-2i\phi} + \bar A_1 (2 +  e^{2i\phi+2i\phi_1})\right] \,.
\eea
$|A_0|\gg |A_1|$ results in $\cI=-2$ while $\cI=0$ when $|A_0|\ll |A_1|$.
$\cI=2$  is not allowed as the first term in the round parenthesis
always dominates over the second.\footnote{
The same can be seen from the Morse counting with, after an additive shift of $X$,
$$
W(X_1,X_2)=\frac13 f (X_1)^3  + g X_1 (X_2)^2 + \frac13 h (X_2)^3 + \cdots \,.
$$
More explicitly, the codimension-one walls are at $f=0$ and $h^2f+4g^3=0$ such that
\bea
&\cI =-2  &\quad 0 < h^2f < -4g^3 \;\;\hbox{or}\;\; -4g^3 < h^2f <0 \,,  \cr\cr
&\cI =0   &\quad  0 < -4g^3 < h^2f \;\;\hbox{or}\; \;\;  h^2f <  -4g^3 <0 \,. \nonumber
\eea}
One can generalize this to higher order polynomials,
\bea
W^{(p+1)}={\rm Re}\left( A_0 \Phi^{p+1}+ A_{1} \Phi^{p}\bar \Phi+\cdots+A_{\lfloor (p+1)/2\rfloor } \Phi^{p+1-{\lfloor (p+1)/2\rfloor }}\bar \Phi^{\lfloor (p+1)/2\rfloor } \right) \,.
\eea
The same reasoning as above tells us that the possible
values of the index in the domains separated by
codimension-one walls are
\bea\label{twoC}
\cI = -p,-p+2,\dots, -p+2\lfloor (p+1)/2\rfloor \,,
\eea
up to a convention-dependent overall sign.
Note that the absolute value of the index, $|\cI|$,
is bounded above by $p$ which is the value in the
holomorphic limit, $A_{k>0}=0$. Note that, in contrast,
the maximum number of non-degenerate critical points
for real $W^{(p+1)}+\cdots$ is a much larger value $p^2$.

What happens when we extend these to a larger number of
matter multiplets? For odd $N$, two general features one can find
are that $\cI=0$ for
odd $p+1$, as already shown in the Section \ref{sec:alternatives}, and
that, for even $p+1$,
\bea
\cI=-2k+1, -2k+3,\cdots, 2k-3, 2k-1 \,,
\eea
for some integer $k>0$. For $p+1=2$, one can see $k=1$
either from the Morse counting or from the winding; e.g.,
the map $X\rightarrow dW^{(2)}$ is linear, so can have
at most $\pm 1$ winding number,
immediately leading us to $\cI=\pm 1$. The sign-symmetric
nature can be argued easily by considering
$W^{(p+1)}$ and $-W^{(p+1)}$
pairwise. For even $N$, no systematic feature appears; for
example, the upper bound on $|\cI|$ is no longer set by
the holomorphic limit for even $N$. Leaving aside all such
complicated patterns, however, the wall-crossing pattern
itself appears to be very simple: For each codimension-one wall,
dictated by the coefficients of the leading power of the
polynomial $W$, the general pattern of the wall-crossing
is very simple with
\bea
\Delta \cI =\pm 2 \,,
\eea
which generalizes (\ref{1dw}) for a single scalar multiplet
or the pattern seen in (\ref{twoC}) of the two scalar multiplet theories.
Generically, a Morse vacuum moves off to the asymptotic
infinity, only to return on the other side of the
wall with the opposite chirality.

Although we confined our attention to codimension-one walls
of marginal stability, it is not difficult to see that
there can be more complicated patterns of the index
when one sits at a codimension-one wall, however.
The leading $W^{(p+1)}$ becomes degenerate
at such a wall, meaning that for some asymptotic
directions, the subleading pieces $W^{(q\le p)}$
can now alter the asymptotic behavior qualitatively.
In other words, when $W^{(p+1)}$ sits on a codimension-one wall,
$\vec\partial W^{(p+1)}$ passes through
the origin such that small ``subleading" corrections
can alter the winding number qualitatively. This means
that there are actually a nested network of marginal stability walls.
In this note, however, we will not consider such possibilities.
When we lift the situation to $d=3$  minimally supersymmetric
theories, for example, there is no non-renormalization
theorem that protects the superpotential, so a fine-tuning
of $W$ does not make much sense.  At least physically there
seems to be no reason for a theory to sit on such a
tightly constrained place in the parameter space.

\subsection{Uplifts to $d=2,3$}

What  happens if we try to extend the computation here straightforwardly
to higher dimensions? One should start by putting $d$ dimensional theory on
$\IT^d$ torus. The supersymmetry content matches two real fermions for one
real boson, so that the determinants of massive sector cancel precisely
due to supersymmetry regardless of the spacetime dimensions. This precise
cancelation is further reinforced by the absence of $R$-charge that commutes
with a supercharge, and as such the Kaluza-Klein modes cancel out precisely.
This means that, as long as we consider massive theories with no continuum
of classical supersymmetric vacua, our computation and the index theorem
are applied to higher $d=2,3$ verbatim. On the other hand, if one starts with
$d=3$ theory with continuum of vacua, the Witten index is ill-defined to
begin with, so only upon some massive deformations that can lift such continua
one can discuss the Witten index. For such theories, again our $d=1$ approach
will work verbatim once the theory is put on $\IT^3$.

Since the wall-crossing phenomenon arises from the asymptotic flat
directions in the matter sector, the wall-crossing uplifts straightforwardly
to higher dimensions as well. This should be contrasted to the more familiar
wall-crossing of $d=1$ $\cN=4$ theories, which is essentially
a Coulomb phase physics and does not uplift to $d>1$. So,
again assuming that we deal with massive theories, our discussion of
the wall-crossing will apply straightforwardly to $d=2$ $\cN=(1,1)$
theories and to $d=3$ $\cN=1$ theories. In fact, the wall-crossing
has been observed directly for some $d=3$ $\cN=1$ theories \cite{Bashmakov:2018wts}.

New ingredients that distinguish different spacetime dimensions will
manifest once we consider gauge theories: The main differences arise
from the would-be zero modes associated with the gauge fields, namely the
gauge holonomies.
In next two Sections, we will explore the smallest of gauged dynamics in $d=1$ and $d=3$ settings,
which will give us some further understanding of how vacuum counting
in these two different spacetime dimensions share some common features
and how they differ from each other. In the end, we will find
index formulae for $d=1$ $\cN=2$ $SO(2)$ theories and for $d=3$ $\cN=1$ $SO(2)$ gauge
theories, which are not that different from (\ref{thm1}) at least in spirit.

\section{$d=1$ $\cN=2$  Abelian Gauge Theories}

Gauging requires further steps. For the same reason as the scalar
multiplet theories, the massive modes cancel out precisely in the
path integral. The zero mode is another matter. The gauge zero modes $a_i$'s
should be integrated over, and so are their partner gaugini, which
alter the procedure qualitatively. Also the trace over the Hilbert
space must be done for the gauge invariant states only, so we include the
projection operator
$\cP_\cG$
\bea
{\rm Tr}\; (-1)^\cF e^{-\beta\,\cH}\cP_\cG \,,
\eea
where the projection operator is given by
\bea
\cP_\cG=\frac{1}{{\rm vol}(\cG)}\int_\cG d^g\Theta\; e^{i\Theta\cdot G} \,,
\eea
with the Gauss constraint $G$ and $g={\rm dim}(\cG)$. This results in
the normalization of the measure for the time-like gauge
zero modes, $\sim \Theta/\beta$, that is a little different
from its space-like counterpart, as we see below.

In particular, while the
standard $\beta\rightarrow 0$ scaling seemingly reduces $\Theta$
into $\IR^g$-valued, replaced by a gauge-field along the time
direction,  its periodic nature must not be forgotten. How this
periodic origin should enter the index quantitatively has been
only very recently understood \cite{Hwang:2017nop,Hwang:2018riu},
surprisingly enough. We will outline it in the next subsection
and encounter it repeatedly in the rest of this note. The same
observation will play an essential role when we uplift the story
to $d=3$ later, where we will compute $d=3$ Witten index
from a $d=1$ perspective.

Another issue for gauge theories is the potentially
flat asymptotic direction along the Coulomb phase.
Such new flat gapless directions can potentially complicate the
problem further in two essential ways. The first is how $\Omega$ could
generically become non-integral due to the continuum
contribution; the second is potential additional wall-crossing
if such flat gapless directions appear during a continuous
deformation. We will encounter some of these issues with
concrete examples below. However, it is worthwhile to note
that the connection to $d=3$  $\cN=1$ theories mitigates some of
these complications. One unexpected result is that
$\Omega$ of a $d=1$  theory obtained from dimensional reduction
of $d=3$  $\cN=1$ theory does not actually experience wall-crossing
due to asymptotically gapless Coulomb directions, contrary
to the general wisdom.

The first problem of non-integral value of $\Omega$ persists,
on the other hand. In particular we will find
that $\Omega$ in such theories is a multiple of $1/2$ in all
examples we discuss. In strict $d=1$ problems, computing
$\cI$ in such situations is not straightforward. With larger
supersymmetries, the continuum part of such non-integral results
have been studied rather generally in recent years and in some
cases classified and understood to the extent that the extraction
of $\cI$ is possible\cite{Lee:2016dbm}. For the current problems,
we do not have such a machinery. On the other hand, when we move on to $d=3$
Chern-Simons theories, where as noted above there are no longer
Coulombic continuum, this non-integral contribution will be a seed to
a very well-known interpretation as the half-integral shift
of the Chern-Simons level due to massive charged fermions.
We come back to this in Section 4.

\subsection{$\Omega$ for  Gauge Theories and Holonomy Saddles}
\label{sec:H-saddle}

The analog of $(\ref{start})$ for gauge theories is, at least naively,
\bea\label{g-start}
\lim_{\beta\rightarrow 0^+}
\frac{\beta^{N/2+g}}{(2\pi)^{N/2+g}{\rm vol}({\cG})}
\int d^NX d^{3g}{\bf u}\; {\rm Pf}(M(X,{\bf u})) \; e^{-\beta V_{\rm bosonic}(X, {\bf u})} \,,
\eea
where $N$ is the number of real scalar multiplets and $\mathbf u$ denotes
collectively the $3g$ number of constant bosonic modes from
the vector multiplet. According to the convention to count real multiplets,
$N_f$ matters charged to the $SO(2)$ gauge group and ${\rm dim}(S)$ singlets lead to
$N=2N_f + {\rm dim}(S)$. The prefactor of the integrand arises from the fermionic
zero-mode integral of $N+g$ copies of the fermion bilinear;
It is the Pfaffian of a rank $2(N+g)$ matrix
\bea \label{pfaffian1dGeneral}
M(X,{\bf u})\;=\;\left( \begin{array}{cc}
u_i^r (T_r^{\rm adj})_{pq}(\epsilon \cdot \sigma^i)_{ab}
& (T_qX)_\nu \epsilon_{ac}  \\ &\\
(X T_p^\dagger )_\mu\epsilon_{db}  & \partial_\mu\partial_\nu W\epsilon_{dc}
+ u_i^r(T_r)_{\mu\nu}(\epsilon\cdot \sigma^i)_{dc}\end{array}\right) \,,
\eea
where $T_p$ is the representation of the Lie Algebra of
$\cG$ on $X$'s, and $T^{\rm adj}_r$ the adjoint representation.
Note that in addition to the spatial zero modes $u_{1,2}$,
its time-like counterpart arises as $u_0\equiv \Theta/\beta$,
which means that on the right hand side of (\ref{g-start})
the integration over $\Theta$ is now converted to a vanishingly
small neighborhood of $\Theta=0$. This is a somewhat trivial
example of the holonomy saddle, to which we come back below.

For now, let us see how the factor $\beta$ can be removed from
the expression. The bosonic potential has three types of terms,
$\sim (\partial W)^2$, $\sim {\bf u}^2 X^2$, and $\sim {\bf u}^4$.
Unless $W$ is purely cubic there is no scaling of variables that remove
the overall factor $\beta$ cleanly from the exponent. However,
since $W$ itself can be scaled without affecting the genuine index
$\cI$, we may as well introduce
\bea
X \leftarrow \beta^{1/4} X\ , \quad
{\bf u} \leftarrow \beta^{1/4} {\bf u} \,,
\eea
and in addition scale $\beta^{3/4}W\rightarrow \tilde W$, so that
\bea
\partial \tilde W \leftarrow \beta^{1/2}\partial W
\eea
is kept finite as a function of $\beta$-rescaled $X$. This in particular
demands appropriate scalings of coefficients in $W$.
Only if $W$ is purely cubic in $X$, no separate scaling of $W$ is necessary.
As we have already noted for matter-only theories, however,
such scaling is harmless away from codimension-one walls of marginal stability.

In terms of the newly rescaled quantities, we have
\bea
M\;=\;\beta^{-1/4}\tilde M \,,
\eea
where $\tilde M$ denotes the same matrix but with the rescaled
$X$'s and ${\bf u}$ in places of the original $X$'s and ${\bf u}$
and with $\tilde W$ in place of $W$.
Hence we have
\bea
{\rm Pf}(M)=\beta^{-(N+g)/4}{\rm Pf}(\tilde M) \,,
\eea
which  cancels the factor of $\beta$ left over after
absorbing $\beta^{1/4}$'s to the integral variables
\bea
\beta^{N/2+g}\int d^NX d^{3g}{\bf u}\;\;\rightarrow\;\;
\beta^{(N+g)/4} \int d^N X d^{3g}{\bf u}
\eea
precisely. Therefore, we find
\bea\label{g-start2}
\frac{1}{(2\pi)^{N/2+g}{\rm vol}({\cG})}
\int d^N X d^{3g}{\bf u}\;\; {\rm Pf}(\tilde M )
\; e^{- \tilde V_{\rm bosonic}( X, {\bf u})} \,,
\eea
which brings the path integral to a finite-dimensional one.
Throughout this note, when the integral involves no
factor of $\beta$ explicitly, the variable should be understood
as those rescaled.

Most generally, however, there could be further additive
contributions to $\Omega$,
similar to (\ref{g-start}) but from the expansion of the gauge
variables of type $\Theta=\Theta_*+\beta u_0$ where
$e^{i\Theta_*\cdot G}\neq 1$. One thus finds a sum over
$\Theta_*$'s,
\bea\label{HS}
\Omega=\sum_{\Theta_*} d_H \cZ_{H} \,.
\eea
where $\cZ_H$ is the partition function of the low-lying theory
$H$ at $\Theta_*$, analogous to the right hand side of
(\ref{g-start2}) but for a different theory $H$, while
the factor $d_H$ is from integrating out heavy modes of
masses $\sim 1/\beta$. Each such $\Theta_*$
contribution arises from a vanishingly small local neighborhood around
$\Theta_*$ in the small $\beta$ limit, and (\ref{g-start2}) is
merely one such at $\Theta_*=0$.

The criterion for the contributing $\Theta_*$ is that the low energy
theory $H$ there has nonzero partition function $\cZ_H$. This in
turn requires that the light sector sitting at such a point $\Theta_*$
admits no decoupled fermions \cite{Hwang:2017nop}. For example,
one obvious class of $\Theta$'s which cannot be a contributing saddle
are those holonomies where the light theory $H$ has
a decoupled Abelian gauge sector; the resulting free gaugini
which kill any reasonable form of the partition function $\cZ$
there in small $\beta$ limit. At generic $\Theta$, all charged
field content tends to acquire a heavy mass of order $\Theta/\beta$,
leaving behind the free Cartan sector, which suffices to tell us
that the contributing $\Theta_*$ would occur at most discretely.

Such pairs $(\Theta_*,H)$ have been dubbed the holonomy saddles in the context
of supersymmetric gauged quantum mechanics \cite{Hwang:2017nop}
as well as for various partition functions of supersymmetric
gauge theories in higher dimensions \cite{Hwang:2018riu}.
Although this phenomenon persists in all spacetime dimensions,
it takes a particularly simple form for $d=1$. In the latter
$d_H$ reduces to a numerical factor whereas, in higher spacetime
dimensions, it can carry further nontrivial physics such as
the Cardy exponent \cite{Hwang:2018riu,DiPietro:2014bca}.

A simplest example of the non-trivial holonomy saddles can be
found with an $SO(2)$ theory with both unit-charged and double-charged
matter multiplets. $\Theta_*=0$ is the naive saddle where
the theory $H$ has the same field content as the original
theory. The other holonomy saddle sits at $\Theta_*=\pi$,
where the $H$ theory is again an $SO(2)$ theory but with
double-charged matter only, and $d_H$ would be from
integrating out the unit-charged matter. For other values
of $\Theta$, the would-be $H$-theory consists of a pure $SO(2)$
gauge theory, whose $\cZ_H$ vanishes identically due to the
decoupled gaugino. Thus the holonomy saddles of this simple
theory consist of $\Theta_*=0$ and $\Theta_*=\pi$. Likewise,
if an $SO(2)$ theory admits a charge $q$ scalar multiplet,
we will find holonomy saddles at $\Theta_*=2\pi n/|q|$ for
$n=0,1,\cdots,|q|-1$.

Another special class of holonomy saddles is when one
finds $\Theta_*\neq 0$ but $e^{i\Theta_*\cdot G}$ acts on
the entire field content trivially. Each such saddle contributes
the same amount as the canonical $\Theta_*=0$ saddle in (\ref{g-start}).
The net effect of these can be incorporated as a numerical
multiplicative factor in front of expressions like (\ref{g-start}).
Alternatively, the same effect is achieved by using the version
of $\cG$ divided by the subgroup defined by a collection of
such mutually equivalent holonomies. For such theories in $d=1$,
the net effect of these can be achieved by resorting to (\ref{g-start})
but taking care to choose $\cG$ in (\ref{g-start}) to be ``minimal"
by modding out by any discrete subgroup that acts trivially on
the field content. For instance, if charged matter $X$ is absent,
$\cG$ would be the one divided by the center. For $d>1$, however,
such shortcut is no longer valid because the multiplicity of such
holonomy saddles depends on spacetime dimension. As such, we must
honestly evaluate the sum (\ref{HS}).

\subsection{A Prototype: $SO(2)$ with Massive Charged Matters }
\label{sec:so2gap}

As a prototype, let us consider a
simple $SO(2)$ theory, now with $N_f$ massive real two-component charged scalars
$\vec X_I = (X_{I(1)} , X_{I(2)}) $ rotating in a pair,
with charge $q_I$ under the $SO(2)$, or equivalently $q_I$-charged
$N_f$ complex scalar multiplets $\Phi_I = (X_{I(1)}+iX_{I(2)})/\sqrt{2}$.
Let us take the superpotential,
\begin{align}
W = \frac12 \sum_{I= 1}^{N_f} m_I |\vec X_I |^2 \,,
\end{align}
The $\beta$-scaled bosonic potential and Pfaffian are given by
\begin{align}
\tilde V_\text{bosonic} &= \sum_{I = 1}^{N_f} \half \left(m_I^2+q_I^2 {\bf u}^2\right) |\vec x_I|^2, \\
\text{Pf} (\tilde M) &= \sum_{I = 1}^{N_f} \left[\prod_{J \neq I}^{N_f} \left(m_J^2+q_J^2 {\bf u}^2\right)\right] \, m_I q_I^2 |\vec x_I|^2.
\end{align}
Although this example can be integrated straightforwardly,
we take an alternate route that will be useful for generalizing
to more general field content and the superpotential.

Let us
define
\begin{align}
\vec y_I &=  \sqrt{m_I^2+q_I^2 {\bf u}^2} \cdot \vec x_I \,,
\end{align}
whereby the bosonic potential is completely written as the sum of their squares.
With the accompanying Jacobian
\begin{align}
\prod_{I = 1}^{N_f} \left(m_I^2+q_I^2 {\bf u}^2\right)^{-1} \,,
\end{align}
we arrive at
\begin{align}
 \Omega^{\Theta_* =0} = \frac{1}{(2 \pi)^{N_f+2}} \int d^3{\bf u} \, \sum_{I = 1}^{N_f} \frac{m_I q_I^2 }{\left(m_I^2+q_I^2 {\bf u}^2\right)^2}\int d^{2N_f}y\; |\vec y_I|^2 e^{-\sum_{J = 1}^{N_f} \half |\vec y_J|^2} \,,
\end{align}
where the matter sector and the gauge sector become factorized.
The universal gauge sector, parameterized by variables $\bf u$, gives
\begin{align}
\int_{\mathbb R^3} d^3{\bf u}  \frac{m_I q_I^2 }{\left(m_I^2+q_I^2{\bf u}^2\right)^2}  = \pi^2  \frac{m_I}{|m_I|} \frac{1}{|q_I|} \,,
\end{align}
so that we end up with
\bea\label{massive0}
 \Omega^{\Theta_*=0}
= \frac{1}{2} \, \sum_{I = 1}^{N_f} \, \frac{m_I}{|m_I|} \frac{1}{|q_I|} \,.
\eea
This simple trick needs to be refined for more general theories,
but this suffices as an illustration for now.

With $|q_I|> 1$, the holonomy saddles enter the story, which is
why we labeled the above by $\Theta_*=0$. The additional saddles
appear at $\Theta_l = \frac{2 \pi l}{|q_I|}$ for $l = 1, \ldots, |q_I|-1$,
since the holonomy saddles are given by
\begin{align}
\left\{ \left. \quad 0 \leq \Theta_l < 2 \pi \quad \right| \quad \exists \, I \quad \text{s.t. } \quad \frac{|q_I| \Theta_l}{2 \pi} \in \mathbb Z \quad \right\}.
\end{align}
At the saddle $\Theta_l$, only the matters whose charge satisfies a condition $ |q_I| \Theta_l \in 2 \pi \mathbb Z$ contribute to the partition function. Thus, the contribution from $\Theta_l$ is
\begin{align}
\Omega^{\Theta_l} = \frac{1}{2} \sum_{I = 1}^{N_f} \delta_{\frac{|q_I| \Theta_l}{2 \pi}} \frac{m_I}{|m_I|} \frac{1}{|q_I|} \,,
\end{align}
where $\delta_{\frac{|q_I| \Theta_l}{2 \pi} }$ is defined by
\begin{align}
\delta_{\frac{|q_I| \Theta_l}{2 \pi}} = \left\{\begin{array}{cc}
1 \,, \quad & {|q_I| \Theta_l} \in 2 \pi \mathbb Z \,, \\
0 \,, \quad & \text{else} \,.
\end{array}\right.
\end{align}
It satisfies the following condition
\begin{align}
\sum_l \delta_{\frac{|q_I| \Theta_l}{2 \pi}} = |q_I| \,,
\end{align}
because each matter multiplet appears $|q_I|$ times over the entire holonomy saddles. Therefore, the total twisted partition function is given by
\begin{align}
\label{eq:massive charged chiral}
\Omega = \sum_l d_l \Omega^{\Theta_l} = \frac{1}{2} \sum_{I = 1}^{N_f} {\rm sgn}(m_I) \,.
\end{align}
Although the holonomy saddles are not all equivalent if $q_I\neq q_J$
for some pairs, the end result is merely to multiply $|q_I|$ to $1/|q_I |$
for each summand in (\ref{massive0}).

The result in \eqref{eq:massive charged chiral} by itself is
not acceptable as the Witten index $\cI$ of the $d=1$ theory; these
fractional answer appears because of the three flat
$\bf u$ directions with $E_{\rm gap}=0$, sitting at $\vec X=0$.
Note that, however, the answer is an integer multiple of $1/2$,
so it cannot be some random continuum contribution either.
In the final section we will make sense of it in the context
of $d=3$ theories.

\subsection{Gaussian Reduction}\label{sec:gaussian}

Recall how the index for matter-only theories reduces to a pure Gaussian
integral in the space of $\partial W(x)$, which can be, under a suitable
condition on the asymptotic data, expressed either as the winding number
of the map $x\rightarrow \partial W(x)$ or the Morse counting. Although
we have found a similar change of variables for the simple example above,
where the charged matter enters $W$ quadratically at most, this turns out
to be not universal enough. When we begin to discuss $d=3$ index,
we will make a good use of these variable choices, for
an additive piece proportional to the Chern-Simon level. For the
current $d=1$ computation, however, we need something a bit more
sophisticated to turn $\Omega$ into Gaussian integrals much like (\ref{thm1}).

Here, we present one such approach for general $SO(2)$ gauge theory
with superpotential being an arbitrary function of the ``radial'' variables
$\rho_I= | \vec X_I |^2 / 2$, and possibly of
singlets $P_A$'s as well. In other words, we precluded $\vec X_I \cdot \vec X_J$ term with
$I \neq J$. What have we lost in doing so? Note that such a restriction
would come about automatically if we impose $SO(2)^{N_f-1}$ flavor
symmetries. Such flavors, however, have been used routinely in
recent supersymmetric path integral evaluations, and also by us in
Section 2 for a contact with the Lefschetz fixed point theorem.

The lesson from these is that restricting the Lagrangian or
adding (mass) terms to reinforce such a symmetry is innocuous
as long as the original dynamics is sufficiently gapped so that
the usual topological robustness of index, or more precisely the
twisted partition function, can be argued. As such our restriction,
\bea
W=W(\rho_I \,, P_A) \,,
\eea
possibly with neutral $P_A$'s, is well-justified as long as $dW$
is sufficiently divergent along all asymptotic directions.

With such $W(\rho_I \,, P_A)$,
we decompose its $\beta$-scaled Pfaffian of fermion bilinear based on the recursive definition of Pfaffian
\begin{align}\label{pfrecursive}
\text{Pf} (A) = \sum_{j=2}^{N}  (-1)^j  a_{1j} \text{Pf} (A_{\hat 1 \hat j}) \,,
\end{align}
of $N \times N$ anti-symmetric matrix $A = (a_{ij}) $, where $A_{\hat 1 \hat j}$ denotes $(N-2) \times (N-2)$ minor of matrix $A$ with both the first and $j$-th rows and columns removed.
Then we seek a series of appropriate parameterizations for minors so that each renders the part of integrand to a Gaussian form.
The Pfaffian of fermion bilinear $\tilde M$ of an Abelian gauge theory with $N_f$ doublets and ${\rm dim}(S)$ singlets is expanded as
\begin{align}\label{pfdecom1}
\text{Pf} ( \tilde{M} ) = 
\sum_{I} \, (Q \vec X_I ) \cdot \text{Pf} ( \tilde{M}_I )  \,,
\end{align}
where ${\rm dim}(S)$ stands for the number of singlets $P$ introduced in the theory. Recall $T_q$, a gauge-symmetry generator in \eqref{pfaffian1dGeneral} is replaced with $Q$, Abelian $SO(2)$ gauge symmetry charge by which $\vec X_I$ transforms.
In the summand, $(Q \vec X_I ) \cdot \text{Pf} ( \tilde{M}_I ) $ stands for a collective expression:
\begin{align}\label{pfdecom1-1}
(Q \vec X_I ) \cdot \text{Pf} ( \tilde{M}_I )  \equiv (- q_I X_{I (2)}) \cdot \text{Pf}(\tilde M_{I(1)} )  + (q_I \vec X_{I(1)}) \cdot \text{Pf} ( \tilde M_{I(2)} ) \,,
\end{align}
where $q_I$ is the charge of $\vec X_I$
and $\tilde M_{I(j)}$ is a fermion bilinear matrix with both ($\lambda$, $\psi_{I(j)}$) columns and rows removed.
Apparently, a singlet $P$ does not appear in \eqref{pfdecom1} due to its charge neutrality $Q P= 0$. However, the Pfaffian depends on $P$ as the second derivative terms $\partial^2 W$ in minor $\tilde M_{I(i)}$ involves $P$ in general.

For the $I$-th doublet with $I \in \{ 1, \cdots, N_f \} $, a set of integration variables are suggested by
\begin{align}\label{GaugeGaussian1}
\begin{split}
&{\bf v} \equiv |q_I| | \vec x_I | {\bf u} \,, \quad  \vec z_I  \equiv \frac{\partial W}{\partial \rho_I} \vec x_I \,, \quad   s_A \equiv \frac{\partial W}{\partial p_A} \\
& \text{and} \quad  \vec y_{J} \equiv \sqrt{ q_J^2 {\bf u}^2   + \left( \frac{\partial W}{\partial \rho_J} \right)^2 } \cdot \vec x_J  \quad (\text{for } J \neq I ) \,,
\end{split}
\end{align}
upon which we have
\begin{align}\label{GGIntegrand1}
(Q \vec X_I ) \cdot \text{Pf} ( \tilde{M}_I ) \, \left| \frac{\partial ( \vec x \,, p \,, {\bf u} )}{\partial (\vec z, \vec y \,, s \,, {\bf v}  )} \right|
=  \frac{1}{|q_I| W_I |\vec x_I| } = \frac{1}{|q_I| \, \text{sgn}(W_I) |\vec z_I |} \,,
\end{align}
for the $(Q \vec X_I )$-led piece \eqref{pfdecom1-1} of the expanded Pfaffian where $W_I \equiv \frac{\partial W}{\partial \rho_I}$. Then the bosonic potential in terms of the $I$-th set of new parametrization reads
\begin{align}\label{Vgaussian1d}
\tilde V_\text{bosonic} = \half {\bf v}^2+\half \vec z_I^{\, 2} + \sum_{J\neq I} \half \vec y_J^{\, 2} + \sum_{A=1}^{{\rm dim}(S)} \half s_A^2 \,.
\end{align}
The full integral formula for twisted partition function $\Omega$ is arranged by the sum of $N_f$ integrations with the holonomy saddle contribution counted.
Assume that a holonomy saddle $\Theta= \Theta_*$ accommodates $N_* (\leq N_f )$ charged scalar fields $\vec X_I$, with $I \in \mathcal N_*$, a set of $N_*$ indices. Then the full integral formula for $d=1$ index is
\begin{align}
\Omega = \sum_{\Theta_* }\sum_{I \in \mathcal N_* } \Omega_I^{ \Theta_* } \,,
\end{align}
where the $(Q \vec X_I )$-led piece of twisted partition function $\Omega_I^{\Theta_*}$ around the holonomy saddle $\Theta_*$ reads
\begin{align} \label{omg1}
\Omega_{I}^{\Theta_*} = \frac{1}{(2 \pi)^{N_*+2+\half {\rm dim}(S) }} \int  d^{\text{dim} (\mathcal{S})} s \, d^3{\bf v} \, d^{2(N_*-1)} y \, d^2z_I
\frac{1}{| q_I | \, \text{sgn}(W_I) |\vec z_I| } e^{-\tilde V^*_\text{bosonic}}\,.
\end{align}
Note the bosonic potential $\tilde V^*_\text{bosonic}$ in the exponential is given by
\bea
\tilde V^*_\text{bosonic} = \half {\bf v}^2+\half \vec z_I^{\, 2} + \sum_{ \substack{ J\neq I \\ J \in \mathcal N_* }} \half \vec y_J^{\, 2} + \sum_{A=1}^{{\rm dim}(S)} \half s_A^2 \,.
\eea
Armed with this, let us briefly revisit the example discussed in Section \ref{sec:so2gap} as well as
other examples with higher order terms in $W$.

\subsubsection*{$N_f$ Massive Charged Matters}

With this new change of variables, let us recompute the
$SO(2)$ theory with charged massive fields.
A quadratic superpotential $W=\sum m_I \rho_I = \sum \half m_I |\vec X_I|^2 $ gives
 \begin{align}
 W_I = m_I \,, \quad W_{I,J} = 0 \quad \text{for any } I, J \in \{1, 2, \cdots, N_f \}  \,.
 \end{align}
 A partial contribution to the twisted partition function is
 \begin{align}\label{massivehalf}
 \Omega_{I}^{(\Theta = 0)} = \frac{\text{sgn}(m_I)}{(2\pi)^{N_f+2}| q_I |} \int d^3{\bf v} \,d^{2(N_f-1)} y\, d^2z_I\frac{1}{|z_I |} e^{-\half v^2 -\half z_I^2- \sum_{J\neq I} \half \vec y_J^2} = \half \frac{\text{sgn}(m_I)}{|q_I |} \,.
 \end{align}
The holonomy saddles with some $|q_I | > 1$ mean effectively a multiplicative factor of $|q_I|$
for $\Omega_{I}^{(\Theta = 0)}$, reproducing \eqref{eq:massive charged chiral},
\begin{align}\label{eq:omega1dmassive}
\Omega = \sum_{I=1}^{N_f} \sum_{l=0}^{|q_I|-1} \Omega_I^{\Theta_l} =  \half \sum_{I = 1}^{N_f}  \text{sgn} (m_I)   \,.
\end{align}
Note that the sum in \eqref{eq:omega1dmassive} results from shuffling
the sum over the holonomy saddle and the sum over charged matters.

\subsubsection*{$\IW\IC\IP$}

One way to avoid Coulombic flat directions, responsible for
the above fractional result, is to
introduce a neutral scalar multiplet, say $P$, which imposes
a symmetry breaking condition among the charged
matters. Consider, for example,
\bea
W=PK(\rho_I)+L(\rho_I) \,,
\eea
where $K$ and $L$ are functions of gauge-invariant ``radial'' variables $\rho_I= | \vec X_I |^2 /2 $.
One can see easily
that, with generic enough $K$ and $L$, the
classical solutions to the supersymmetry condition would
be isolated points in the space spanned by $\{P;\rho_I\}$.
Each of such points will generate contribution of $\pm 1$.

As such, let us consider a $\mathbb{WCP}^{N_f-1}$ model,
whose charge-dependent superpotential $W(\rho, P)$ is given by
\begin{align}
W = C P \left( \sum_{I=1}^{N_f} q_I \frac{| \vec X_I| ^2}{2} - \xi \right) = C P \left( \sum_I q_I \rho_I - \xi \right) \,,
\end{align}
requires the change of variables for each $I= 1 \,, \cdots \,, N_f$ ,
\begin{align}
\begin{split}
&  {\bf v} \equiv |q_I| | \vec x_I | {\bf u} \,, \quad \vec z_I  \equiv  c p q_I \vec x_I \,, \quad  s \equiv c\left( \sum_J q_J \rho_J - \xi \right)  \\
& \text{and} \quad  \vec y_K \equiv \sqrt{q_K^2 c^2 p^2 + q_K^2 {\bf u}^2  } \cdot \vec x_K  \quad (\text{for } K \neq I ) \,,
\end{split}
\end{align}
which brings us to
\begin{align}
\Omega_{I}^{\Theta_* = 0} = \frac{\text{sgn}(q_I c)}{(2\pi)^{N_f+2+1/2} |q_I |} \int ds \, d^3{\bf v} \, d^{2(N_f-1)}y\, d^2z_I \left[ \frac{1}{|\vec z_I |} - \left(-\frac{1}{|\vec z_I |} \right) \right] e^{-\tilde V^{\Theta_*=0}_{\rm bosonic} }\ ,
\end{align}
with $s \in \text{sgn}(c q_I) ( + \infty \, , \, - \text{sgn}(q_I) |c| \xi ) $
coming from a half range of $p>0$; the integration over $p<0$ is reflected
in the second term in the square bracket. As before, we have a quadratic exponent
\bea
\tilde V_{\rm bosonic}^{\Theta_*=0} = \half {\bf v}^2 + \half s ^2 +\half \vec z_I^{\, 2} + \half \sum_{K\neq I} \vec y_K^{\, 2} \,,
\eea
and the integral is entirely Gaussian. Note how the integral is
split into two parts separated by $\partial_{\rho_I} W=0$. This
will be perhaps the most important lesson we learn here.

Therefore,
the partial contribution to the index $\Omega_I$ at trivial saddle is,
\bea
\Omega_I^{\Theta_*=0}
&=& \left\{\begin{array}{lr} -\frac{ 1}{| q_I |} &\quad  q_I \xi >0 \\0& q_I \xi <0 \end{array}\right. \,.
\eea
Note how the sign $\text{sgn}(c q_I)$ canceled out in the end.
Collecting $|q_I |$  holonomy saddle contributions, we find
\begin{align}\label{wcpomega1d}
\Omega =  \sum_{I=1}^{N_f} \sum_{l=0}^{|q_I|-1} \Omega_I^{\Theta_l} = \left\{\begin{array}{lr} -N_f &\quad  q_I \xi >0 \\0& q_I \xi <0 \end{array}\right. \,,
\end{align}
where we have assumed that all charges are of the same sign.
This model has no asymptotic gapless directions, hence we find
an integral index in the end.

\subsubsection*{Multiple Charged Scalars and No Neutrals}

Generalizing the previous example,
let us turn to a theory of multiple charged scalars with a general
radial superpotential $W=W(\rho_1, \cdots, \rho_{N_f})$.
Recall the index is expressed by the sum of at most $(2N_f + 3)$-dimensional integrals \eqref{omg1}.
Each term is defined per charged matter at given holonomy saddle.
Take a specific $(Q \vec X_I)$-led term $\Omega_I$ at the trivial saddle, for example,
\begin{align}\label{omg1-sec3.4}
\Omega_{I}^{\Theta_*=0} = \frac{1}{(2 \pi)^{N_f+2 }} \int  \, d^3{\bf v} \, d^{2(N_f-1)} y \, d^2 z_I
\frac{\text{sgn}(W_I)}{| q_I | \, |\vec z_I| } e^{-\tilde V_\text{bosonic}^{\Theta_*}} \,,
\end{align}
where the change of variables is noted by \eqref{GaugeGaussian1} for specific $I \in \{ 1, \cdots, N_f \} $.

An important point on the integration variables and the domains thereof
should be addressed. Given how the superpotential is a function of the
radial variables $\rho_I$'s rather than individual $\vec X_I$, it is
important to consider \eqref{omg1-sec3.4} as an integral over $\rho_I$. This means
that the domain is really $\IR_+^{N_f}$, which we will split as
\bea
\mathbb R_+^{N_f} = \cup_{a} \Delta_I^{(a)} \,,
\eea
separated by the locus  $0=W_I \equiv \partial_{\rho_I}W$,
so that $W_I $ has a definite sign on each subdomain $\Delta_I^{(a)}$.
With this, \eqref{omg1-sec3.4} can be more precisely written as
\begin{align} \label{omg1-decomposed}
\Omega_I^{\Theta_*=0} = \sum_{a} \frac{\text{sgn}(W_I(\Delta_I^{(a)}))}{(2\pi)^{N_f+2} |q_I| } \int_{\mathbb R^3} d^3{\bf v} \int_{Y_{\Delta_I^{(a)}}} d^{2(N_f-1)}y \, d^2 z_I  \frac{1}{|\vec z_I |} e^{-\tilde V_\text{bosonic}} \,,
\end{align}
where $Y_{\Delta_I^{(a)}}$ is the image of $\Delta_I^{(a)}$ under the map \eqref{GaugeGaussian1}.
Let us introduce following variables parametrizing the image $Y_{\Delta_I^{(a)}}$,
\bea
\lambda_I = \half \vec z_I^{\, 2} \,, \quad \sigma_{J (\neq I)} = \half \vec y_J^{\, 2} \,,
\eea
which span $\tilde{\mathbb R}_+^{N_f}$ at most.
Our claim is that integral \eqref{omg1-decomposed} vanishes for most of these domains,
leaving behind a single unbounded domain for each $I$, to be denoted by $\Delta^\infty_I$
with the image $Y_{\Delta_I^{\infty}}$.

Note that the boundary $\partial \Delta_I^{(a)}$ is composed of
\begin{align}
W_I =0 \,, \quad \rho_I = 0 \,, \quad \rho_{J(\neq I)} =0 \,,
\end{align}
and the boundary at infinity. The points satisfying one of the first two equations
are mapped to the $\lambda_I = 0$ hyperplane by the reparameterization \eqref{GaugeGaussian1}
while the third kind is mapped to the $\sigma_{J (\neq I)} =0$ hyperplane.
This restricts $Y_{\partial \Delta_I^{(a)}}$, the image of boundary $\partial \Delta_I^{(a)}$,
within $ \lambda_I = 0 $ or within $\sigma_{J(\neq I)}=0$ hyperplane.

\begin{figure}[tbp]
\begin{center}
\resizebox{0.9\hsize}{!}{
\includegraphics[height=5cm]{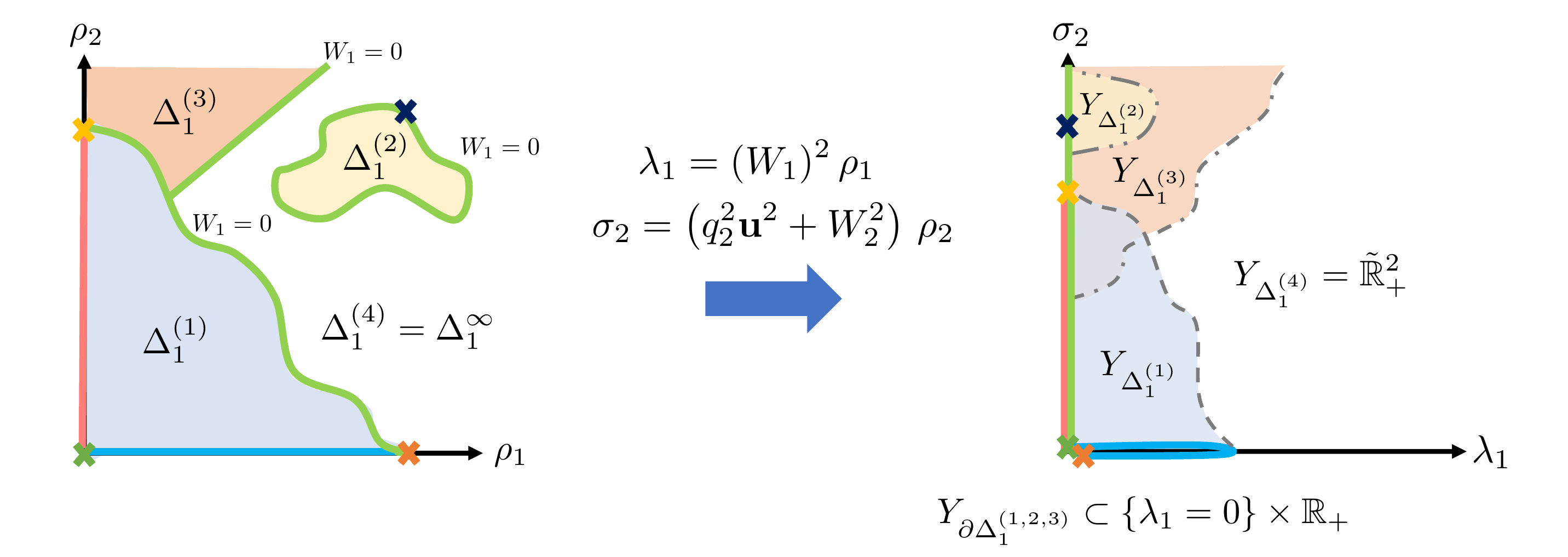}
}
\caption{
The domain decomposition and mapping under reparameterization \eqref{GaugeGaussian1} for $\Omega_{I=1}$ of the $N_f=2$ theory.
The angular part is projected out to make the drawing simple.
The image of boundary is different from the boundary of the image, i.e.,
$Y_{\partial \Delta_I^{(a)}} \neq \partial  Y_{\Delta_I^{(a)}}$, generally, which
underlies how most domains cancel out in the integral.
$Y_{\Delta^\infty_{I=1}}=Y_{\Delta_1^{(4)}}$, whose projection to this $(\lambda_1,\sigma_2)$ plane
equals the entire quadrant $\tilde{\mathbb R}_+^2$, is the only domain that can contribute
to $\Omega_{I=1}^{\Theta_*=0}$. \label{fig:omg0-radial}}
 \end{center}
 \end{figure}

Figure \ref{fig:omg0-radial} briefly depicts the splitting and mapping of the matter
integration domain for $N_f=2$ theory, for $\Omega_{I=1}$. The green curve in
the left quadrant is $W_1 = 0$. Note how, for exmple, all components of
$\partial \Delta_1^{(1)}$, the bold colored lines, are all mapped to the
two axes in the right figure. The dashed line on the right
quadrant is the part of the boundary $\partial Y_{\Delta_1}^{(1)}$
that doesn't belong to the image of $\partial \Delta_1^{(1)}$. Here, the image of
$\Delta_1^{(1)}$  folds itself, which means that the integral cancels out
between the two sides. Such a folding could happen multiple times with
the same result. This means
\bea
\frac{\text{sgn} (W_1 (\Delta_1^{(1)}))}{(2\pi)^{N_f+2} |q_I| } \int_{\mathbb R^3} d^3 {\bf v} \int_{Y_{\Delta_1^{(1)}}}
d^{2(N_f-1)}y \, d^2 z_I \frac{1}{|\vec z_I|} e^{-\tilde V_{\text{bosonic}}} = 0 \,.
\eea

The same mechanism works for most of the domains, bounded or unbounded.
What matters is how the part of the boundary $\partial \Delta^{(a)}_I$ sitting at the locus
$W_I=0$ is mapped to the axis $\lambda_I=0$. The other two possible boundaries
are either $\rho_I=0$ or $\rho_I=\infty$. We will call $\Delta_I^\infty$
the domain that has the latter as part of the boundary. For all the other $\Delta_I^{(a)}$'s,
the boundary is mapped to $\lambda_I =0$ or $\sigma_{J \neq I}=0$ in such a way
$Y_{\Delta_I^{(a)}}$ cancels itself out in the Gaussian integration
(\ref{omg1-decomposed}) due to orientation flips.
$\Delta_I^\infty$ is an exception to this, however, since one side of
the boundary goes off to $\rho_I \rightarrow +\infty$, which is clearly
not mapped to $\lambda_I=0$, so the cancelation of the integral need not happen.
Rather, its image $Y_{\Delta_I^\infty}$ fully expands $\mathbb R_+^{N_f}$.

Hence, the surviving part of the integral is one over a single copy of
$Y_{\Delta_I^\infty}$
\begin{align}
 \frac{\text{sgn}(W_I^\infty) }{(2\pi)^{N_f+2} |q_I| } \int_{Y_{\Delta_I^\infty}} d^{2(N_f-1)}y \, d^2 z_I \int_{\mathbb R^3} d^3 {\bf v} \, \frac{1}{|\vec z_I|} e^{-\tilde V_{\text{bosonic}}} \,,
\end{align}
where $\text{sgn}(W_I^\infty)$ is defined by the limit along specific direction
\begin{align}
\text{sgn}(W_I^\infty) \equiv  \text{sgn}(W_I(\Delta_I^\infty))
=\lim_{\substack{ \rho_I \rightarrow \infty \\ \rho_{J(\neq I)} \rightarrow 0} } \text{sgn} \left( \frac{\partial W}{\partial \rho_I} \right) \,.
\end{align}
Since the domain $Y_{\Delta_I^\infty}$ is always identical to $\tilde{\mathbb R}_+^{N_f}$,
a theory with the radial superpotential always gives rise to the index of
following form at the trivial holonomy saddle
\bea\label{1d basic}
\Omega_I^{\Theta_* = 0} = \half \frac{\text{sgn} (W_I^\infty  )}{|q_I |} \,.
\eea

At each non-trivial holonomy saddle, a similar logic makes the computation simple:
a bounded subdomain never contributes to the index, while a specific unbounded domain generates a non-zero value,
depending on the asymptotics of the superpotential.
A straightforward holonomy saddle counting for charge-$q_I$ matters yields
\bea \label{index-radial-infty}
\Omega =\sum_{I=1}^{N_f} \sum_{l=0}^{|q_l|-1}  \Omega_I^{\Theta_l} = \half \sum_I \text{sgn}(W_I^\infty) \,,
\eea
for generic radial $W$ without asymptotic flat directions. This completes the proof.

If we further assume that $W=c_W \cdot G(\rho_1,\cdots, \rho_{N_f})$ where
\bea
\partial_I G(\rho) \rightarrow +\infty
\eea
for each $I$, along all asymptotic directions in the $\rho$ space, we have a simplified version of \eqref{index-radial-infty}. Again $G$ is an analytic function
with no singular behavior, especially at the origin. For instance this
is guaranteed if $G$ is a polynomial with top degrees forming
a non-degenerate homogenous polynomial with all positive coefficients
since $\rho_I$'s all live in the positive half lines.
Then \eqref{index-radial-infty} is translated into
\bea\label{Nf charged}
\Omega = \frac{N_f}{2} \text{sgn}(c_W) \,.
\eea
A simple example of this is the $N_f=2$ Abelian gauge theory with
quartic superpotential $W$,
\begin{align} \label{quarticNf2}
W(\rho_1, \rho_2) = \frac{c_W}{4} (\rho_1 + \rho_2 - v)^2  \,.
\end{align}
Given a positive real constant $v$, the integration domain for $\Omega_1$ can be split as follows,
\begin{align}
\Delta_1^{(1)} = \{ (\rho_1, \rho_2) \, | \,  \rho_1 + \rho_2 - v < 0 \} \,, \quad  \Delta_{1}^{\infty} = \{ (\rho_1, \rho_2) \, | \, \rho_1 + \rho_2 -v > 0 \}  \nn \,.
\end{align}
In a similar way, the domain for the $\Omega_2$ integral is split as follows,
\begin{align}
\Delta_2^{(1)} =\{ (\rho_1, \rho_2) \, | \,  \rho_1 + \rho_2 - v < 0 \} \,, \quad \Delta_{2}^{\infty} =  \{ (\rho_1, \rho_2) \, | \, \rho_1 + \rho_2 -v > 0 \}  \nn \,.
\end{align}
The exchange symmetry $\rho_1 \leftrightarrow \rho_2 $ in \eqref{quarticNf2} makes the domain split in the same way for $\Omega_1$ and for $\Omega_2$.

Following the above argument, it turns out that $\Omega_{I \in \{ 1,2\}}$ integrals over $\Delta_{\{I=1,2\}}^{(1)}$ vanish. On the other hand, the integration over the unbounded domain converges to $\half \text{sgn} (c_W) / |q_I|$ since each domain covers $\mathbb R_+^2 \times \mathbb R^3 $ in $\{ \lambda_1 , \sigma_2 , \vec v \}$ and $\{ \sigma_1 , \lambda_2 , \vec v \} $ parametrizations, respectively. Thus
\begin{align}
\Omega^{\Theta_*=0}_I =  \frac{1}{2|q_I|}\cdot \text{sgn} (c_W)\,.
\end{align}
at the saddle at the origin. Computation at other saddles would be
isomorphic to this; the only difference would be that the set of
charged matters which enter the computation is selected by $\Theta_*$ and
is in general a subset of those at $\Theta_*=0$.
Summing over $|q_I |$  holonomy saddle contributions for each $I$,
we have the index of \eqref{quarticNf2}
\bea
\Omega =\left(\frac{|q_1|}{2|q_1|}+\frac{|q_2|}{2|q_2|}\right)  \text{sgn} (c_W) =  \text{sgn} (c_W) \,,
\eea
consistent with (\ref{Nf charged}).

\subsection{Why Is Coulombic Wall-Crossing Absent?}

Let us close the section with an observation.
Consider the following three prototypes of theories
we discussed above,
\bea
W_1=\frac12 m\,|\vec X|^2\ , \qquad
W_2= P\left( \half |\vec X|^2-  \xi\right) \ , \qquad
W_3=\frac12 \left (|\vec X|^2-\xi\right)^2 \,,
\eea
with neutral and real $P$ and a unit-charged
two-component field $\vec X$. The direct computation of the indices
produced,
\bea
\Omega_1=\left\{\begin{array}{rr} 1/2 &\quad m>0 \\ -1/2 &\quad m<0\end{array}\right. \ , \qquad
\Omega_2=\left\{\begin{array}{rr} -1 &\quad \xi>0\\ 0 &\quad \xi<0\end{array}\right. \ , \qquad
\Omega_3=\half \,,
\eea
so that the former two experiences wall-crossing
while the last does not.

At $\xi=0$ all three theories
admit gapless flat directions, a necessary condition
for wall-crossing, so what is it that makes the
difference?
At $m=0$ or $\xi=0$,  the classical vacuum
manifolds are, respectively,
\bea
\{ {\bf u} \}\cup \{\vec X\}\ , \qquad \{ {\bf u} \}\cup \{P\}\ , \qquad \{{\bf u} \} \,,
\eea
so one immediate difference is the presence/absence
of the flat direction associated with the matter field.
In the last theory with no wall-crossing, at most the
Coulombic direction become asymptotically flat. Is there
a way to understand this phenomenon in general term?
It turns out that there is a simple explanation, not from
$d=1$ physics but rather from the possibility of embedding
such theories as the lowest Kaluza-Klein sector of a $d=3$
Chern-Simons theory.

Consider the above theories as coming from $\IT^2$ compactification
of $d=3$ $\cN=1$ gauge theories. One would compute
the Witten index of such a theory by performing the
path integral on $\IT^3$. Since the size of the
torus cannot enter the twisted partition function of
the gapless theory, one could imagine sending the spatial
radius to zero first and thus reduce the path integral to
that of a $d=1$ $\cN=2$ theory. If we were considering $d=3$
theories, therefore, the Coulombic direction is no longer
asymptotic as the bosonic variables live in a dual torus,
hence cannot produce asymptotic flat directions.
Is this why there is no wall-crossing in $d=1$ as well
when the asymptotic direction is entirely Coulombic?

Actually this is too rash.
Consider the well-known wall-crossing story with the twice
supersymmetries, where the $d=1$ $\cN=2$ vector multiplet
is accompanied by one more real adjoint scalar multiplet, call it
$\Sigma$. It is well-known that this system generically
wall-crosses due to wavefunctions that leak out to the
asymptotic Coulomb phase including the $\Sigma$ directions.
The Coulombic physics is notoriously dimension-dependent,
which actually underlies why $\cN=4$ quantum mechanics
wall-crosses while $d=2$ $\cN=(2,2)$ GLSM does not. As such the
above naive argument that $\bf u$'s originate from periodic
variables does not immediately preclude wall-crossing due to
the Coulombic asymptotic infinity in strict $d=1$ theories.

We believe that the difference comes from how the
Chern-Simons terms, when a $d=1$ gauge theory is lifted
to $d=3$, control infrared issues on the Coulombic side.
If one can embed a given $d=1$ $\cN=2$  gauge theory to
a $d=3$ Chern-Simons theory with arbitrary nonzero level $\kappa$,
the latter gaps out $d=1$ $\cN=2$ Coulombic physics by
turning it into a Landau problem, as we will see in next section.
Even in the lowest Landau level, infinitely
degenerate in the infinitesimal $\IT^2$ limit, individual
states are enumerative and of fixed size controlled by the
effective magnetic field and angular momentum in the $u_{1,2}$
plane. The runaway behavior of ground states necessary for
wall-crossing is impossible, and this makes wall-crossing
due to the Coulombic flat directions impossible.

One might wonder if this explanation makes any sense since
$d=1$ limit of the Chern-Simons level is subtle. With any finite
$\kappa$, one ends up with a Landau problem with infinite total
flux when a strict $d=1$ limit is taken. Thus we are effectively
studying a dimensional reduction of $\kappa=0$ theory. However,
as we  will see in the next section, $\kappa$-independent part
of the $d=3$ twisted partition functions, which we will call $\Omega_0$,
share the common building blocks with $d=1$ $\Omega$. These are
nothing but $\Omega_I^{\Theta_*=0}$'s above, and the difference
between $d=3$ $\Omega_0$ and $d=1$ $\Omega$ is merely in the
details of the holonomy saddles.

This means that $d=1$ wall-crossing behavior, which would be
entirely due to discontinuity of $\Omega_I^{\Theta_*=0}$, is
connected to $d=3$ wall-crossing of this $\kappa$-independent
$\Omega_0$.
The latter's insensitivity to $\kappa$ tells us, on the other hand,
that nothing special happens at $\kappa=0$ with $\Omega_0$, and
thus with $\Omega_I^{\Theta_*=0}$. Therefore, the absence of
the $d=1$ wall-crossing from the Coulombic continuum can be
argued with a mere logical possibility of $d=3$ Chern-Simons lift,
even though we are effectively studying dimensional reduction
of the $\kappa=0$ version.

With such a Chern-Simons uplift, the wall-crossing that
does happen in $d=1$ $\cN=4$ will occur due to the asymptotics
of $\Sigma$, the real adjoint matter that enlarges the vector
multiplet to the one with the twice supersymmetries. In a more
conventional localization computation, where one introduces
an $R$-charge chemical potential as an infrared regulator,
again it is the $\Sigma$ direction that is left unaffected and
responsible for the wall-crossing. In our $d=1$
$\cN=2$ viewpoint, however, $\Sigma$ is one of real
scalar multiplets. We conclude that in $d=1$ $\cN=2$ (gauge)
theories, a wall-crossing occurs only if there are asymptotic
gapless directions that emerge along the scalar multiplet side.

\section{$d=3$ $\cN=1$ Chern-Simons-Matter}\label{sec:csm}

In this final section, we will explore  $d=3$ version of
the same problem for $SO(2)$ Chern-Simons-Matter theories.
Witten indices of $d=3$ massive theories can be computed
by compactification on $\IT^2$, which would bring us to
an extension of $d=1$ computations, which we wish to exploit
here.

\subsection{$d=3$ $\cN=2$ Chern-Simons-Matter Revisited}\label{sec:3dN=2}

Before we delve into $d=3$ $\cN=1$ theories, it is worthwhile
to recall how vacuum countings work for $d=3$ $\cN=2$
theories \cite{Lee:1990it}.
For massive $d=3$ theories, the quantum vacuum counting
at a symmetric phase can be most easily done by integrating
out the matter fields first. For such contributions, one can
resort to the previous countings by Witten of pure Chern-Simons
theories \cite{Witten:1999ds}. When vacua with the gauge symmetry
broken are present, one must also consider them, so generally we have
\bea
\cI= \cI_{\rm symmetric}+\cI_{\rm broken} \,.
\eea
For $d=3$ $\cN=2$ theories with real masses, Intriligator and
Seiberg gave a beautiful demonstration \cite{Intriligator:2013lca}
of how to count these vacua and to classify them, and how
the sum remains robust although individual contribution might vary
with parameters of the theory.

Let us take a simple $\cN=2$ theory, say, an $SO(2)_\kappa$ theory with
a chiral multiplet $\Phi=(X_1+iX_2)/\sqrt{2}$ of charge $q>0$ for an
illustration. The bosonic potential takes the form
\bea\label{DV}
{\cal V}_{\cN=2} = \frac{1}{2} \left( q|\vec X |^2 -
\xi -\kappa\,\Sigma \right)^2 +
2 \, q^2\Sigma ^2|X|^2 \,,
\eea
modulo the electric coupling $e^2$, where the would-be $\cN=1$
scalar multiplet $\Sigma$ now belongs to the $\cN = 2$ vector
multiplet. Let us assume $|\kappa| > q^2/2$, just to avoid the clutter
coming from one-loop shift of $\kappa$. The potential admits two types of vacua,
\bea
\left\{\Sigma=0,\, q|\vec X|^2=\xi\right\},\qquad
\left\{\Sigma=-\xi/\kappa,\, |\vec X|^2=0\right\},\qquad
\eea
of which the former exists only if $\xi>0$. The existence of a
symmetric vacuum is a universal feature of $\cN=2$ Chern-Simons
theories for large enough $|\kappa|$, which will not be the case for $d=3$ $\cN=1$ theories.

Here we will follow a reasoning which is similar to \cite{Intriligator:2013lca}, but
in an $\cN=1$ theory language, with a care given to the  sign
convention for the Witten index. In our $\cN=1$ notation, the
would-be  $D$-term potential above
arises from the superpotential
\bea
W= \Sigma\left(q|\vec X|^2 - \xi -\frac12\kappa\,\Sigma\right) \,.
\eea
Note that the singlet $\Sigma$ comes with mass $-\kappa$. Let us
expand this in the two phases as
\bea
\delta W_{\rm broken}(\Sigma,\delta |\vec X|)&\simeq & + \sqrt{q\xi}
\; \Sigma\, \delta |\vec X| -\frac12\,\kappa\,\Sigma^2\ ,\cr\cr
\delta W_{\rm symmetric}(\delta\Sigma, \vec X)&\simeq &
-(q\xi/\kappa)\, |\vec X|^2 -\frac{\kappa}{2}\left(\delta\Sigma\right)^2 \,.
\eea
For instance, the broken vacuum for $\xi>0$ will contribute
\bea
\cI_{\rm broken}= \left\{\begin{array}{cc}
-\,q^2  \qquad & \xi>0 \\
0  \qquad & \xi<0 \end{array}\right.
\eea
The $q^2$ degeneracy can be understood from
$q^2$ many discrete gauge holonomies along spatial $\IT^2$,
remnants from the partial symmetry-breaking $SO(2)\rightarrow
\IZ_q$. Note that this is one more manifestation of
the holonomy saddle, although now in the context of
the canonical approach, as opposed to the path integral
approach. On the other hand, the minus sign arises from the
determinant of the $2\times 2$
mass matrix in $\delta W_{\rm broken}$; the contribution from these
two gauge singlet real fields $\Sigma$ and $\delta |\vec X|$ can be found
by compactification on $\IT^3$ and thus by our $d=1$ computation
easily.

In the symmetric phase, contributions from $\vec X$ and $\delta\Sigma$
play different roles. $\vec X$ can be first integrated out and shift
the Chern-Simons level to
\bea
\kappa_{\rm eff}=\kappa+\frac12\,{\rm sgn}(-\xi/\kappa)\,q^2 \,.
\eea
On the other hand, $\delta\Sigma$ factorizes out from the
Chern-Simons dynamics and merely affects the overall sign of
the index such that the contribution to the index from the symmetric vacuum
is
\bea
\cI_{\rm symmetric}={\rm sgn}(-\kappa)\,\kappa_{\rm eff} =-|\kappa|+\frac12\,{\rm sgn}(\xi)\;q^2 \,,
\eea
where we invoked the well-known index, $\kappa$,
for the pure $SO(2)_\kappa$ theory. Therefore, for either sign
of $\xi$, we find
\bea\label{N2I}
\cI=\cI_{\rm symmetric}+\cI_{\rm broken}= -|\kappa| - \frac12\,q^2 \,,
\eea
and that there is no wall-crossing at $\xi=0$,
consistent with the absence of any asymptotic flat
direction appearing.

One thing we wish to note here is that all of the above was
possible only thanks to the very rigid  $D$-term potential,
although we recast it into the $\cN=1$ superpotential language.
So, what could be done if we really consider
general $d=3$ $\cN=1$ theories with a generic and not-so-simple
superpotential $W$? Keeping track of the symmetric and the
broken vacua may not be so feasible in such general theories.
Recall that the Witten index in Section 2 only requires
a winding number information of $W$ and has little to do
with details at finite $X$. If we wish to find an analog
of the above counting there, if any, it would be the Morse
theory interpretation. On the other hand, we are not aware
of how to extend the Morse theory to the gauged dynamics.

What should be done is of course to go back to the $d=3$
$\cN=1$  path integral and ask whether there is a way to reduce
it to ordinary Gaussian integrals and eventually to the
winding number information in $W$, on par with what we have
seen for $d=1$ $\cN=2$ theories in Section 2. The rest
of this note is devoted to this question.

\subsection{$d=1$ Approach for $d=3$ Chern-Simons-Matter}\label{sec:CSindexthm}

Suppose we want to compute the same Witten index such as
(\ref{N2I}) for general $d=3$ $\cN=1$ Chern-Simons theories,
entirely by compactifying the theory on $\IT^3$ and performing
the Euclidean path integral. This will bring us to the kind
of computation in the two previous sections. The Kaluza-Klein
sector cancels out neatly in the end, so the reduction to $d=1$
is achieved naturally. The new ingredient is the
remnant of the Chern-Simons term in the $d=1$ Lagrangian,
\bea
\sim - \frac{\kappa L_1L_2}{2\pi} u_1\partial_t u_2 \,,
\eea
as well as its accompanying mass term for gaugino $\lambda$.
The holonomies $u_a$'s are of $2\pi/L_a$ periods, where
$L_{1,2}$ are the lengths of the two spatial circles in $\IT^3$.

This means that the kinetic part of the vector multiplet is
the same as that of the Landau problem with the magnetic field $\kappa L_1L_2/2\pi $
and the total flux $\Phi_B=2\pi \kappa$.  Note that since we will
compute this on arbitrarily small torus, $\kappa$ here is the bare
Chern-Simon level.	Since the magnetic field couples to $u_{1,2}$
at quadratic level with a single time derivative, the heat kernel expansion
should be in principle worked out from scratch. In Appendix A, we
derive the 0-th order heat kernel for the Landau problem, necessary
for the heat kernel expansion. As observed there,
the small $\beta$ expansion, combined with the insertion of $(-1)^F$,
is such that the computation we have performed so far in the absence of
the external magnetic field is applicable. The only change in the
end is the extra fermion bilinear $\sim (\kappa L_1L_2)\lambda\lambda$
which affects the Pfaffian part of the computation.

The path integral has a somewhat different decomposition
than the counting based on $d=3$ vacuum analysis above. For
rank 1 theories, we find
\bea\label{3d}
\Omega^{3d} = \Omega_0 + \cZ_\kappa\int
\frac{du_1du_2 du_3}{(2\pi)^3}\; \beta L_1L_2 \kappa= \Omega_0+\kappa\cdot\cZ_\kappa \,,
\eea
where $\Omega_0$ is part of the path integral independent of $\kappa$.
Here, we displayed variables prior to the $\beta$-scaling, so that
$u_3$ is the holonomy along the time direction of period $2\pi/\beta$,
and similarly $u_{1,2}$ have the periods $2\pi/L_{1,2}$.  For higher rank theories,
we similarly expect higher order pieces in $\kappa$'s to show up
as well, although here we confine ourselves to $SO(2)$ theories.

The second term in (\ref{3d}) can be understood, from the $d=1$ perspective,
easily since the supersymmetric Landau level problem comes
with the ground state degeneracy of $|\Phi_B|/2\pi=|\kappa|$,
which equals the usual level-by-level degeneracy for the
purely bosonic Landau problem. For the lowest Landau level,
the $SO(2)$ gaugino merely shifts the ground
state energy to zero and does not affect this degeneracy.
$\cZ_\kappa$ takes into a further degeneracy factor, with sign,
that, for example, might come from a decoupled neutral sector,
tensored with this Landau problem.

As far as such a path integral on $\IT^3$ goes, $\Omega_0$ and
${\cal Z}_\kappa$ are computed blindly,
regardless of the classical vacuum structure. This means that the
$d=3$ index has two different decompositions,
\bea
\cI= \cI_{\rm symmetric}+\cI_{\rm broken} =\kappa\cdot\cZ_\kappa + \Omega_0 \,.
\eea
While $\cI_{\rm symmetric}$ includes the $\kappa$-dependent pieces,
it also has the additive one-loop shift of $\kappa$ from charged
matters, so the two decompositions are not the same
in the presence of charged matters. $\Omega_0$ can and will
generally contribute to both $\cI_{\rm symmetric}$ and $\cI_{\rm broken}$.

For $d=3$ $\cN=1$ theories, the classical vacua are determined
by the more flexible superpotential $W$, instead of the rigid
$D$-term potential \eqref{DV} above. Also, unlike $\cN=2$ theories,
where the $D$-term potential with large enough $|\kappa|$ always admits a symmetric
vacuum, one can easily imagine $W$ that does not admit any symmetric
vacuum even for large enough $|\kappa|$. Since, from $d=3$ physics, it is clear that the $\kappa$-dependent
part can get a contribution only from a symmetric vacuum, we find immediately
that
\bea
 \cZ_\kappa=0 \,,
\eea
in theories with no classical symmetric vacuum. When an asymptotically
flat direction is present, one may need to understand this as
\bea
\lim  \cZ_\kappa=0 \,,
\eea
where the limit is taken to scale up the finite asymptotic
gap, as usual. On the other hand, if the theory admits exactly one
symmetric vacuum classically, we should expect $|\cZ_\kappa|$, or the
above limit thereof if needed, to equal 1.

Such relatively simple behaviors of $\cZ_\kappa$ will manifest in
the path integral computation as well; we will see presently that this quantity is
really determined by the neutral sector of the matter content. In particular,
we will see that, for $\cN=1$ $SO(2)$ theories with no neutral scalar multiplets,
$\cZ_\kappa=1$ has to hold universally for in our sign convention.
By the same token, we end up finding $\cZ_\kappa=\pm 1, 0$ at most,
with the above limit taken if necessary and with wall-crossing possibilities,
if a single neutral scalar multiplet is present.

\subsubsection*{Reduction of the Path Integral}

Let us now turn to the evaluation of
\begin{align}
\Omega^{3d} = \kappa \cdot \mathcal Z_\kappa + \Omega_0 \,,
\end{align}
for a general superpotential.
The two terms $\Omega_0$ and $\cZ_\kappa$ are computed from the
two additive pieces in the Pfaffian of the fermion bilinear,
which for the current Abelian problem has the simple form,
\bea
M(X,u)\;=\;\left( \begin{array}{cc} -\frac{\kappa L_1L_2}{2\pi}\epsilon_{ab} & (QX)_\nu\epsilon_{ac}  \\
&\\
(X Q^\dagger )_\mu\epsilon_{db}  & \partial_\mu\partial_\nu W\epsilon_{dc}
+ u_i Q_{\mu\nu}(\epsilon\cdot \sigma^i)_{dc}\end{array}\right) \,.
\eea
$\Omega_0$ results from the Pfaffian of this matrix with
$\kappa L_1 L_2 \rightarrow 0$ while ${\cal Z}_\kappa$ is from the piece
linear in the Chern-Simons mass term on the upper-left corner.
Therefore, the matrix relevant for $\cZ_\kappa$ is much simpler,
\bea
(M_{\kappa})_{\mu\nu} \;\equiv\; \partial_\mu\partial_\nu W\epsilon_{dc}
+ u_i Q_{\mu\nu}(\epsilon\cdot \sigma^i)_{dc} \,,
\eea
which is the mass matrix for the matter fermions only;
the $\kappa$-linear piece saturated the gaugino zero
modes.  A useful decomposition of the entire Pfaffian is
\begin{align}\label{pfdecomCS}
\text{Pf} ( \tilde{M} ) = - \frac{\tilde \kappa}{2 \pi} \cdot \text{Pf} (\tilde M_\kappa) + \sum_{I} 
\, (Q \vec X_I) \cdot \text{Pf} ( \tilde{M}_I )  \,,
\end{align}
where the latter sum leads to $\Omega_0$ being expressed
as a sum over the flavor index $I=1,\cdots, N_f$. We
refer the readers to Section \ref{sec:gaussian} for
the complete details on the latter sum.

In the end, we find that both  $\Omega_0$ and $\cZ_\kappa$
are expressed as simple and factorized Gaussian integrals
with all the nontrivial information encoded in the integration
domain. In other words, we find an analog of (\ref{thm1})
for the $d=3$ Chern-Simons gauge theory. 
One important point we need to emphasize is that, for
$d=3$ theories, $\kappa$ which enters our path integral
approach is the UV one, as various examples below will
show. This is natural since we are evaluating the index
in the small torus $\IT^3$ limit, which effectively cuts
off the infrared end of the path integral.

\subsubsection*{Gaussian Reduction for $\cZ_\kappa$}

The simpler form of $M_\kappa$ suggests a further simplification
for $\cZ_\kappa$. As we show in Appendix \ref{sec:pf-jac-proof}, $\bf u$'s are decoupled
completely from the rest, for the computation of $ \cZ_\kappa$,
upon the following universal change of variables,
\bea\label{chart1}
{\bf v} = {\bf u} \,,\quad s_A = \partial_{p_A}W \,, \quad
\vec y_I =\sqrt{ q_I^2 {\bf u}^2   + \left( \partial_{\rho_I}  W \right)^2 } \cdot \vec x_I \,,
\eea
for $\beta$-scaled two-component charged matter $\vec x_I$ of charge $q_I$
and neutral real scalar $p_A$, while the gauge
holonomies ${\bf u}$ are left intact. This results in
a clean decoupling of $\bf u$ from the matter integral,
leading to the factorized form of the second
term in (\ref{3d}).
Note that this change of variables is the same
one as in Section \ref{sec:so2gap} where the Witten index
$\Omega$ of
$d=1$ massive gauge theory is computed.

The Jacobian to \eqref{chart1} turns out to be equal
precisely to minus the Pfaffian of matter-only minor
$\tilde M_\kappa$,
\begin{align}
\label{eq:cancelation0}
\text{Pf} ( \tilde M_ \kappa ) =
 - \text{Det} \left ( \frac{\partial ( \vec y_I , {\bf v} ) }{ \partial ( \vec x_I, {\bf u}) } \right) \,,
\end{align}
the proof of which can be found in Appendix \ref{sec:pf-jac-proof}.
Simultaneously, the bosonic potential is written in terms of new variables \eqref{chart1}
\begin{align}\label{Vgaussian}
\tilde V_\text{bosonic} = \half \sum_{A=1}^{\text{dim}(S)} s_A^2+\half \sum_{I=1}^{N_f} \vec y_I^{\, 2} \,.
\end{align}
The bosonic potential $\tilde V$ is now independent of new three-dimensional holonomy variable ${\bf v}$.
This leads to
\begin{align} \label{Z}
\mathcal Z_\kappa  &= \frac{1}{(2\pi)^{N_f+\frac{{\rm dim}(S)}{2}}}\int  d^{2N_f} y \, d^{\, {\rm dim}(S)}s
\, \exp \left(- \half \sum_A s_A^2 - \half\sum_{I} \vec y_I^{\, 2} \right) \\
&= \frac{1}{(2\pi)^{\frac{{\rm dim}(S)}{2}}}\int  d^{\, {\rm dim}(S)}s \, \int \prod_I d(\vec y_I^{\,2}/2)
\, \exp \left(- \half \sum_A s_A^2 - \sum_{I}  (\vec y_I^{\, 2}/2) \right) \nn
\end{align}
because the integration domains of $\vec y_I$ and $s_A$ are also independent of $\bf v$ in the end.
Please note that, for $\cZ_\kappa$, the holonomy saddle is
irrelevant. No saddle appears because the integral is free
along the $\bf v$ directions and the contribution from
the holonomy torus enters via its volume, as in \eqref{3d}.

In fact, this Gaussian integration for $\cZ_\kappa$ can be
universally reduced to a reduced theory with neutral scalar
multiplets only on par with (\ref{thm1}). If no neutral
scalars are present, on the other hand, in particular,
we will see that $\cZ_\kappa=1$ universally.  See the next
subsection for a derivation of this.

\subsubsection*{Gaussian Reduction for $\Omega^{\Theta_*=0}_I$}

For $\Omega_I^{\Theta_*}$ with a general superpotential, the computation
proceeds exactly in the same way as for those that entered $d=1$
$\Omega$ in Section \ref{sec:gaussian}. Here we merely reiterate
results there for the sake of completeness, and shortly turn to $\Omega_0$
which must be built up from $\Omega_I^{\Theta_*}$ with the
holonomy saddles taken into account. We will come back to this in
the last part of this subsection.

For  $\Omega^{\Theta_*=0}_I$, we must resort to the alternate
coordinate systems, introduced in Section \ref{sec:gaussian}.
These can be summarized as $N_f$
many sets of distinct coordinates, labeled by $I$,
\bea\label{chart2}
&&{\bf v}= |q_I| |\vec x_I| {\bf u} \,, \quad s_A = \partial_{p_A}W \ ,\cr\cr
&&\vec z_{I} = (\partial_{\rho_I} W) \, \vec x_I \,,
\quad\vec y_{K(\neq I)} =  \sqrt{ q_K^2 {\bf u}^2   + \left( \partial_{\rho_K} W \right)^2 } \cdot \vec x_K \,,
\eea
to be used upon decomposing the Pfaffian for $\Omega_0$
into $N_f$ many summands. As was mentioned previously,
the assumption $W=W(\rho_I,P_A)$ can be justified given
$SO(2)^{N_f-1}$ flavor symmetries, which we believe to
be innocuous for the purpose of computing the index for
massive theories.

For each doublet label $I \in \{ 1, \cdots, N_f \} $, the same set of integration variables is suggested as in
\eqref{GaugeGaussian1}, so that its Jacobian simplifies the $\tilde \kappa$-independent piece of Pfaffian, denoted by $(Q \vec X_I) \cdot \text{Pf} (\tilde M_I ) $, as follows:
\begin{align} \label{eq:cancelation1}
 (Q \vec X_I) \cdot \text{Pf} ( \tilde{M}_I ) \,  \left|  \frac{\partial ( \vec x \,, p\,,{\bf u} )}{\partial (\vec z, \vec y \,, s\,, {\bf v} )} \right|
=  \frac{1}{|q_I| W_I |\vec x_I| } = \frac{1}{|q_I| \,\text{sgn}(W_I) |\vec z_I |} \,.
\end{align}
We are yet to find a general analytical proof of this but have checked
its veracity for all examples in this note up to
$N_f=3$.

Also, unlike $\cZ_\kappa$, the holonomy saddles become relevant and contribute additively to
$\Omega_0$, the $\kappa$-independent piece of the partition function. For instance, the $I$-th contribution of
the saddle at the origin $\Omega_{I}^{{\Theta_*}=0}$ reads,
\begin{align} \label{OM}
\Omega_{I}^{{\Theta_*}=0} &= \frac{1}{(2 \pi)^{N_f+2+\frac{{\rm dim}(S)}{2}}} \int d^3{\bf v} \, d^{2(N_f-1)} y \, d^2z_I\, d^{{\rm dim}(S)}s \, \frac{1}{|q_I| \, \text{sgn}(W_I) |\vec z_I| }\, e^{-\tilde V_{\rm bosonic}} \cr
& = \frac{1}{(2 \pi)^{2+\frac{{\rm dim}(S)}{2}}} \int d^3{\bf v} \,  d^{{\rm dim}(S)}s \int d(\vec z_I^2/2)\prod_{J\neq I} d(\vec y_J^2/2)\, \frac{1}{|q_I |\, \text{sgn}(W_I) |\vec z_I| }\, e^{-\tilde V_{\rm bosonic}} \,,
\end{align}
with
\bea
\tilde V_{\rm bosonic} = \half {\bf v}^2 + \half\sum_{A=1}^{\text{dim}(S)} s_A^2 +\half \vec z_I^{\, 2}+ \half \sum_{J\neq I} \vec y_J^{\, 2} \,,
\eea
where again everything reduces to a Gaussian integral and all the
nontrivial information of the theory is transferred to the
integration domain, on par with (\ref{thm1}).

\subsubsection*{Building Up $\Omega_0$ from Holonomy Saddles on $\IT^3$}

If there were no holonomy saddle except at ${\Theta_*}=0$, i.e.
if $|q_I|=1$ for all $I$, we would have
\bea
\Omega_0= \sum_I \Omega^{{\Theta_*}=0}_I \,.
\eea
More generally, however, we need to take account of
holonomy saddles. We follow the same logic  of Section \ref{sec:so2gap}
by taking into account the additional holonomy  saddles due to the
spatial $\IT^2$ as well. For a $d=3$  theory, the holonomy saddles
are labeled by three angles on the dual torus of $\mathbb T^3$:
${\Theta_*} = \Theta_{l,m,n} = (\theta_l,\theta_m,\theta_n)$. One can have an
additive $\Omega_K^{\Theta_*}$ contribution only if
\bea
\theta_l q_K \in 2\pi \IZ\ , \quad
\theta_m q_K \in 2\pi \IZ\ , \quad
\theta_n q_K \in 2\pi \IZ\ ,
\eea
for all three holonomies, determining the charged scalar multiplets $\vec X_K$
that enter the path integral contribution at $\Theta_*$.

Let us denote the collection of such $K$'s, $\cN_*$ or $\cN_{l,m,n}$ to be precise.
We then proceed in the same way for the reduced theory sitting at
$\Theta_*$,  and find additive contributions,
\bea
\sum_{K\in \mathcal N_*} \Omega^{{\Theta_*}}_K \,,
\eea
where the superscript ${\Theta_*}$ indicates the low energy
theory at the holonomy saddle, obtained by removing all
charged scalar multiplets  such that $q_J \theta_i/{2\pi}$ is non-integral
at least for one of $i=l,m,n$.  In fact, a necessary condition for $\Theta_*$
to be a contributing
holonomy saddle  for Abelian theories is that $\cN_*$ is not empty.
As such, we have the general routine,
\bea\label{HS0}
\Omega_0= \sum_{\Theta_*} \sum_{K\in \mathcal N_*} \Omega^{{\Theta_*}}_K =
\sum_{l,m,n} \sum_I \Omega^{\Theta_{l,m,n}}_I\delta_ {l,m,n}(q_I) \,,
\eea
where the formal presence of  the ill-defined
$\Omega^{\Theta_{l,m,n}}_I$ for $I\notin \cN_{l,m,n}$ is immaterial since
\begin{align}
\delta_ {l,m,n}(q_I)=   \delta_{\frac{q_I\theta_l}{2 \pi}} \delta_{\frac{q_I \theta_m}{2 \pi}} \delta_{\frac{q_I \theta_n}{2 \pi}}
\end{align}
is designed to vanish for such $I$'s, and is 1 if $I=K\in \cN_{l,m,n}$

This symbol also allows us to exchange the two summations as
\begin{align}\label{sumsum}
\Omega_{0} =\sum_I \sum_{l,m,n}  \Omega^{\Theta_{l,m,n}}_I\delta_ {l,m,n}(q_I) \,.
\end{align}
For $SO(2)$ theories, there is a further simplification due to the fact that
the holonomy saddles share the common gauge group: the value of (\ref{OM})
is by and large independent  of the full charged scalar content, other than
$\vec X_I$, so that
 \bea\label{saddles}
\Omega_{K\in \mathcal N_*}^{\Theta_*}= \Omega^{{\Theta_*}=0}_K \,,
\eea
as we will see this more explicitly in Section \ref{sec:index-cs-matter} by evaluating these
Gaussian integrals. This happens because, in the end, $\Omega_I^{\Theta_*}$'s
are such that  of all charged fields, only one charged scalar multiplet $X_I$
meaningfully enters the result.

These two mean that we can formerly and safely replace
$\Omega^{\Theta_{l,m,n}}_I$ by $ \Omega^{\Theta_*=0}_I$ in (\ref{sumsum})
for all $I$, factor out the $(l,m,n)$ summation for the obvious identity,
\begin{align}
\sum_{l,m,n} \delta_ {l,m,n}(q_I)
= |q_I|^3\ ,
\end{align}
and arrive at
\bea\label{3dsum}
\Omega_0= \sum_{I=1}^{N_f} \;|q_I|^3 \,\Omega^{{\Theta_*}=0}_I \,,
\eea
regardless of details of the $SO(2)$  theory.

\subsection{Index Theorem for $SO(2)$ Chern-Simons-Matter\label{sec:index-cs-matter}}

In (\ref{Z}) and (\ref{OM}), we have successfully reduced
the path integrals of $d=3$ $\cN=1$ $SO(2)$ Chern-Simons-Matter theories
to ordinary Gaussian integrals, while (\ref{3dsum}) has given a simple
way to build up the matter contribution $\Omega_0$ from holonomy
saddles on $\IT^3$. This means that the essence of the Witten index
is transferred again to the matter of understanding the images
of the maps, (\ref{chart1}) and (\ref{chart2}), much as in (\ref{thm1}).
These are still much cumbersome, so we will discuss how charged-matter
integrals can be performed universally, leaving behind only the
neutral scalar images $s_A$'s both for $\cZ_\kappa$ and for $\Omega_0$.

For this, we will restrict our attention to $W(\rho_J,P_A)$ which is
a polynomial of $\rho_J=|\vec X_J| ^2/2$'s, and find
 \bea\label{3dthm}
 \cI^{3d}=\kappa\cdot \Omega_{\rm neutral}
 + \sum_I \left( \frac{q_I^2}{2}\;\sum_a n_I^{(a)}\cdot{\rm sgn}(W_I(\Delta^{(a)}_I))\right) \,.
\eea
Let us explain various quantities in this formula.
$\Omega_{\rm neutral}$ is the $d=1$ Witten index or the
twisted partition function for the subsector consisting of
the neutral scalar multiplets only, in the sense of Section 2. In the  double
sum on the right hand side, the inner sum is associated with
the decomposition of the integral domains into $\cup_a \Delta^{(a)}_I$,
 split by $0=W_I\equiv \partial_{\rho_I} W$, and $n^{(a)}_I$ is an
 effective winding number associated with the neutral part of
 the Gaussian variable $s_A=\partial_{p_A}W$ in the domain $\Delta_I^{(a)}$.
The precise definition of $n^{(a)}_I$ is a little more involved
than $\Omega_{\rm neutral}$, which we will discuss presently.

If the neutral sector is absent altogether, in particular, the end result
can be obtained by a decoupling limit of theories with the neutral sector, such
that the index simplifies to
\bea
\label{3dthm2}
\cI^{3d}=\kappa + \sum_I \left( \frac{q_I^2}{2}\cdot{\rm sgn}(W_I^\infty)\right) \,,
\eea
where effectively only one ``last'' domain $\Delta_I^\infty$ in the $a$-sum
contributes with the unit winding number. Note that we have already derived
$d=1$ version of the second term, in Section \ref{sec:gaussian}, which differed by
the absence of $q_I^2$ in the summand. We will dedicate  this
subsection to a proof of these two formulae.

\subsubsection*{Evaluation of $\cZ_\kappa$}

Despite the apparently factorized forms of (\ref{Z}) and (\ref{OM}),
the subtlety lies with the integration domain, determined entirely by $W$.
Fortunately, the $\vec y_J$ variables are such that for generic $\bf u$,
$$
\sigma_J \equiv  \left((\partial_{\rho_J}W)^2+q_J^2{\bf u}^2\right) \rho_J =| \vec y_J|^2/2
$$
spans $[0,+\infty)$, as $\rho_J$ spans $[0,+\infty)$, effectively once
if we take the orientation into account.
The biggest difference of the gauged version from (\ref{thm1})
is the absence of winding number in the charged sector. Although $\vec x_J\rightarrow \vec y_J$
 in (\ref{chart1}) looks like a map $\IR^{2N_f}\times \cdots \rightarrow \tilde\IR^{2N_f}\times\cdots$,
it is in reality a map $\rho_J\rightarrow \sigma_J$ between $\IR^{N_f}_+\times\cdots$ and $\tilde\IR^{N_f}_+\times\cdots$ since
$W=W(\rho_I=|\vec X_I|^2/2,\cdots)$. Furthermore, the boundary
$\cup_I\{|\vec x_I|=0\}$ is mapped to the boundary
$\cup_I\{|\vec y_I|=0\}$.

This means that, regardless of how complicated might be the image of map $\sigma_J$ in
the wedge $\tilde\IR^{N_f}_+$, the Gaussian integral over them eventually always reduces to
\bea
\prod_J \int_0^\infty d(\vec y_J^2/2) e^{-(\vec y_J)^2/2} =\prod_J \Gamma(1)=1 \,.
\eea
In particular, this happens regardless of details of $W$, as long as the latter
is smooth and at least linearly divergent at large $\rho_J$'s.

What about the integration domain of $s_A$'s? Consider the limit of very
large $p_A$, while $\vec y_J$ are held fixed. Decomposing
\bea
W=W_{\bf C}(\rho) + W_{\bf CN}(\rho;p)+W_{\bf N}(p)\ ,
\eea
and concentrating on the order of magnitude only, we find
\bea\label{SP}
s_A = \partial_{p_A} W = \partial_{p_A} W_{\bf N} +O(1/|p|)
\eea
at large $p$'s and finite $\vec y$'s, for polynomial $W$. The
latter suppressed term can be understood as,
we have
\bea
|\partial_p (W_{\bf CN})|\;\; \sim\;\; \frac{|W_{\bf CN}|}{|p|}\;\; \sim \;\;
\frac{|\vec y|}{|p|}\cdot \frac{W_{\bf CN}}{\sqrt{|\partial_{\vec x}(W_{\bf C}+W_{\bf CN})|^2+ {\bf u}^2|\vec x|^2}}
\eea
at large $p$ and finite $\vec y$. So, regardless of the potential
complications from $W_{\bf CN}$, only $W_{\bf N}$ matters for the
contributing $s$ domains. We therefore conclude that the $s$ integrals
collapse to a form as in (\ref{thm1}) for a reduced theory
involving $\{P_A\}$ only.

These observations translate to one of the following two possibilities:
{\bf (1)} If the Chern-Simons theory contains no neutral field, we must find
\bea\label{C}
\cZ_\kappa = 1 \,.
\eea
This, in particular, implies that in theories with no neutral scalar
multiplet, there is always an $SO(2)$ symmetric phase in $d=3$, since
otherwise the Witten index cannot have $\kappa$ contribution at all.
{\bf (2)} If the Chern-Simons theory comes with neutral matter multiplets,
$\cZ_\kappa$ is computed by the same formula as in (\ref{thm1}) such that
\bea\label{N}
\cZ_\kappa = \Omega_{\rm neutral} \,,
\eea
which is obtained by removing the gauge sector and charged matter entirely.
If the latter does not give an integral result due to an
asymptotic flat direction, one must again scale up $W$ by an
arbitrary large and positive constant $C$, which in the end
will give an integral result as
\bea\label{NC}
\cZ_\kappa = \cI_{\rm neutral}= \lim_{C\rightarrow \infty}\Omega_{\rm neutral}\biggr\vert_{W\rightarrow C\cdot W} \,.
\eea
For example, a $d=3$ $\cN=2$ $SO(2)$ Chern-Simons theory,
with the single real $\cN=1$ scalar multiplet $\Sigma$ embedded in
the $\cN=2$ vector and with charged $\cN=2$ chirals as
the matter sector, one finds $\cZ_\kappa =-{\rm sgn}(\kappa)$,
leading to (\ref{N2I}). Integrating out this auxiliary
$\Sigma$, while leaving the $\cN=2$ matter sector intact,
brings us back to the canonical $\cZ_\kappa =1$ as in (\ref{C}).
In Section \ref{sec:CSgaussian}, we will confirm
(\ref{NC}) explicitly for more examples.

In fact, these results can be understood also from $d=3$ Chern-Simons
physics. Recall that the $\kappa$-linear contributions come from the
pure gauge dynamics at symmetric phases and also that our $\kappa$ is the UV
Chern-Simons level. At such symmetric phases, the charged chirals
can contribute to the index but only via an additive shift of
$\kappa$, which, for our $SO(2)$ theories, can enter only via
$\Omega_0$, however, cannot possibly affect $\cZ_\kappa$. This
explains (\ref{C}). On the other hand, the neutral matter multiplets
are decoupled from the gauge sector at the symmetric phase, once we
integrate out charged matters. This means the ground state sector
of such neutral sector will have to be tensored with those from
the gauge sector, which in turn explains why one has to find (\ref{N}).

\subsubsection*{Evaluation of $\Omega_0=\sum \Omega_K^{\Theta_*}$}

What can we say for $\Omega_0$, or more precisely for $\Omega_I^{{\Theta_*}=0}$,
where $N_f$ $\vec y_{J}$'s are replaced by $\vec z_I$ and $N_f-1$ $\vec y_{K\neq I}$?
The variable $\vec z_I$ behaves very similarly as $\vec y_{K\neq I}$'s except that
the inverse image of $\vec z_I=0$ includes the loci $\{ \vec x_J\vert W_I\equiv \partial_{\rho_I}W=0\}$,
which split $\IR^{N_f}_+\times \IR^{{\rm dim}(P)}$ into several domains,
\bea
\label{eq:decomposition}
\IR^{N_f}_+\times \IR^{{\rm dim}(P)} =\cup_a \Delta_I^{(a)} \,,
\eea
with a definite sign of $W_I \equiv\partial_I W$ in each such subdomain.
Whether or not such a domain contributes to $\Omega_I^{{\Theta_*}=0}$
depends on how its boundary is mapped into the target
$\tilde \IR^{N_f}_+\times \tilde \IR^{{\rm dim}(S)={\rm dim}(P)}$
obviously spanned by
\bea
\lambda_I=\vec z_I^{\, 2}/2 \ , \qquad \sigma_{K\neq I}=\vec y_K^{\, 2}/2, \qquad s_A \,.
\eea
We may safely ignored ${\bf v}=|q_I||\vec x_I|{\bf u}$ in the following discussion.

For each domain, the boundary $\partial \Delta_I^{(a)}$ sitting at finite
values of the matter fields consists of $\rho_I=0$ or of $\partial_I W=0$, so
the map in (\ref{chart2}) from $(\rho_J,p_A)$ to
$(\lambda_I, \sigma_{K\neq I}, s_A)$  induces a map $f_I^{(a)}$
such that
\bea
f_I^{(a)} \; :\; \partial\Delta_I^{(a)} \qquad\hookrightarrow \qquad \Upsilon_I\equiv \left\{\lambda_I=0, \sigma_{K\neq I}, s_A\right\} \,,
\eea
since $\lambda_I=(\partial_IW)^2\rho_I$. Note that each of $\Upsilon_I$ can
be regarded as a copy of $\tilde \IR^{N_f-1}_+\times \tilde \IR^{{\rm dim}(S)}$.
Here we are freezing  ${\bf v}=|q_I||\vec x_I|{\bf u}$ to a generic constant value,
as its precise value does not affect the classification below
and the result of the integral below.

There are two logical possibilities where domain $\Delta_I^{(a)}$
can contribute to $\Omega_0$: (i) $f_I^{(a)}$ covers the entire
$\Upsilon_I$ $n_I^{(a)}$ times, or (ii) $f_I^{(a)}$ does not cover
the entire $\Upsilon_I$ but does so $n_I^{(a)}$ times if
we scale up $W\rightarrow C\cdot W$ with an arbitrarily large
positive number $C$. When either of this happens, assuming
polynomial form of $W$, one can see that the image of $\Delta_I^{(a)}$
under the map (\ref{chart2}) is the entire $\tilde \IR^{N_f}_+\times
\tilde \IR^{{\rm dim}(S)}$, so each Gaussian integral over $\Delta_I^{(a)}$
contributes to $\Omega_0$ an additive piece
\bea
\frac{1}{2|q_I|} n_I^{(a)}\cdot {\rm sgn}(W_I^{(a)})\ ,
\eea
although this requires the scaling in case of (ii). Summation over
the domains give
\bea
\Omega^{{\Theta_*}=0}_I =\frac{1}{2|q_I|}\sum_a n_I^{(a)}\cdot{\rm sgn}(W_I^{(a)})\ ,
\eea
while the entire $\Omega_0$ is constructed via (\ref{3dsum}) as
\bea
\Omega_0 =\sum_I \left( \frac{q_I^2}{2}\;\sum_a n_I^{(a)}\cdot{\rm sgn}(W_I^{(a)})\right) \,.
\eea
This concludes the proof of (\ref{3dthm}).

Note that each and every term in the sum is a multiple
of $1/2$; this is consistent with the fact that the Chern-Simons level
shifts in unit of $1/2$ due to the one-loop correction by matters. Such a would-be one-loop
contribution is embedded in $\Omega_0$ in this path-integral computation,
as was commented earlier.

What are the other possibilities other than (i) and (ii) above?
The typical reason why $\Delta_I^{(a)}$ fails to contribute is
that the map $f_I^{(a)}$ folds its image in $\Upsilon_I$ in such
a way $n^{(a)}_I=0$ effectively.
Note that $\rho_{K\neq I}=0$ always maps to $\sigma_K=0$. This implies that
there is no winding number to consider in the charged matter sector.
Instead, the winding number comes from how $p_A$ gets mapped to $s_A$,
along $\partial \Delta_I^{(a)}$. Therefore, even though things are
a little more complicated compared to the $\cZ_\kappa$ computation,
$\Omega_0$ is again dictated by the winding numbers which are
by and large determined by the neutral matter sector. In fact,
if the neutral sector is absent, at most one $\Delta_I^{(a)}$
for each $I$ may contribute because of this, as we will see below.

When there is only one $P$, on the other hand, the only possible winding
number is either $n^{(a)}_I =\pm 1$ or $n^{(a)}_I =0$. To determine $n_I^{(a)}$,
all one has to do is to trace how $s$ grows as one follows the two
disjoint asymptotic ends of $\partial \Delta_I^{(a)}$. If $s$ grows
to the two opposite infinities (after $C$ scaling if necessary),
we have $n^{(a)}_I =\pm 1$ and if not $n^{(a)}_I =0$.
In Section \ref{sec:CSgaussian}, we will deal with a class of examples of this
kind, and confirm what we described here somewhat abstractly by
a more explicit computation.

If the neutral field $P$'s are absent altogether, the images
$(\lambda_I,\sigma_{K\neq I})$ of these domains simplify further.
With ${\bf u}^2$ present, the inverse image of $\sigma_{K\neq I}=0$ is
simply $\rho_{K\neq I}=0$. This means that most of the above domains
$\Delta^{(a)}_I$ are such that two adjacent locus of $\partial_I W=0$
are both mapped to $\lambda_I=0$ and that the image folds itself
in $\tilde\IR^{N_f}_+$. Once we take into account the orientation,
the integration in (\ref{OM}) over such $\Delta_I^{(a)}$ would
vanish; and the only outermost domain, with one
boundary mapped to $\lambda_I=\infty$, contributes to the
integral $\Omega_I^{{\Theta_*}=0}$.

Calling this last domain $\Delta_I^\infty$, we have the universal contribution,
$$
\frac{1}{2|q_I|\cdot{\rm sgn}(W_I^\infty)} \,,
$$
where the sign is unambiguously determined on the same outermost
domain in space $\IR^{N_f}_+$, spanned by $\rho_J$. This in turn
leads to
\bea
\Omega_0=\sum_I \frac{q_I^2}{2}\cdot{\rm sgn}(W_I^\infty) \,,
\eea
as claimed in (\ref{3dthm2}).

\subsection{Examples, Wall-Crossings, and Dualities \label{sec:CSgaussian}}

Here we will give a cursory look at a few examples. The goal is to solidify
the general formula by performing the explicit integration of the Gaussian
formulae (\ref{Z}) and (\ref{OM}), and in particular to give a little more concrete
picture of the index formula (\ref{3dthm}) by studying explicit examples.
We will close with 3d $\mathcal N = 1$ duality pairs by showing how the index counting is
consistent with the claimed dualities.

A word of caution is in order. Recall that the Witten index has an inherent
sign ambiguity. Sometimes, we may use $(-1)^F\rightarrow (-1)^J$ with $J$
an unambiguously defined integral fermion number, but no such is naturally
available in our class of theories with Majorana supercharges. Instead,
we have adopted a sign convention consistent with the standard Morse theory
counting or the natural winding number counting. This leads to $\cZ_\kappa=1$ for
a pure $\cN=1$ $SO(2)$ gauge theory, in particular. It is worthwhile to note that
extending the same sign convention to $d=3$ $\cN=2$  Chern-Simons theories, with
a single real $\cN=1$ scalar multiplet $\Sigma$ as part of the $\cN=2$ vector multiplet,
we have $\cZ_\kappa=-{\rm sgn}(\kappa)$. For example, the pure $\cN=2$ $SO(2)$
Chern-Simons theory would have the index $-|\kappa|$ in our convention.

\subsubsection*{$SO(2)$ without Neutral Matters}

For $SO(2)$ theories with charged matters only and a generic
superpotential of the form $W(\rho_I)$, the Gaussian integrals to be
performed for $\Omega^{\Theta_*=0}_I$ are in fact identical to the ones already
computed  for $d=1$ in Section \ref{sec:gaussian}. For $N_f$ matters,
$\Omega_I^{\Theta_*=0}$ was originally computed in
the $d=1$ context, which we quote here again
\bea
\Omega_I^{\Theta_*=0} = \frac{1}{2|q_I|}\cdot {\rm sgn}(W_I^\infty) \,,
\eea
and which went into the theorem (\ref{3dthm2}). The simplest theory is
the quadratic $W=\sum_{J = 1}^{N_f} m_J \rho_J$, which leads to
\bea
\Omega_I^{\Theta_*=0}=  \frac{1}{2|q_I|}\cdot {\rm sgn}(m_I) \,.
\eea
Another example is the quartic $W=c_W(\rho-\xi)^2/2$ with a single
charge $q$ complex scalar multiplet. We have
\bea
\Omega_I^{\Theta_*=0}=  \frac{1}{2|q|}\cdot {\rm sgn}(c_W) \,,
\eea
for this example.

What do change are the holonomy saddles, which are more numerous
in $d=3$ than in $d=1$; the net effect of them is now
\bea
\Omega^{1d}=\sum_I \;|q_I|\;\Omega_I ^{\Theta_*=0} \qquad\rightarrow\qquad
\Omega_0=\sum_I\; |q_I|^3\,\Omega_I^{\Theta_*=0}
\eea
with the common quantities $\Omega_I^{\Theta_*=0} $ shared by $d=1$ and $d=3$.
This counting (\ref{3dsum}) has been incorporated into the formulae (\ref{3dthm})
and (\ref{3dthm2}). The truly new part for $d=3$ theories is $\cZ_\kappa$, which
in the previous subsection we have shown to be 1 universally if the neutral
matters entirely are absent.

Combining these, one finds, for example,
\begin{align}\label{eq:index_quartic}
\cI^{3d}=\kappa + \half |q|^2 \text{sgn}(c_W)
\end{align}
for the above quartic theory with a single charge $q$ complex scalar multiplet,
regardless of the sign of $\xi$. Therefore, there is a wall-crossing at
$c_W=0$ but none at $\xi=0$. We can view this result in the $d=3$ perspective
as follows.  For $\xi>0$ where we have both the symmetric and the broken
vacua, this result can be more physically written as
\begin{align}\label{eq:index_rho-v1}
\cI^{3d}_{(\xi>0)}= \left(\kappa - \half |q|^2 \text{sgn}(c_W \xi)\right) + |q|^2 \text{sgn}(c_W\xi) \,,
\end{align}
where the last term comes from the Higgs vacuum at $\rho=\xi$; $q^2$ is due to
so-many holonomy choices along the spatial $\IT^2$. The first two terms
are interpreted due to the quantum vacua sitting at the symmetric phase
$\rho=0$. The local matter multiplet has the mass $-c_W \xi$, hence the
Chern-Simons level is shifted in the infrared by $-q^2\text{sgn}(c_W \xi)/2$.
With $\xi<0$, where only symmetric vacua survive, the shift of the
Chern-Simons level is now $ - \half |q|^2 \text{sgn}(c_W \xi)= \half |q|^2 \text{sgn}(c_W) $.
The entire index arises from the infrared Chern-Simons level,
\begin{align} \label{eq:index_rho-v2}
\cI^{3d}_{(\xi<0)} = \kappa + \half |q|^2 \text{sgn}(c_W) \, ,
\end{align}
as expected.

We can in fact expand this discussion to theories with an arbitrary number
of charged matters, equipped with a radial superpotential $W(\rho_I)$.
The index we compute by a direct path integral would be
\bea\label{multi-index}
\cI^{3d}  =\kappa + \sum_I \frac{q_I^2}{2}\cdot {\rm sgn}(W_I^\infty) \,,
\eea
where the sign of $W_I=\partial_{\rho_I}W$ is measured at the corner of
$\rho_I\rightarrow \infty, \rho_{J(\neq I)} \rightarrow 0^+ $. Again we can write this as
\bea
\cI^{3d} =\left(\kappa + \sum_I \frac{q_I^2}{2}\cdot{\rm sgn}(m_I)\right)+
\sum_I \frac{q_I^2}{2}\cdot\left({\rm sgn}(W_I^\infty)-{\rm sgn}(m_I)\right) \,,
\eea
where the first captures the shift of level $\kappa$ in the symmetric phase,
with the effective mass $m_I$ of field $X_I$ at the symmetric vacuum.
Among other things, this implies that an $SO(2)$ symmetric vacuum
always exists, regardless of the details of $W(\rho_I)$. For generic
$W$, the second piece, which is always integral, may be attributed to
additional Higgs vacua.

These $m_I$'s, which parameterize $W$,
affect the infrared Chern-Simons level, which in turn affects
the contribution to the index from the symmetric phase.
On the other hand, the same parameters can affect where and
how the broken vacua appear, and therefore also affect the
contribution from Higgs vacua. This latter effect is reflected
in the second part of the above index formula. In the end, however,
these parameter-dependences of individual contributions cancel out,
much as in the above quartic model, and the index is independent of
these $m_I$'s. Instead, the wall-crossing happens only when the asymptotics
of the above superpotential changes, in a very limited sense via ${\rm sgn}(W_I)$'s.
This should be contrasted to (\ref{thm1}) where one can in principle have
an arbitrarily high winding number as the index. This is the consequence
of the fact that the effective domain of matter fields is $\IR_+^{N_f}$
rather than $\IR^{2N_f}$, as far as the map $dW$ induced by the superpotential goes.

\subsubsection*{$SO(2)$ with Neutral Matters }

With neutral matters present, we have seen in the previous subsection that the index formula
changes substantially in such a way certain winding numbers enter the
final formula. Here we will take one last example of type with
a single neutral matter $P$ and the superpotential
\begin{align}
\label{eq:generalizedWCP}
W = P \cdot K(\rho)+L(\rho)+F(P) \,.
\end{align}
This may be regarded as a generalization of the $\mathbb{WCP}^{N_f-1}$
model in Section \ref{sec:gaussian}, which corresponds to the choice
\begin{align}
K(\rho) = \sum_{I = 1}^{N_f} q_I \rho_I \,, \qquad L(\rho) = 0 \,, \qquad  F(P) = -\xi P \,.
\end{align}
In $d=3$, even if we start with the $\mathbb{WCP}$ model, the one-loop corrections to the superpotential
will typically generate quadratic terms in $F(P)$, as we will see at the end of this subsection. Thus, here we consider
a generic polynomial $K(\rho), \, L(\rho)$ and $F(P)$, which can include such one-loop corrections. With this general class of examples,
we aim at understanding quantities that appear in the index theorem (\ref{3dthm}) a little
more concretely.

As before, the Witten index consists of two parts: $\kappa \cdot \mathcal Z_\kappa$ and $\Omega_0 = \sum_{\Theta_*} \sum_I \Omega_I^{\Theta_*}$. We demonstrate the explicit computation of those two contributions and confirm the index theorem in the previous subsection. First let us look at $\kappa \cdot \mathcal Z_\kappa$, given by
\bea
\kappa\cdot\cZ_\kappa   = \frac{\kappa}{(2\pi)^{N_f+\frac{7}{2}}} \int d^3{\bf v} \, ds \, d^{2 N_f} y \, \exp \left(- \half s^2 - \half \sum_J \vec y_J^2\right),
\eea
where
\begin{align}
\begin{aligned}
\label{eq:gaussian_var}
\mathbf v = \mathbf u \, ,\quad s  = K(\rho)+F'(p)\ , \quad
\vec y_J =  \sqrt{ q_J^2 {\bf u}^2  + (p K_J(\rho)+L_J(\rho))^2 } \cdot \vec x_J  \,,
\end{aligned}
\end{align}
with $K_J = \frac{\partial K}{\partial \rho_J}$, $L_J = \frac{\partial L}{\partial \rho_J}$ and $F'(p) = \frac{dF}{dp}$. We note that the slice of constant $\sigma_J = \vec y_J^2/2$ and constant ${\bf v}^2$ is determined by
\bea
\label{eq:repar}
\sigma_J = \left(q_J^2 {\bf v}^2  + \left(p K_J(\rho)+L_J(\rho)\right)^2\right) \rho_J \,, \quad \text{for } J = 1, \dots, N_f \,,
\eea
in the $(\rho,p)$-space.

For the moment, we assume that, for given $\sigma_J$ and ${\bf v}^2$, \eqref{eq:repar} has the unique solution $\rho_J = f_J \left(p;\sigma,{\bf v}^2\right)$ for $J = 1, \cdots, N_f$. Then $s$ is solely a function of $p$:
\begin{align}
s = K \left(f \left(p;\sigma,{\bf v}^2\right)\right)+F'(p) \,,
\end{align}
which determines the image of the integration domain of $p \in (-\infty,\infty)$ on the $s$-space as follows:
\begin{align}
s \in \left(\lim_{p \rightarrow -\infty} \left[K \left(f \left(p;\sigma,{\bf v}^2\right)\right)+F'(p)\right],\lim_{p \rightarrow +\infty} \left[K \left(f \left(p;\sigma,{\bf v}^2\right)\right)+F'(p)\right]\right).
\end{align}
Since $\rho_J = f_J \left(p;\sigma,{\bf v}^2\right)$ remains finite when $p$ goes to $\pm \infty$, $K(\rho)$, a polynomial of $\rho_J$, is also finite and negligible compared to $F'(p)$. Thus, we have the following integration domain of $s$ for fixed $\sigma_J$ and ${\bf v}^2$:
\begin{align}
\label{eq:s}
s \in \left(\lim_{p \rightarrow -\infty} F'(p),\lim_{p \rightarrow +\infty} F'(p)\right)\ ,
\end{align}
which is nothing but a consequence of \eqref{SP}.
In general, \eqref{eq:repar} does not give rise to the unique solution $\rho_J = f_J \left(p;\sigma,{\bf v}^2\right)$ because the codimension-$N_f$, i.e., one-dimensional, subspace defined by \eqref{eq:repar} in the $(\rho,p)$-space is rather arbitrary. Nevertheless, \eqref{eq:gaussian_var} defines a map from this one-dimensional space to the Gaussian variable $s$, whose image is given by, with $\sigma_J,\, {\bf v}^2 $ fixed,
\bea
s \in \left\{ \, K(\rho)+F'(p) \, | \, \sigma_J = \left(q_J^2 {\bf v}^2  + \left(p K_J(\rho)+L_J(\rho)\right)^2\right) \rho_J \, \right\}\ .
\eea
Since \eqref{eq:repar} restricts $\rho_J$, and $K(\rho)$ accordingly, to be finite, the image on $s$ again spans \eqref{eq:s}.

Note that \eqref{eq:s} is independent of the fixed values of $\sigma_J$ and ${\bf v}^2$. Thus, the entire path integral decomposes into the decoupled integrals over $\sigma_J, \, {\bf v}$ and $s$. The integration domain of each $\sigma_J$ is nothing but $\mathbb R_+$ while that of ${\bf v}$ is $\mathbb T^3$ of volume $(2 \pi)^3$. Therefore, we are left with the index of the single neutral matter,
\begin{align}
\label{eq:pKLF-Z_kappa}
\kappa\cdot\cZ_\kappa  &= \frac{\kappa}{\sqrt{2 \pi}} \int_{F'(p \rightarrow -\infty)}^{F'(p \rightarrow \infty)} ds \, e^{-\frac{1}{2} s^2} \nonumber \\
&= \left\{\begin{array}{rc}
+\kappa \, \,, \qquad &  \lim\limits_{p \rightarrow -\infty} F'(p) \rightarrow -\infty \quad  \text{ and} \quad \lim\limits_{p \rightarrow \infty} F'(p) \rightarrow +\infty \,, \\
0 \, \,, \qquad &  \lim\limits_{p \rightarrow -\infty} F'(p) \rightarrow \pm \infty \quad \text{ and} \quad \lim\limits_{p \rightarrow \infty} F'(p)\rightarrow \pm \infty \,, \\
-\kappa \, \,, \qquad & \lim\limits_{p \rightarrow -\infty} F'(p) \rightarrow +\infty \quad \text{ and} \quad \lim\limits_{p \rightarrow \infty} F'(p) \rightarrow -\infty \,,
\end{array}\right.
\end{align}
which agrees with the index theorem \eqref{NC}.

Next, we move on to $\Omega_0 = \sum_{\Theta_*} \sum_I \Omega_I^{\Theta_*}$, which can be inferred entirely from
\begin{align}
\Omega_I^{{\Theta_*} = 0} = \frac{1}{(2 \pi)^{N_f+\frac{5}{2}}} \int d^3{\bf v} \, ds \, d^2z_I \, d^{2 (N_f-1)} y \, \frac{1}{|q_I| \, \text{sgn}(W_I) |\vec z_I| } \exp(-\tilde V_{\rm bosonic})
\end{align}
with $\tilde V_{\rm bosonic}=({\bf v}^2 + s^2 + \vec z_I^2 +  \sum_{J \neq I} \vec y_J^2)/2$ as before, where
\begin{align}
\begin{aligned}
\label{eq:gaussian_var2}
&{\bf v} = |q_I| | \vec x_I | {\bf u} = |q_I| \sqrt{2 \rho_I} {\bf u} \,, \quad s  = K(\rho)+F'(p) \,, \\
&\vec z_I  = (p K_I(\rho)+L_I(\rho)) \vec x_I \,, \quad
\vec y_{J (\neq I)}  = \sqrt{ q_J^2 {\bf u}^2  + (p K_J(\rho)+L_J(\rho))^2 } \cdot \vec x_J  \, .
\end{aligned}
\end{align}
Again the slice of constant $\lambda_I = \vec z_I^2/2, \, \sigma_{J (\neq I)} = \vec y_J^2/2$ and ${\bf v}^2$ in the $(\rho,p)$-space is determined by
\begin{align}
\lambda_I  &= (p K_I(\rho)+L_I(\rho))^2 \rho_I \,, \label{eq:repar_lambda} \\
\sigma_{J (\neq I)} &= \left( \frac{q_J^2}{q_I^2} \frac{{\bf v}^2}{2 \rho_I}  + (p K_J(\rho)+L_J(\rho))^2 \right) \rho_J \,. \label{eq:repar_sigma}
\end{align}
In particular, \eqref{eq:repar_lambda} is equivalent to the following two equations
\begin{align}
p K_I(\rho)+L_I(\rho) = \pm \sqrt \frac{\lambda_I}{\rho_I} \,,
\end{align}
each of which, together with \eqref{eq:repar_sigma}, defines the one-dimensional subspace in the $(\rho,p)$-space. Let us denote those one-dimensional spaces by $\mathcal S_{\lambda_I,\sigma_J}^{(\pm)}$. For $(\rho,p) \in \mathcal S_{\lambda_I,\sigma_J}^{(\pm)}$, $\pm W_I = \pm (p K_I+L_I) > 0$ respectively. \eqref{eq:gaussian_var2} maps $(\rho,p) \in \mathcal S_{\lambda_I,\sigma_J}^{(\pm)}$ to the Gaussian variable $s$ as follows:
\begin{align}
\label{eq:image}
s \in \left\{ \, K(\rho)+F'(p) \, | \, (\rho,p) \in \mathcal S_{\lambda_I,\sigma_J}^{(\pm)} \right\}, \qquad \lambda_I, \, \sigma_{J (\neq I)}, \, {\bf v}^2 \text{ fixed}.
\end{align}
In the previous subsection, we argued that the domain of $(\rho,p)$: $\mathbb R_+^{N_f} \times \mathbb R$ is split into several subdomains $\cup_a \Delta_I^{(a)}$, each of which has a definite sign of $W_I$. Indeed, each $\mathcal S_{\lambda_I,\sigma_J}^{(\pm)}$ with extra directions from $\lambda_I$ and $\sigma_{J (\neq I)}$ spans each $\Delta_I^{(\pm)}$; i.e.,
\begin{align}
\Delta_I^{(\pm)} = \left\{ \, (\rho, p) \, | \, (\rho, p) \in \mathcal S_{\lambda_I,\sigma_J}^{(\pm)}  \text{ for } \lambda_I \in \mathbb R_+, \, \sigma_{J (\neq I)} \in \mathbb R_+ \, \right\}\ ,
\end{align}
whose boundary is given by
\begin{align}
\partial \Delta_I^{(\pm)} = \lim_{\lambda_I \rightarrow 0} \Delta_I^{(\pm)} = \left\{ \, (\rho, p) \, | \, (\rho, p) \in \mathcal S_{0, \sigma_J}^{\pm} \text{ for } \sigma_{J (\neq I)} \in \mathbb R_+ \, \right\}\ .
\end{align}
To determine the image \eqref{eq:image}, we need to examine where the asymptotic regions of $\mathcal S_{\lambda_I,\sigma_J}^{(\pm)}$ are mapped to.

\begin{figure}[tbp]
\begin{center}
\resizebox{\hsize}{!}{
\includegraphics[height=10cm]{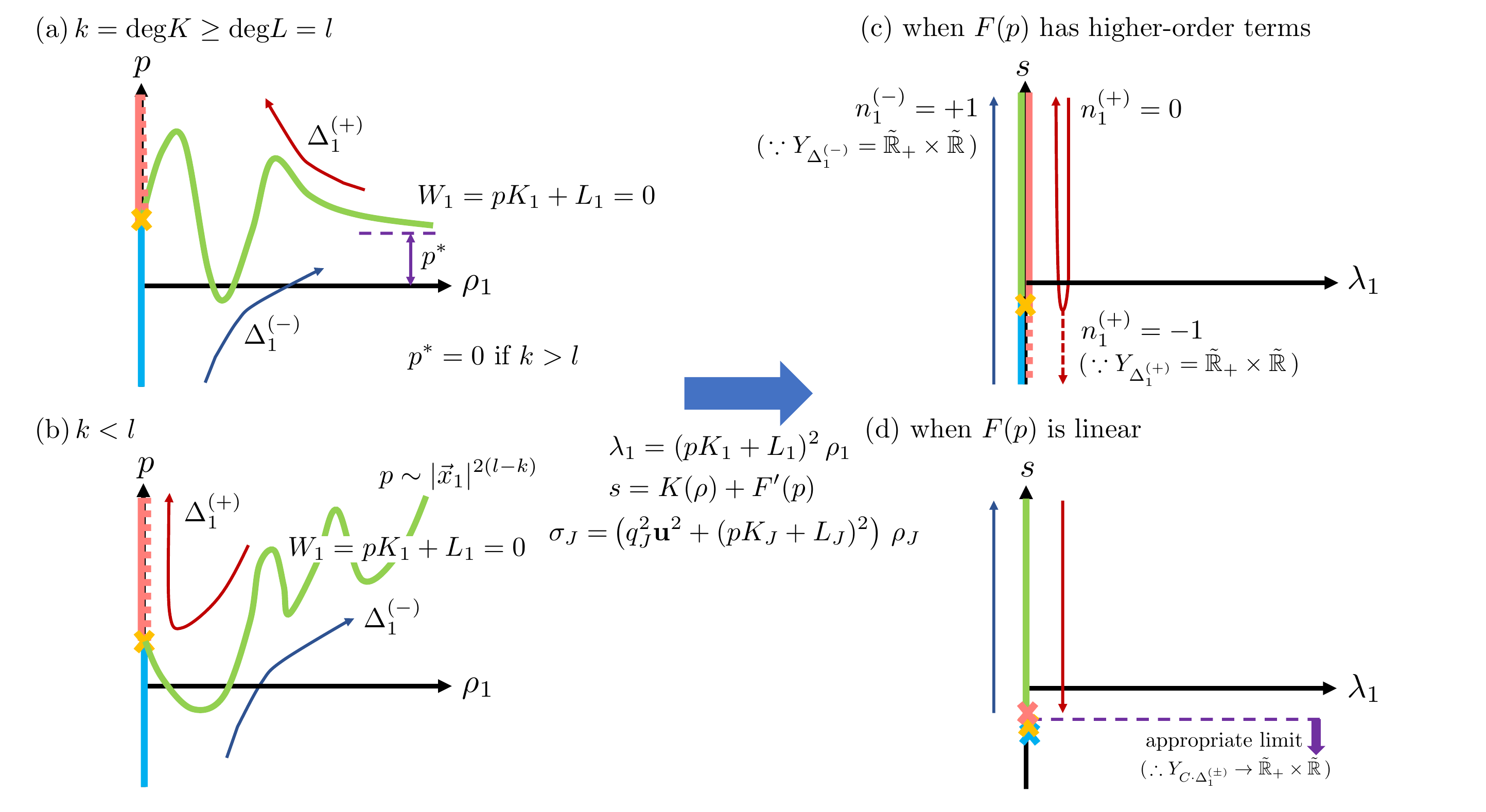}
}
\caption{
The domain decomposition and mapping under reparameterization \eqref{chart2} for $\Omega_{I=1}$ of $W=P\cdot K(\rho)+L(\rho)+F(P)$ theory. The angular part and other directions such as $\rho_{I\geq2}$,
$\sigma_{I \geq 2} $ are ignored to avoid the clutter. \label{fig:OM-pKLF}}
 \end{center}
 \end{figure}

We first note that, for fixed $\sigma_{J (\neq I)}$ and ${\bf v}^2$, $\rho_{J (\neq I)}$ may diverge if
\begin{align}
\rho_I \rightarrow \infty \,, \qquad W_J = p K_J(\rho)+L_J(\rho) \rightarrow 0
\end{align}
simultaneously, which does not happen unless $K(\rho)$ and $L(\rho)$ are fine-tuned. Thus, from now on, we assume that the potential is generic such that $\rho_{J (\neq I)}$ is always finite in $\mathcal S_{\lambda_I,\sigma_J}^{(\pm)}$. Then the asymptotic regions of $\mathcal S_{\lambda_I,\sigma_J}^{(\pm)}$ are characterized by either $\rho_I$ or $p$ being infinity. Indeed, from \eqref{eq:repar_lambda}, one can see that there are two asymptotic regions of each $\mathcal S_{\lambda_I,\sigma_J}^{(\pm)}$:
\begin{align}
\label{eq:asymp}
\left\{\begin{array}{ll}
\rho_I \rightarrow 0 \,, & \quad p \rightarrow s_I^{(\pm)} \infty \,, \\
\rho_I \rightarrow \infty \,, & \quad p \rightarrow p^*
\end{array}\right.
\end{align}
where $s_I^{(\pm)}$ is defined by
\begin{align}
s_I^{(\pm)} = \pm \text{sgn} \lim_{\rho_I \rightarrow 0} \left(\left.K_I(\rho)\right|_{S_{\lambda_I,\sigma_J}^\pm}\right).
\end{align}
$p^*$ is either finite or infinite depending on the details of $K(\rho)$ and $L(\rho)$. In particular, assuming
\begin{align}
K(\rho) \approx \alpha \rho_I^k \,, \qquad L(\rho) \approx \beta \rho_I^l \,,
\end{align}
for large $\rho_I$, i.e., the highest order terms in $\rho_I$ is independent of the other $\rho_{J (\neq I)}$, $p^*$ is given by
\begin{align}
\label{eq:p*}
p^* = \lim_{\rho_I \rightarrow \infty} \left(\left.p\right|_{\mathcal S_{\lambda_I,\sigma_J}^{(\pm)}}\right) \approx -\frac{l \beta}{k \alpha} \rho_I^{l-k} \,,
\end{align}
which diverges if $l > k$ and remains finite if $k \geq l$. See Figure \ref{fig:OM-pKLF} (a) and (b).

The image on the $s$-space is thus given by
\begin{align}
s &\in \left\{ \, K(\rho)+F'(p) \, | \, (\rho,p) \in \mathcal S_{\lambda_I,\sigma_J}^{(\pm)} \right\} \nonumber \\
&= \left(\lim_{p \rightarrow -\infty} F'(p),\lim_{\rho_I \rightarrow \infty} K(\rho)+\lim_{p \rightarrow p^*} F'(p)\right) \cup \left(\lim_{\rho_I \rightarrow \infty} K(\rho)+\lim_{p \rightarrow p^*} F'(p),\lim_{p \rightarrow +\infty} F'(p)\right).
\end{align}
with $\text{sgn}(W_I) = s_I^{(-)}$ or $\text{sgn}(W_I) = s_I^{(+)}$ for each segment respectively. Again this is independent of the fixed values of $\lambda_I, \, \sigma_{J (\neq I)}$ and ${\bf v}^2$. Thus, the entire path integral decomposes into the decoupled Gaussian integrals over $\vec z_I, \, \vec y_{J (\neq I)}, \, {\bf v}$ and $s$. In the end, we are left with
\begin{align}
\Omega_I^{{\Theta_*} = 0} = \frac{s_I^{(-)}}{2 |q_I|} \frac{1}{\sqrt{2 \pi}} \int_{F'(p \rightarrow -\infty)}^{K(\rho_I \rightarrow \infty)+F'(p \rightarrow p^*)} d s \, e^{-\frac{1}{2} s^2} + \frac{s_I^{(+)}}{2 |q_I|} \frac{1}{\sqrt{2 \pi}} \int_{K(\rho_I \rightarrow \infty)+F'(p \rightarrow p^*)}^{F'(p \rightarrow +\infty)} d s \, e^{-\frac{1}{2} s^2}
\end{align}
where each integral with prefactor $\frac{1}{\sqrt{2 \pi}}$ gives the winding number of the map
\begin{align}
f_I^{(\pm)} \; :\; \partial\Delta_I^{(\pm)} \qquad\hookrightarrow \qquad \left\{ \, \lambda_I = 0, \, \sigma_{J (\neq I)}, \, s \, \right\}
\end{align}
induced from \eqref{eq:gaussian_var2}, which is either $\pm 1$ or 0 in this case.
See Figure \ref{fig:OM-pKLF} (c). Taking the holonomy saddles into account, we have
the following result for entire $\Omega_0$:
\begin{align}
\label{eq:pKLF-Omega0}
\Omega_0 = \sum_{I = 1}^{N_f} \frac{q_I^2}{2} \sum_{a = \pm} s_I^{(a)} n_I^{(a)}
\end{align}
where
\begin{align}
\begin{aligned}
n_I^{(+)} &= \frac{1}{\sqrt{2 \pi}} \int_{K(\rho_I \rightarrow \infty)+F'(p \rightarrow p^*)}^{F'(p \rightarrow +\infty)} d s \, e^{-\frac{1}{2} s^2} \,, \\
n_I^{(-)} &= \frac{1}{\sqrt{2 \pi}} \int_{F'(p \rightarrow -\infty)}^{K(\rho_I \rightarrow \infty)+F'(p \rightarrow p^*)} d s \, e^{-\frac{1}{2} s^2} \,,
\end{aligned}
\end{align}
which confirms the index theorem \eqref{3dthm}.

So far we have considered a linear coupling of $P$ to a polynomial of $\rho$, which however can be generalized to higher order interactions. In the above linear interaction example, the domain of the original integration splits into two region divided by a codimension-one space defined by $W_I = 0$. On the other hand, with the generalized superpotential
\begin{align}
W = \sum_{m \leq M} P^m \cdot K^{(m)} (\rho)+F(p),
\end{align}
now the original integration domain splits in a much more complicated way. For instance, $W_I = 0$ has at most $M$ solutions of $P$ for fixed $\rho_I$. Those solutions can merge, split and cross each other along the way that $\rho_I$ changes, and the trajectories of such solutions define the codimension-one spaces that divide the domain into several subdomains. Note that, among those subdomains, only those unbounded along the $\rho_I \rightarrow \infty$ direction can give nontrivial contributions to the integral; i.e., any bounded subdomain gives a vanishing contribution because its image under the map \eqref{chart2} is folded and effectively vanishes. Thus, we only need to consider unbounded subdomains and the boundaries between them stretched to $\rho_I \rightarrow \infty$. There are at most $M$ such infinitely stretched boundaries. Their asymptotic values of $p$ for $\rho_I \rightarrow \infty$ can be denoted by $p^{(n)}$ with $n = 1, \dots, N \leq M$ where $N+1$ is the number of unbounded subdomains. Each subdomain contributes to the integral by its winding number $\pm 1$ if
\begin{gather}
\begin{gathered}
\lim_{\substack{\rho_I \rightarrow 0, \\ p \rightarrow -\infty}} \text{sgn}(s) \neq \lim_{\substack{\rho_I \rightarrow \infty, \\ p \rightarrow p^{(1)}}} \text{sgn}(s), \\
\lim_{\substack{\rho_I \rightarrow \infty, \\ p \rightarrow p^{(N)}}} \text{sgn}(s) \neq \lim_{\substack{\rho_I \rightarrow 0, \\ p \rightarrow \infty}} \text{sgn}(s)
\end{gathered}
\end{gather}
for the first and the last subdomain respectively and
\begin{align}
\lim_{\substack{\rho_I \rightarrow \infty, \\ p \rightarrow p^{(n-1)}}} \text{sgn}(s) \neq \lim_{\substack{\rho_I \rightarrow \infty, \\ p \rightarrow p^{(n)}}} \text{sgn}(s)
\end{align}
for the others.

Lastly, we conclude the discussion by demonstrating an explicit example with definite winding numbers. For example, let us choose $K(\rho), \, L(\rho)$ and $F(p)$ as follows:
\begin{align}
\label{eq:WCP}
K(\rho) = \sum_{I = 1}^{N_f} q_I \rho_I \,, \quad L(\rho) = \sum_{I = 1}^{N_f} m_I \rho_I \,, \quad F(P) = -\xi P+\frac{1}{2} m P^2+\frac{1}{2} \alpha P |P| \,.
\end{align}
Although $F(p)$ here, strictly speaking, is not a polynomial, the above discussion is still valid for this example. Using the results obtained above, we find that $\kappa \cdot \mathcal Z_k$ and $\Omega_I^{{\Theta_*} = 0}$ for this model are as follows:
\begin{align} \label{eq:SpecialOmega_0}
\kappa\cdot\cZ_\kappa  = \left\{\begin{array}{rc}
+\kappa \,, \qquad & |\alpha| < m \,, \\
0 \,, \qquad & -|\alpha| < m < |\alpha| \,, \\
-\kappa \,, \qquad & m < -|\alpha| \,,
\end{array}\right.
\end{align}
and
\begin{align} \label{eq:SpecialOmega_I}
\Omega_I^{{\Theta_*} = 0} = \left\{\begin{array}{rc}
-\frac{1}{2} \frac{1}{|q_I|} \,, \qquad & |\alpha| < m \,, \\
0 \,, \qquad & -\text{sign}(q_I) \alpha < m < \text{sign}(q_I) \alpha \,, \\
-\frac{1}{|q_I|} \,, \qquad & \text{sign}(q_I) \alpha < m < -\text{sign}(q_I) \alpha \,, \\
-\frac{1}{2} \frac{1}{|q_I|} \,, \qquad & m < -|\alpha| \,.
\end{array}\right.
\end{align}
\eqref{eq:SpecialOmega_0} and \eqref{eq:SpecialOmega_I} with the holonomy saddles lead to the following result for the 3d index:
\begin{align}
\label{eq:index_WCP}
\cI^{3d} = \left\{\begin{array}{rc}
\kappa-\frac{1}{2} \sum_I q_I^2 \,, \qquad & |\alpha| < m \,, \\
-\sum_{I \in \hat I_\pm} q_I^2 \,, \qquad & \pm \alpha < m < \mp \alpha \,, \\
-\kappa-\frac{1}{2} \sum_I q_I^2 \,, \qquad & m < -|\alpha|
\end{array}\right.
\end{align}
where $\hat I_\pm$ are sets of the indices having positive/negative charges respectively; i.e.,
\begin{align}
\hat I_\pm = \{I| \pm q_I > 0\} \,.
\end{align}
We note that there are wall-crossings at $m = \pm \alpha$. On the other hand, the index is independent of $\xi$.

In particular, if we set
\begin{align}
\alpha = -\sum_{I = 1}^{N_f} \frac{q_I |q_I|}{2} \,,
\end{align}
this example captures the index of the $\mathbb{WCP}^{N_f-1}$ model with massive $P$ and the one-loop corrected superpotential. For the $\mathbb{WCP}^{N_f-1}$ model, in fact, the one-loop correction to the superpotential is given by
\begin{align}
-\frac{1}{2} \sum_{I = 1}^{N_f} (q_I P+m_I) |q_I P+m_I| \,,
\end{align}
which leads to a more complicated vacuum equation. However, we have seen that only the asymptotics of $F(p)$ is important as long as we compute the Witten index. Hence, the index of $\mathbb{WCP}^{N_f-1}$ can be captured by more simpler one like \eqref{eq:WCP}.

One can easily check \eqref{eq:index_WCP} for the $\mathbb{WCP}^{N_f-1}$ model with massive $P$ in some limiting cases. First we assume $m_I$ are non-zero but very small compared to other parameters. For $|m| \gg |\alpha|$, a vacuum sits either at $P = \frac{\xi}{m}$ or at $P = 0$. At $P = \frac{\xi}{m}$, massive $P$ is integrated out and its vacuum expectation value induces mass of a charged fermion, whose sign is determined by $\text{sgn}(\frac{q_I \xi}{m})$. Thus, the effective IR theory is the Chern-Simons theory of level
\begin{align}
\kappa+\text{sgn}\left(\frac{\xi}{m}\right)\sum_{I = 1}^{N_f} \text{sgn} (q_I) \frac{q_I^2}{2} \,.
\end{align}
Its index contribution is the level multiplied by $\text{sgn}(m)$, which is from the fermion partner of $P$ integrated out:
\begin{align}
\label{eq:top}
\text{sgn}(m) \kappa+\text{sgn}(\xi) \sum_{I = 1}^{N_f} \text{sgn}(q_I) \frac{q_I^2}{2} \,.
\end{align}
On the other hand, at $P = 0$ with small $m_I$ assumed, we have a set of Higgs vacua around it; we have $\sum_{I \in \hat I_+} q_I^2$ of them for $\xi > 0$ and $\sum_{I \in \hat I_-} q_I^2$ for $\xi < 0$. Their index contribution is multiplied by -1 due to our sign convention. Combined with \eqref{eq:top}, the total index is given by
\begin{align}
\label{eq:index_WCP2}
\mathcal I^{3d} = \text{sgn}(m) \kappa-\sum_{I = 1}^{N_f} \frac{q_I^2}{2} \,,
\end{align}
which is independent of $\xi$.

For $|m| \ll \frac{N_f}{2}$, a vacuum sits either at $P = 0$ or at $|P| = \frac{\xi}{\alpha}$. For $\frac{\xi}{\alpha} < 0$, we have only Higgs vacua around $P = 0$, which give the index
\begin{align}
\label{eq:Higgs}
\mathcal I^{3d} = -\sum_{I \in \hat I_\pm} q_I^2, \qquad \pm \xi > 0
\end{align}
or equivalently
\begin{align}
\mathcal I^{3d} = -\sum_{I \in \hat I_\pm} q_I^2, \qquad \mp \alpha > 0
\end{align}
since $\frac{\xi}{\alpha} < 0$.
For $\frac{\xi}{\alpha} > 0$, on the other hand, we have extra topological vauca at $P = \pm \frac{\xi}{\alpha}$, described by Chern-Simons theories of level
\begin{align}
\kappa \pm \sum_{I = 1}^{N_f} \text{sgn}(q_I) \frac{q_I^2}{2}
\end{align}
respectively. Their index contributions are multiplied by $\pm \text{sgn}(\alpha)$ due to the massive fermion partner of $P$:
\begin{align}
\text{sgn}(\alpha) \left(\pm \kappa+\sum_{I = 1}^{N_f} \text{sgn}(q_I) \frac{q_I^2}{2}\right) \,.
\end{align}
Combined with \eqref{eq:Higgs}, the total index is again given by
\begin{align}
\mathcal I^{3d} = -\sum_{I \in \hat I_\pm} q_I^2, \qquad \mp \alpha > 0
\end{align}
regardless of $\xi$. This, with \eqref{eq:index_WCP2}, confirms the result \eqref{eq:index_WCP}. Our result also reproduces that of \cite{Gaiotto:2018yjh} for $N_f = 1$, up to extra overall -1.

\subsubsection*{Some Dualities of $d=3$ $\CN=1$ Gauge Theories}

Now we explore dualities between 3d $\CN=1$ gauge theories
with the computation technique of indices we have used throughout the paper.
Before proceeding, let us remark that we focus on dualities each side of which is either a $SO(2)$ gauge theory or a Wess-Zumino model \cite{Benini:2018umh}.
Earlier discussion on $d=3$ $\cN=1$ dualities has been made in \cite{Gremm:1999su, Gukov:2002es, Armoni:2009vv} for a last few decades.
For recent discoveries of wider variety of $d=3$ $\CN=1$ dualities involving more complicated Wess-Zumino type interactions or non-Abelian gauge groups, we refer the readers
to \cite{Benini:2018bhk, Bashmakov:2018wts, Gaiotto:2018yjh, Choi:2018ohn}.

We start from the following duality between an $\mathcal N = 1$ QED and a Wess-Zumino model:
\begin{align}
\label{eq:duality0}
\begin{split}
\begin{tabular}{@{}c@{}} \text{$SO(2)_{\frac1 2}$ with $1$ flavor $\vec Q$}  \, \\ $W = \frac{m}{2} |\vec Q|^2-\frac{1}{4} |\vec Q|^4$  \, \end{tabular}
\longleftrightarrow
\begin{tabular}{@{}c@{}}  \text{\, WZ model with $P, \, \vec X$}  \\  \, $ W = P \left(|\vec X|^2+m\right)  - \frac{1}{3} P^3 $
\end{tabular}
\end{split}
\end{align}
where $m$ is the relevant deformation parameter.
For the QED with the quartic superpotential,
we have a general expression of the index in \eqref{eq:index_quartic}, which leads to the vanishing index for this example:
\bea
\cI^{3d}_{\text{quartic } SO(2)_{1/2}}= \kappa + \half |q|^2 \text{sgn}(c_W) = \half - \half = 0 \,.
\eea
Although this result is independent of the relevant deformation parameter $m$, it is interpreted as the Witten index of the theory only when $m > 0$ because we have the $S^1$ moduli space of vacua when $m < 0$.
On the right hand side, the index of the WZ can be obtained in various ways as explained in section \ref{sec:matter}. We have
\bea
\cI^{3d}_{\text{WZ}} = 0 \,,
\eea
which is also independent of $m$.

It was discussed in \cite{Benini:2018umh} that one can derive three different dualities involving Abelian gauge theories from the duality \eqref{eq:duality0}. The theories in \eqref{eq:duality0} have a $SO(2)$ global symmetry, which can be gauged with an extra Chern-Simons interaction. Turing on the Chern-Simons level $l = 1, \, 0, \, -1$, one has three different dualities as follows.

For $l = 1$, we have the following $\CN=1$ duality involving a pair of $d=3$ Abelian gauge theories, which have {\it quartic} and {\it critical superpotentials}\footnote{This name came from \cite{Benini:2018umh} whose authors were inspired by $d=3$ bosonization, which maps the critical scalar theory (with $\phi^4$ quartic interaction) to regular fermions. Note that the superpotential of (RHS) in \eqref{eq:duality1} induces quartic scalar interaction $|\vec x|^4$ to its bosonic potential.} respectively.
\begin{align}
\label{eq:duality1}
\begin{split}
\begin{tabular}{@{}c@{}} \text{$SO(2)_{-\frac1 2}$ with $1$ flavor $\vec Q$}  \, \\ $W= \frac{m}{2} |\vec Q|^2-\frac{1}{4} |\vec Q|^4$  \, \end{tabular}
\longleftrightarrow
\begin{tabular}{@{}c@{}}  \text{\, $SO(2)_{\frac 1 2}$ with $1$ flavor $\vec X$ and a singlet $P$}  \\  \, $ W = P \left(|\vec X|^2+m\right)  - \frac{1}{3} P^3 $  \,. \end{tabular}
\end{split}
\end{align}
The semi-classical vacuum structure was drawn in \cite{Benini:2018umh} and it is consistent with the path-integral computation below. For the quartic theory, we again use the formula \eqref{eq:index_quartic}, which reads
\bea
\label{eq:duality_quartic}
\cI^{3d}_{\text{quartic } SO(2)_{-1/2}}= \kappa + \half |q|^2 \text{sgn}(c_W) = - \half - \half = -1 \,.
\eea
For the critical theory on the right hand side, the index is given by
\begin{align}
\label{eq:CS}
\cI^{3d} = \kappa \cdot \mathcal Z_\kappa+\Omega_0
\end{align}
where one can obtain each term from \eqref{eq:pKLF-Z_kappa} and from \eqref{eq:pKLF-Omega0} respectively.
As a result, we have the index of the critical theory as follows:
\bea
\cI^{3d}_{\text{critical } SO(2)_{1/2}} = 0 \cdot \kappa + \left(-\half\right) - \half = -1 \,,
\eea
which agrees with \eqref{eq:duality_quartic}.

For $l = 0$, we have a duality between a gapped theory of massive fields and an Abelian gauge theory:
\begin{align}
\label{eq:duality2}
\begin{split}
\begin{tabular}{@{}c@{}} \text{Massive field $\vec Q$}  \, \\ $W= \frac{m}{2} |\vec Q|^2$  \, \end{tabular}
\longleftrightarrow
\begin{tabular}{@{}c@{}}  \text{\, $SO(2)_{-\frac 1 2}$ with $1$ flavor $\vec X$ and a singlet $P$}  \\  \, $ W = P \left(|\vec X|^2+m\right) + \frac{1}{2} P^2 $  \,. \end{tabular}
\end{split}
\end{align}
Note that the superpotential on the right hand side is not the critical one; it has a quadratic interaction for $P$. Indeed, those theories have the $\mathcal N = 2$ supersymmetry. For $m = 0$, we have a non-compact moduli space of vacua, which does not have a well-defined Witten index. Hence, we assume $m \neq 0$ for the computations below. The index of a single massive field is $\text{sgn}(m)$. Assuming $m \neq 0$, the index of the left hand side is given by
\begin{align}
\cI^{3d}_{\text{massive}} = \text{sgn}(m) \times \text{sgn}(m) = 1
\end{align}
because we have two massive fields. On the right hand side, we have to take into account the one-loop correction to the superpotential. The one-loop corrected superpotential is given by
\begin{align}
W_{\text{one-loop}} = P \left(|\vec X|^2+m\right) + \frac{1}{2} P^2 - \frac{1}{2} P |P| \,.
\end{align}
Using the language of $W = P K(\rho)+L(\rho)+F(P)$, the index is determined by the following asymptotic values:
\begin{align}
\begin{aligned}
\lim_{\rho \rightarrow \infty} K(\rho) &= \infty \,, \\
\lim_{P \rightarrow +\infty} F'(P) &= m \,, \\
\lim_{P \rightarrow -\infty} F'(P) &= -\infty
\end{aligned}
\end{align}
with $\lim_{p \rightarrow p^*} F'(p)$ negligible. In order to have an integer index, as before, we have to scale $m \rightarrow C m$ with $C \rightarrow \infty$. For $m > 0$, we read the index from the formula \eqref{eq:CS}:
\bea
\cI^{3d}_{\text{quadratic } SO(2)_{1/2}} = 1 \cdot \kappa + 0 - \half = -1 \,.
\eea
For $m < 0$, we have
\bea
\cI^{3d}_{\text{quadratic } SO(2)_{1/2}} = 0 \cdot \kappa + \left(-\half\right) - \half = -1 \,.
\eea
Thus, the index of the right hand side is -1 regardless of $m$. The Witten indices of the duality pair agree with each other up to sign. This sign discrepancy is explained by the fact that, in reaching the dual pairs, we have integrated out
an $SO(2)$ gauge multiplet on the left hand side of \eqref{eq:duality2}. The indices before and after this step
differ by a sign due to the extra gaugino in this example and the one below. Alternatively,
we can choose to keep this $SO(2)$ and match the sign as well. Recall that something similar
happens when we start with $\cN=2$ Chern-Simons and choose to integrate out
the $\Sigma$ part of the $\cN=2$ vector multiplet.

The last duality with $l = -1$ is the following:
\begin{align}
\begin{split}
\begin{tabular}{@{}c@{}} \text{$SO(2)_{\frac3 2}$ with $1$ flavor $\vec Q$}  \, \\ $W= \frac{m}{2} |\vec Q|^2-\frac{1}{4} |\vec Q|^4$  \, \end{tabular}
\longleftrightarrow
\begin{tabular}{@{}c@{}}  \text{\, $SO(2)_{- \frac 3 2}$ with $1$ flavor $\vec X$ and a singlet $P$}  \\  \, $ W = P \left(|\vec X|^2+m\right)  - \frac{1}{3} P^3 $   \,. \end{tabular}
\end{split}
\end{align}
For the quartic theory on the left hand side, we have
\bea
\cI^{3d}_{\text{quartic } SO(2)_{3/2}} = \kappa + \half |q|^2 \text{sgn}(c_W) = \frac{3}{2} - \half =1 \,.
\eea
The index of of the critical theory on the right hand side reads
\bea
\cI^{3d}_{\text{critical } SO(2)_{-3/2}} = 0 \cdot \kappa + \left(-\half\right) - \half = - 1 \,.
\eea
The Witten indices agree with each other, again up to sign which is due to extra $SO(2)_{-1}$ integrated out.
\\

\vskip 1cm \centerline{\large \bf Acknowledgment} \vskip 0.5cm
We would like to thank Vladimir Bashmakov for useful discussions.
CH is grateful to KIAS for the kind hospitality where part of this work was done. CH is partially supported by the ERC-STG grant 637844-HBQFTNCER and by the INFN.
This work is also supported in part by KIAS Individual Grants (PG071301 for DG and PG005704 for PY) at Korea Institute for Advanced Study.

\appendix

\section{Heat Kernel for a Landau Problem}

Let us compute the analog of
\bea
\langle X'\vert e^{s\partial^2/2}\vert X\rangle=\frac{1}{2\pi s}e^{-(X-X')^2/2s}
\eea
for $\IR^2$, when a uniform magnetic field is turned on, i.e., for
\bea
-\frac12 \partial^2\quad\rightarrow\quad H_B\equiv \frac12\left((i\partial_x -yB/2)^2+(i\partial_y + xB/2)^2\right)
\eea
with $B$ positive. The Hamiltonian can be rewritten as
usual as
\bea
H_B= B\left(a^\dagger a+\frac12\right), \qquad [a,a^\dagger]=1\ ,
\eea
via a Harmonic oscillator,
\bea
a\equiv i\sqrt{\frac{2}{B}}\left( \partial_{\bar z} +Bz/4\right),\quad
a^\dagger \equiv i\sqrt{\frac{2}{B}}\left( \partial_{z} - B\bar z/4\right)
\eea
where $z=x+iy$ and $\partial_z=(\partial_x-i\partial_y)/2$ etc.
The lowest Landau level wavefunctions at $H_B=B/2$ are
\bea
\Psi_m^{(0)}(x,y) &=&\sqrt{\frac{(B/2)^{m+1}}{\pi m!}} z^m e^{-Bz\bar z/4} \cr\cr
&=& \sqrt{\frac{(B/2)^{m+1}}{\pi m!}} \left(-\frac{4}{B}\partial_{\bar z}\right)^m e^{-Bz\bar z/4}
\eea
labeled by the eigenvalues $m\ge 0$ of the conserved angular
momentum,
\bea
L= -ix\partial_y+iy\partial_x \ .
\eea
Higher Landau level wavefunctions are generated as
\bea
\Psi^{(n)}_m(x,y) = \frac{1}{\sqrt{n!}}(a^\dagger)^n\Psi_m^{(0)}(x,y) \ ,\quad n\ge 0
\eea
with energy eigenvalues $(n+1/2)B$.

The  heat kernel of $H_B$, necessary for the computation of the path integral, is
\bea
\langle W\vert e^{-s H_B}\vert Z\rangle&=&\sum_{n=0}^\infty e^{-s(n+1/2)B} \Delta^{(n)}(W,Z) \cr\cr
&&\Delta^{(n)}(W,Z)\equiv \sum_{m=0}^\infty \Psi^{(n)}_m(W)^*\Psi^{(n)}_m(Z) \ .
\eea
The first contribution from $n=0$ is easy enough,
\bea
\Delta^{(0)}(W,Z)
= \frac{B}{2\pi}  e^{ 8\partial_{\bar z}\partial_w/B} e^{ -B(z\bar z+w\bar w)/4}=
\frac{B}{2\pi} e^{ -B(z\bar z+w\bar w-2z\bar w)/4} \ .
\eea
For higher levels $n\ge 1$, we find
\bea
\Delta^{(n)}(W,Z)= \frac{1}{n!}\left(\frac{2}{B}\right)^n (\partial_z-B\bar z/4)^n(\partial_{\bar w}-B w/4)^n \Delta^{(0)}(W,Z)
\eea
which brings us to
\bea
\langle W\vert e^{-s H_B}\vert Z\rangle
=\left[\sum_{k=0}^\infty
\frac{1}{(k!)^2}\cdot\left(\frac{e^{-sB}}{e^{-sB}-1}\cdot\frac{B|z-w|^2}{2}\right)^k \right]\cdot\frac{\Delta^{(0)}(W,Z)}{2\sinh(sB/2)} \ .
\eea
The sum in the square bracket can be traded off in favor
of a single integral. It reduces to, in the small $s$ limit which
is relevant for the heat kernel expansion,
\bea
\sim \frac{1}{2\pi}\int_0^{2\pi} d\theta \;e^{2i\cos\theta\;  \sqrt{|z-w|^2/2s}}\ .
\eea
playing the role of $\sim e^{-|X-X'|^2/2s}$ of flat $\IR^2$.
The other factor similarly replaces $1/2\pi s$.

These expressions say that, again, the heat kernel expansion
becomes an expansion in positive powers of $s$, and the coincident limit
is all that matters for the index-like computation. In the coincident
limit $Z=W$, on the other hand, the  heat kernel collapses to
\bea
\langle Z\vert e^{-s H_B}\vert Z\rangle 
\;\;=\;\;\frac{B}{2\pi}\cdot \frac{1}{2\sinh(sB/2)}\ ,
\eea
which in the small $s$ limit again reduces to
\bea
\simeq \frac{1}{2\pi s}\ ,
\eea
as is necessary for the usual heat kernel expansion for $\Omega$
to be applicable.

\section{Pfaffian vs. Jacobian}\label{sec:pf-jac-proof}

In this appendix, we derive the cancelation between the Pfaffian and the Jacobian given in \eqref{eq:cancelation0}. Note that, as a matter of notational ease, we now use $\vec x_I = (x_{I(1)}, x_{I(2)} ) \rightarrow (x_I,x'_I)$ and similarly $\vec y_I =  (y_I, y'_I)$ for Gaussian integral variables.

First, let us consider the case of no neutral field. We want to show
\begin{align}\label{eq:pfjac-id1}
\text{Pf}(\tilde M_\kappa) = -\text{Det} \left(\frac{\partial (y_I, y'_I,{\bf v})}{\partial (x_I,x'_I,{\bf u})}\right).
\end{align}
for a radial superpotential.
Let us examine the Pfaffian of fermion bilinear $\text{Pf}(\tilde M_\kappa)$ on the left hand side of \eqref{eq:pfjac-id1}. We first note that $\tilde M_\kappa$ is written in the following block form:
\begin{align}
\tilde M_\kappa = \left(\begin{array}{cc}
L & Q \\
-Q^T & N
\end{array}\right).
\end{align}
Each block is further decomposed into $2 \times 2$ blocks, each of which is labelled by doublet labels $I, J \in \{ 1, \cdots, N_f \} $;
\footnote{
Throughout this appendix, we choose as a matter of computational convenience
$$\left[ \mathcal D \psi \cD \lambda \right]^{\text{(appendix)}} \equiv \left( d\lambda^a  d\lambda_a \right)
\left( \prod_{I=1}^{N_f} d\psi_{I(1)}^{b} d\psi_{I(2)}^{c} \right)
\left( \prod_{J=1}^{N_f} d\psi_{J(1),b} d\psi_{J(2),c} \right) $$
where super- and subscript $a \,, b$ and $c$'s are spinor indices.
This differs from our convention in the main text,
$$\left[ \cD \psi \cD \lambda  \right]^{\text{(main text)}} \equiv - \left( d \lambda^a d\lambda_a \right) \left( \prod_{I=1}^{N_f}  d \psi^b_{I(1)} d \psi_{I(1),b} d \psi^c_{I(2)} d \psi_{I(2),c} \right) $$
which translates to a relative sign factor $(-1)^{N_f+1}$.}
For example, the top-left block $L$ is written as $L = (L_{IJ})$ where $L_{IJ}$ is a $2 \times 2$ matrix defined by
\begin{align}
L_{IJ} =  \delta_{IJ} \left(\begin{array}{cc}
0 & q_I (u_x+i u_y) \\
-q_I (u_x+i u_y) & 0
\end{array}\right)  .
\end{align}
In the same manner, $Q = (Q_{IJ})$ and $N = (N_{IJ})$ are also given by
\begin{align}
Q_{IJ} &= \left(\begin{array}{cc}
\partial_I \partial_J W & -\delta_{IJ} q_I u_z+\partial_I \partial'_J W \\
\delta_{IJ} q_I u_z+\partial'_I \partial_J W & \partial'_I \partial'_J W
\end{array}\right), \\
N_{IJ} &= \delta_{IJ} \left(\begin{array}{cc}
0 & - q_I (u_x-i u_y) \\
q_I (u_x-i u_y) & 0
\end{array}\right).
\end{align}
where $\partial_I \equiv \frac{\partial}{\partial x_I}$ and $\partial'_I \equiv \frac{\partial}{\partial x'_I}$.

Since $L$ is an $ (2N_f) \times (2N_f) $ invertible matrix for generic $\bf u$,
its inverse also reads in a block-diagonal form $L^{-1} = \left((L^{-1})_{IJ}\right) $,
\begin{gather}
(L^{-1})_{IJ} = \delta_{IJ} \left(\begin{array}{cc}
0 & -\frac{1}{q_I (u_x+i u_y)} \\
\frac{1}{q_I (u_x+i u_y)} & 0
\end{array}\right).
\end{gather}
Now the Pfaffian of bilinear $\text{Pf}(\tilde M_\kappa)$ can be decomposed
\begin{align} \label{pf-minor}
\text{Pf}(\tilde M_\kappa) = \text{Pf}(L) \, \text{Pf}(N+Q^T L^{-1} Q) \,.
\end{align}
An auxiliary matrix $\hat N$ is defined
\begin{align}
\hat N = (u_x+i u_y) \left(N+Q^T L^{-1} Q\right) \,,
\end{align}
which is also in a block form $\hat N = (\hat N_{IJ})$, whose component for ``radial'' superpotential is given by
\begin{align} \label{Nauxil}
\hat N_{IJ} = \left(\begin{array}{cc}
\begin{array}{l}
\left(\frac{1}{q_J} x_I x'_J W_J-\frac{1}{q_I} x'_I x_J W_I\right) W_{IJ}
\end{array} &
\begin{array}{l}
-\delta_{IJ} \frac{1}{q_I} \left(q_I^2 {\bf u}^2 +W_I^2\right) \\
-\left( \frac{1}{q_J} x_I x_J W_J+\frac{1}{q_I} x'_I x'_J W_I\right) W_{IJ}
\end{array} \\
\begin{array}{l}
\delta_{IJ} \frac{1}{q_I} \left(q_I^2 {\bf u}^2 +W_I^2\right) \\
+\left(\frac{1}{q_J} x'_I x'_J W_J+\frac{1}{q_I} x_I x_J W_I\right) W_{IJ}
\end{array}
&
\begin{array}{l}
\left(\frac{1}{q_I} x_I x'_J W_I-\frac{1}{q_J} x'_I x_J W_J\right) W_{IJ}
\end{array} \end{array} \right).
\end{align}
In \eqref{Nauxil} we used the abbreviation for derivatives of superpotential: $W_I = \frac{\partial W}{\partial \rho_I}$ and $W_{IJ} = \frac{\partial^2 W}{\partial \rho_I \partial \rho_J}$.

According to \eqref{pf-minor} and the fact that the Pfaffian of $L$ is simply $\left(\prod_{I = 1}^{N_f} q_I (u_x+i u_y)\right)$, the Pfaffian of $\tilde M_\kappa$ is given by
\begin{align}
\label{eq:MN}
\text{Pf}(\tilde M_\kappa) = \left(\prod_{I = 1}^{N_f} q_I\right) \text{Pf} (\hat N) \,.
\end{align}

Next let us move on to the Jacobian $\text{Det}\left(\frac{\partial (y_I, y'_I,{\bf v})}{\partial (x_I,x'_I,{\bf u})}\right)$ in the right hand side of \eqref{eq:cancelation0}. Recall that the new variables $\{ y_I, \, y'_I, \, {\bf v} \}$ are defined in terms of the original variables as follows:
\begin{align}
\begin{gathered}
y_I = x_I \sqrt{q_I^2 {\bf u}^2 +W_I^2} \,, \\
y'_I = x'_I \sqrt{q_I^2 {\bf u}^2 +W_I^2} \,, \\
{\bf v} = {\bf u}.
\end{gathered}
\end{align}
Its Jacobian matrix is written in a block form,
\begin{align}
J = \frac{\partial (y_I, y'_I,{\bf v})}{\partial (x_I,x'_I,{\bf u})} = (J_{IJ}) \,,
\end{align}
where each $J_{IJ}$ is a $2 \times 2$ matrix,
\begin{align}
\label{eq:jac}
J_{IJ} = \left(\begin{array}{cc}
\delta_{IJ} \sqrt{q_I^2 {\bf u}^2 +W_I^2}+\frac{x_I x_J W_I W_{IJ}}{\sqrt{q_I^2 {\bf u}^2 +W_I^2}} & \frac{x_I x'_J W_I W_{IJ}}{\sqrt{q_I^2 {\bf u}^2 +W_I^2}} \\
\frac{x'_I x_J W_I W_{IJ}}{\sqrt{q_I^2 {\bf u}^2 +W_I^2}} & \delta_{IJ} \sqrt{q_I^2 {\bf u}^2 +W_I^2}+\frac{x'_I x'_J W_I W_{IJ}}{\sqrt{q_I^2 {\bf u}^2 +W_I^2}}
\end{array}\right).
\end{align}

Another auxiliary matrix $\hat J$ is defined by
\begin{align}
\hat J = \hat N J^{-1} \,.
\end{align}
Surprisingly, $\hat J$ is given by $\hat J = (\hat J_{IJ})$ with
\begin{align}
\hat J_{IJ} = \frac{1}{q_J} \left(\begin{array}{cc}
\frac{x_I x'_J W_J W_{IJ}}{\sqrt{q_J^2 {\bf u}^2 +W_J^2}} & -\delta_{IJ} \sqrt{q_J^2 {\bf u}^2 +W_J^2}-\frac{x_I x_J W_J W_{IJ}}{\sqrt{q_J^2 {\bf u}^2 +W_J^2}} \\
\delta_{IJ} \sqrt{q_J^2 {\bf u}^2 +W_J^2}+\frac{x'_I x'_J W_J W_{IJ}}{\sqrt{q_J^2 {\bf u}^2 +W_J^2}} & -\frac{x'_I x_J W_J W_{IJ}}{\sqrt{q_J^2 {\bf u}^2 +W_J^2}}
\end{array}\right).
\end{align}
Furthermore, one can figure out how to relate $\hat J$ and $J$,
\begin{align}
\hat J = J^T S
\end{align}
where $ S = (S_{IJ})$ is a block matrix whose component is
\begin{gather}
S_{IJ} = \frac{\delta_{IJ}}{q_J} \left(\begin{array}{cc}
0 & -1 \\
1 & 0
\end{array}\right).
\end{gather}
Hence, the determinant of $\hat J$ is given by
\begin{align}
\text{Det} (\hat J) = \text{Det}(J) \, \text{Det} (S) = \frac{\text{Det}(J)}{\prod_{I = 1}^{N_f} q_I^2} \,,
\end{align}
Using this relation, one can also obtain the determinant of $\hat N$:
\begin{align}
\text{Det}(\hat N) = \text{Det}(\hat J) \, \text{Det}(J) = \left(\frac{\text{Det}(J)}{\prod_{I = 1}^{N_f} q_I}\right)^2
\end{align}
and moreover the Pfaffian:
\begin{align}
\label{eq:NJ}
\text{Pf}(\hat N) = \pm \frac{\text{Det}(J)}{\prod_{I = 1}^{N_f} q_I} \,,
\end{align}
where the sign is fixed by the recursive definition of Pfaffian.
By combining \eqref{eq:MN} and \eqref{eq:NJ}
\begin{align}
\text{Pf}(\tilde M_\kappa) = (-1)^{N_f} \text{Det}(J) \,.
\end{align}
Taking into account the convention of Pfaffian, we complete the proof of \eqref{eq:cancelation0}
\bea
\label{eq:cancelation}
\text{Pf} ( \tilde M_\kappa)^{\text{(main text)}} = (-1)^{N_f+1} \text{Pf} ( \tilde M_\kappa)^{\text{(appendix)}} = - \text{Det} (J) \,.
\eea

 Now let us move on to the case with a single neutral field $P$. In that case, the Pfaffian matrix becomes
\begin{align}
\tilde M_\kappa = \left(\begin{array}{cc|c}
L & Q & R \\
-Q^T & N & S \\
\hline
-R^T & -S^T & O
\end{array}\right)
\end{align}
where $R = (R_I)$ and $S = (S_I)$ are $2 N_f \times 2$ matrices defined by
\begin{align}
R_I &= \left(\begin{array}{cc}
0 & \partial_I W_p \\
0 & \partial'_I W_p
\end{array}\right), \\
S_I &= \left(\begin{array}{cc}
-\partial_I W_p & 0 \\
-\partial'_I W_p & 0
\end{array}\right)
\end{align}
with $I = 1, \dots, N_f$; and $O$ is a $2 \times 2$ matrix:
\begin{align}
O = \left(\begin{array}{cc}
0 & W_{pp} \\
-W_{pp} & 0
\end{array}\right).
\end{align}
Also $W_p$ and $W_{pp}$ are defined by $W_p = \frac{\partial W}{\partial p}$ and $W_{pp} = \frac{\partial^2 W}{\partial p^2}$. Note that we have chosen a basis of the matrix such that the terms involving the neutral field are placed at the last part of the columns and rows.

Assuming non-zero $W_{pp}$, which can be taken to be zero in the end, one can apply the identity \eqref{pf-minor}, with some exchanges of columns and rows understood, to this new $\tilde M_\kappa$. The Pfaffian of $\tilde M_\kappa$ is then decomposed into
\begin{align}
\text{Pf} (\tilde M_\kappa) = W_{pp} \times \text{Pf} \left(\begin{array}{cc}
L & Q-\hat O \\
-Q^T+\hat O^T & N
\end{array}\right).
\end{align}
where $\hat O = (\hat O_{IJ})$ is given by
\begin{align}
\hat O_{IJ} = -R_I O^{-1}_{IJ} S_J^T = \frac{1}{W_{pp}} \left(\begin{array}{cc}
\partial_I W_p \partial_J W_p & \partial_I W_p \partial'_J W_p \\
\partial'_I W_p \partial_J W_p & \partial'_I W_p \partial'_J W_p
\end{array}\right).
\end{align}
One can see that the presence of neutral field $P$ affects the above derivation for $\text{Pf} (\tilde M_\kappa)$ without the neutral field in such a way that $Q$ is just replaced by $Q-\hat O$; or equivalently
\begin{align}
W_{IJ} \qquad \rightarrow \qquad W_{IJ}-\frac{W_{pI} W_{pJ}}{W_{pp}} \,.
\end{align}

Next, the Jacobian matrix with the neutral field $P$ is given by
\begin{align}
J = \left(\begin{array}{c|c}
J^{(0)} & G \\
\hline
H & W_{pp}
\end{array}\right)
\end{align}
where $J^{(0)}$ is the Jacobian matrix without the neutral field; $G = (G_I)$ is a $2 N_f \times 1$ matrix:
\begin{align}
G_I = \left(\begin{array}{c}
\frac{x_I W_I W_{pI}}{\sqrt{q_I^2 {\bf u}^2+W_I^2}} \\
\frac{x'_I W_I W_{pI}}{\sqrt{q_I^2 {\bf u}^2+W_I^2}}
\end{array}\right)
\end{align}
and $H = (H_J)$ is a $1 \times 2 N_f$ matrix:
\begin{align}
H_J = \left(\begin{array}{cc}
x_J W_{pJ} & x'_J W_{pJ}
\end{array}\right)
\end{align}
with $I,J = 1, \dots, N_f$. Using the identity
\begin{align}
\text{Det}\left(\begin{array}{cc}
A & B \\
C & D
\end{array}\right) = \text{Det}(D) \, \text{Det}(A-BD^{-1} C) \,,
\end{align}
we obtains the Jacobian as follows:
\begin{align}
\text{Det} (J) = W_{pp} \times \text{Det} \left(J^{(0)}-\hat K\right)
\end{align}
where $\hat K = (\hat K_{IJ})$ is given by
\begin{align}
\hat K_{IJ} = \left(\begin{array}{cc}
\frac{x_I x_J W_I W_{pI} W_{pJ}}{W_{pp} \sqrt{q_I^2 {\bf u}^2+W_I^2}} & \frac{x_I x'_J W_I W_{pI} W_{pJ}}{W_{pp} \sqrt{q_I^2 {\bf u}^2+W_I^2}} \\
\frac{x'_I x_J W_I W_{pI} W_{pJ}}{W_{pp} \sqrt{q_I^2 {\bf u}^2+W_I^2}} & \frac{x'_I x'_J W_I W_{pI} W_{pJ}}{W_{pp} \sqrt{q_I^2 {\bf u}^2+W_I^2}}
\end{array}\right).
\end{align}
Recall $J^{(0)}$, which is given by \eqref{eq:jac}. One can see that again the presence of the neutral field replaces $W_{IJ}$ by $W_{IJ}-\frac{W_{pI} W_{pJ}}{W_{pp}}$ in the derivation of the Jacobian without the neutral field. Therefore, the proof of \eqref{eq:cancelation} above still holds with this replacement understood. With more than one neutral fields, one can perform the same procedure, which reduces the columns and rows involving neutral fields one by one.




\end{document}